\documentclass{amsart}
\usepackage{dsfont}
\usepackage{bm}
\usepackage[latin1]{inputenc}
\usepackage{graphicx}
\usepackage{pstricks}

\newcommand{\norm}[1]{ \left\| #1 \right\|}
\newcommand\R{{\ensuremath {\mathbb R} }}
\newcommand\C{{\ensuremath {\mathbb C} }}
\newcommand\N{{\ensuremath {\mathbb N} }}
\newcommand\Z{{\ensuremath {\mathbb Z} }}

\newcommand\1{{\ensuremath {\mathds 1} }}
\newcommand\gH{\mathfrak{H}}
\newcommand\cC{\mathcal{C}}
\newcommand\cP{\mathcal{P}}
\newcommand\cD{\mathcal{D}}
\newcommand\cE{\mathcal{E}}
\newcommand\ii{\infty}
\newcommand{\gS}{\mathfrak{S}}
\newcommand{\alp}{\bm{\alpha}}
\newcommand\be{\begin{equation}}
\newcommand{\beq}{\begin{eqnarray}}
\newcommand{\eeq}{\end{eqnarray}}
\newcommand{\bq}{\begin{equation}}
\newcommand{\eq}{\end{equation}}
\newcommand{\Sum}{\displaystyle \sum}

\newcommand{\Int}{\displaystyle \int}
\newcommand{\Frac}{\displaystyle \frac}

\newcommand{\Sup}{\displaystyle \sup}

\newcommand{\tr}{\text{tr}}
\renewcommand\phi{\varphi}
\newcommand\BbC{\C}
\newcommand\BbR{\R}
\newcommand\un{\1}

\def\square{\sqcup\!\!\!\!\sqcap}

\newtheorem{theorem}{Theorem}
\newtheorem{lemma}[theorem]{Lemma}
\newtheorem{proposition}[theorem]{Proposition}

\newtheorem{remark}[theorem]{Remark}

\theoremstyle{definition}

\begin{document}

\title[Variational methods in relativistic quantum mechanics]{Variational methods in relativistic quantum mechanics}
\date{January 8, 2008. Final version to appear in \textit{Bull. Amer. Math. Soc.}}

\author{Maria J. ESTEBAN}
\address{CNRS and Ceremade (UMR 7534), Universit{\'e} Paris-Dauphine, Place
du Mar{\'e}chal de Lattre de Tassigny, 75775 Paris Cedex 16 - France}
\email{esteban@ceremade.dauphine.fr}

\author{Mathieu LEWIN}
\address{CNRS and Laboratoire de Mathématiques (UMR 8088), Universit{\'e} de Cergy-Pontoise, 2, avenue Adolphe Chauvin, 95 302 Cergy-Pontoise Cedex - France}
\email{Mathieu.Lewin@math.cnrs.fr}

\author{Eric S\'ER\'E}
\address{Ceremade (UMR 7534), Universit{\'e} Paris-Dauphine, Place
du Mar{\'e}chal de Lattre de Tassigny, 75775 Paris Cedex 16 - France}
\email{sere@ceremade.dauphine.fr}



\begin{abstract}
This review is devoted to the study of stationary solutions of linear and nonlinear equations from relativistic quantum mechanics, involving the Dirac operator. The solutions are found as critical points of an energy functional. Contrary to the Laplacian appearing in the equations of nonrelativistic quantum mechanics, the Dirac operator has a negative continuous spectrum which is not bounded from below. This has two main consequences. First, the energy functional is strongly indefinite. Second, the Euler-Lagrange equations are linear or nonlinear eigenvalue problems with eigenvalues lying in a spectral gap (between the negative and positive continuous spectra). Moreover, since we work in the space domain $\R^3$, the Palais-Smale condition is not satisfied. For these reasons, the problems discussed in this review pose a challenge in the Calculus of Variations. The existence proofs involve sophisticated tools from nonlinear analysis and have required new variational methods which are now applied to other problems.

In the first part, we consider the \emph{fixed eigenvalue problem} for models of a free self-interacting relativistic particle. They allow to describe the localized state of a spin-$1/2$ particle (a fermion)  which propagates without changing its shape. This includes the Soler models, and the Maxwell-Dirac or Klein-Gordon-Dirac equations.

The second part is devoted to the presentation of min-max principles allowing to characterize and compute the eigenvalues of linear Dirac operators with an external potential, in the gap of their essential spectrum. Many consequences of these min-max characterizations are presented, among them a new kind of Hardy-like inequalities and a stable algorithm to compute the eigenvalues.

In the third part we look for \emph{normalized solutions} of nonlinear eigenvalue problems. The eigenvalues are \emph{Lagrange multipliers}, lying in a spectral gap.  We review the results that have been obtained on the Dirac-Fock model which is a nonlinear theory describing the behavior of $N$ interacting electrons in an external electrostatic field. In particular we  focus on the problematic definition of the ground state and its nonrelativistic limit. 

In the last part, we present a more involved relativistic model from Quantum Electrodynamics in which the behavior of the vacuum is taken into account, it being coupled to the real particles. The main interesting feature of this model  is that the energy functional is now bounded from below, providing us with a  good definition of a  ground state. 
\end{abstract}

\maketitle

\tableofcontents

\section*{Introduction}
In this paper, we present various recent results  concerning some linear and nonlinear variational problems in relativistic quantum mechanics,  involving the Dirac operator.

Dirac derived his operator in 1928 \cite{Dirac-28}, starting from the usual classical expression of the energy of a free relativistic particle of momentum $p\in\R^3$ and mass $m$
\begin{equation}
E^2=c^2|p|^2+m^2c^4
\label{classique}
\end{equation}
($c$ is the speed of light), and imposing the necessary relativistic invariances. 
By means of the usual identification 
$$p\longleftrightarrow-i\hbar\nabla$$
where $\hbar$ is Planck's constant, he found that an adequate observable for describing the energy of the free particle should therefore be a self-adjoint operator $D_c$ satisfying the equation
\begin{equation}
(D_c)^2=-c^2\hbar^2\Delta+m^2c^4.
\label{carre}
\end{equation}
Taking the locality principle into account, Dirac proposed to look for a local  operator which is first order with respect to $p=-i\hbar\nabla$:
\begin{equation}
D_c \ = -ic\hbar\; \alp\cdot\nabla + mc^2\beta = \ - ic\hbar\
\sum^3_{k=1} {\bf \alpha}_k \partial _k \ + \ mc^2{\bf \beta},
\label{def_Dirac}
\end{equation}
where $\alpha_1$, $\alpha_2$, $\alpha_3$ and $\beta$ are hermitian matrices which have to satisfy the following anticommutation relations:
\begin{equation} \label{CAR}
\left\lbrace
\begin{array}{rcl}
 {\alpha}_k
{\alpha}_\ell + {\alpha}_\ell
{\alpha}_k  & = &  2\,\delta_{k\ell}\,\un,\\
 {\alpha}_k {\beta} + {\beta} {\alpha}_k
& = & 0,\\
\beta^2 & = & \un.
\end{array} \right. \end{equation}
It can be proved \cite{Thaller-92} that the smallest dimension in which \eqref{CAR} can take place is 4 (i.e. $\alpha_1$, $\alpha_2$, $\alpha_3$ and $\beta$ should be $4\times4$ hermitian matrices), meaning that $D_c$ has to act on $L^2(\R^3,\C^4)$. The usual representation in $2\times 2$ blocks is given by 
$$ \beta=\left( \begin{matrix} I_2 & 0 \\ 0 & -I_2 \\ \end{matrix} \right),\quad \; \alpha_k=\left( \begin{matrix}
0 &\sigma_k \\ \sigma_k &0 \\ \end{matrix}\right)  \qquad (k=1, 2, 3)\,,
$$
where the Pauli matrices are defined as
$$\sigma _1=\left( \begin{matrix} 0 & 1
\\ 1 & 0 \\ \end{matrix} \right),\quad  \sigma_2=\left( \begin{matrix} 0 & -i \\
i & 0 \\  \end{matrix}\right),\quad  \sigma_3=\left( 
\begin{matrix} 1 & 0\\  0 &-1\\  \end{matrix}\right) \, .$$

By \eqref{classique}, the time-dependent Dirac equation describing the evolution of a free particle is 
\begin{equation} 
i\hbar\frac{\partial}{\partial t} \Psi = D_c \Psi.
\label{dir-evol}\end{equation} 
This equation has been successfully used in physics to describe relativistic particles having a spin $1/2$. 

The main unusual feature of the Dirac equation is the spectrum of $D_c$ which is not bounded from below:
\begin{equation}
\sigma(D_c)=(-\infty, -mc^2]\cup[mc^2, \infty).
\label{spectre_Dirac}
\end{equation}
Compared with non-relativistic theories in which the Schrödinger operator $-\Delta/(2m)$ appears instead of $D_c$,  property \eqref{spectre_Dirac} leads to important physical, mathematical and numerical difficulties. Indeed, if one simply replaces $-\Delta/(2m)$ by $D_c$ in the energies or operators which are commonly used in the non-relativistic case, one obtains energies which are not bounded from below.

Although there is no observable electron of negative energy, the negative spectrum plays an important role in physics. Dirac himself suspected that the negative spectrum of his operator could generate new interesting physical phenomena, and he proposed in 1930 the following interpretation \cite{Dirac-30,Dirac-34a,Dirac-34b}:
\begin{quotation}
``We make the assumption that, in the world as we know it, nearly all the states of negative energy for the electrons are occupied, with just one electron in each state, and that a uniform filling of all the negative-energy states is completely unobservable to us." \cite{Dirac-34b} 
\end{quotation}
Physically, one therefore has to imagine that the vacuum (called the \emph{Dirac sea}) is filled with infinitely many virtual particles occupying the negative energy states. With this conjecture, a real free electron cannot be in a negative state due to the Pauli principle which forbids it to be in the same state as a virtual electron of the Dirac sea.

With this interpretation, Dirac was able to conjecture the existence of ``holes" in the vacuum, interpreted as ``anti-electrons" or positrons, having a positive charge and a positive energy. The positron was discovered in 1932 by Anderson \cite{Anderson-33}. Dirac also predicted the phenomenon of  vacuum polarization: in the presence of an electric field, the virtual electrons are displaced, and the vacuum acquires a nonconstant density of charge. All these phenomena are now well known and well established in physics. They are direct consequences of the existence of the negative spectrum of $D_c$, showing the crucial role played by Dirac's discovery.

Actually, in practical computations it is quite difficult to deal properly with the Dirac sea. As a consequence the notion of ``ground state" (state of ``lowest energy" which is supposed to be the most ``stable" for the system under consideration) is problematic for many of the models found in the literature. Numerically, the unboundedness from below of the spectrum is also the source of important practical issues concerning the convergence of the considered algorithms, or the existence of spurious (unphysical) solutions.

Dirac's interpretation of the negative energies will be an implicit assumption in all this review in the sense that we shall (almost) always look for positive energy solutions for the electrons. In the last section, we present a model from Quantum Electrodynamics (QED) in which the nonlinear behavior of the Dirac sea will be fully taken into account.

\medskip

Mathematically, most of the energy functionals that we shall consider are strongly indefinite: they are unbounded from below and all their critical points have an infinite Morse index.
Note that the mathematical methods allowing to deal with strongly indefinite functionals have their origin in the works of P. Rabinowitz concerning the study of nonlinear waves \cite{Rabinowitz-78} and also the study of periodic solutions of Hamiltonian systems \cite{Rabinowitz-78bis}.
Many results have followed these pioneering works, and powerful theories have been devised, in particular in the field of periodic orbits of Hamiltonian systems: the linking theorem of Benci-Rabinowitz \cite{Benci-Rabinowitz-79}, Clarke-Ekeland's dual action functional \cite{Clarke-Ekeland-80}, Conley-Zehnder's relative index \cite{Conley-Zehnder-83}, Floer's homology \cite{Floer-87}...

Another difficulty with the models presented in this review is the lack of compactness: the Palais-Smale condition is not satisfied due to the unboundedness of the domain $\R^3$. Variational problems with lack of compactness also have been extensively studied. Let us mention the work of Sacks-Uhlenbeck \cite{Sacks-Uhlenbeck-80} on harmonic maps, Lieb's Lemma \cite{Lieb-83}, Brezis-Nirenberg's study of elliptic PDEs with critical exponents \cite{Brezis-Nirenberg-83}, the concentration-compactness method of P.-L. Lions \cite{Lions-84}, Bahri-Coron's critical points at infinity \cite{Bahri-Coron-88} and more recently Fang-Ghoussoub's Palais-Smale sequences with Morse information \cite{Fang-Ghoussoub-92,Ghoussoub-93}.

The combination of the two above types of difficulties poses a challenge in the Calculus of Variations. To prove the results presented in this review, it has been necessary to adapt some of the sophisticated tools mentioned above and to introduce new ideas. The novel variational methods that we have designed can be applied in general situations and in particular in the study of crystalline matter (nonlinear Schr\"odinger equations with periodic potentials).

\medskip

The review contains four different parts which are almost independent. The common feature of all the problems addressed in this review is the variational study of linear and nonlinear eigenvalue problems with eigenvalues in spectral gaps. In the nonlinear case, there are two different classes of problems. Either we fix the eigenvalue and let the $L^2$-norm of the solutions free. Or we look for normalized solutions, the eigenvalue is then a Lagrange multiplier which has to stay in the spectral gap. 

In the first section, we describe the results that have been obtained for some models describing one self-interacting free relativistic spin-$1/2$ particle. The simplest case is when the interaction is ``local", i.e. represented by a nonlinear function $F$ of the spinor $\psi(t,x)$ of the particle. The general form for the equations that we consider in this part is:
$$D_c\psi-\omega\psi=\nabla F(\psi).$$
These models are phenomenological. A Lorentz-invariant interaction term $F(\psi)$ is chosen in order to find a model of the free localized electron (or on another spin $1/2$ particle), which fits with experimental data (see \cite{Ranada-82}).

At the end of the first section, we present two nonlocal models: the Maxwell-Dirac and the Klein-Gordon-Dirac equations in which the electron interacts with its own electromagnetic field. The Maxwell-Dirac equations take the following form:
$$\left\{\begin{array}{l}
(D_c+v-\alp\cdot A)\psi=\omega\psi,\\
 -4\pi\Delta v  =  |\psi|^2,\\
 -4\pi\Delta A_k  =  (\psi,\alpha_k\psi)\,,\quad k=1,2,3.\\
\end{array}\right.$$
From a mathematical viewpoint, the equations considered in the first section are nonlinear eigenvalue problems, in which the eigenvalue is fixed in a spectral gap, but the $L^2$ norm of the solution is not known. They are the Euler-Lagrange equations of a strongly indefinite functional. Moreover, this functional does not satisfy the Palais-Smale condition and the classical Benci-Rabinowitz linking theorem \cite{Benci-Rabinowitz-79} cannot be applied. The solutions are obtained by a ``noncompact" linking argument inspired by the works of Hofer-Wysocki \cite{Hofer-Wysocki-90} and S\'er\'e \cite{Sere-95} on homoclinic orbits of Hamiltonian systems. An additional difficulty is that the nonlinearity can vanish even for very large values of $\psi$, and this makes the a priori bounds on Palais-Smale sequences very delicate.

The second section is devoted to the study of min-max principles allowing to characterize the eigenvalues of Dirac operators with an external potential $V$ in the gap of their essential spectrum. Such operators are commonly used to describe the dynamics of an electron which is subject to the action of an external electrostatic field with associated potential  $V$ (for instance an electron in the field created by a nucleus). It can also be used to describe many \emph{non-interacting} electrons. For  potentials $V$ satisfying appropriate assumptions, the spectrum of the perturbed Dirac operator $\,D_c+V\,$  takes the form
$$\sigma(D_c+V)=(-\infty, -mc^2]\cup\{\varepsilon_i\}_{i\in \N} \cup[mc^2, \infty)$$
where the $\,\varepsilon_i$'s are eigenvalues of finite multiplicity in $(-mc^2, mc^2)$, which can  only accumulate at the thresholds $-mc^2$ or $mc^2$ (see \cite{Berthier-Georgescu-83, Berthier-Georgescu-87}).
The min-max formulas presented in this section furnish a useful variational characterization of the $\varepsilon_i$'s and of the associated eigenfunctions:
$$(D_c+V)\phi_i=\varepsilon_i\phi_i.$$
The min-max formulas are general and can be used in other settings where eigenvalues in a gap of the essential spectrum have to be characterized or computed. Many consequences of the min-max principles are derived in this section, including an algorithm for the computation of the eigenvalues.

In Section 3, we present results concerning the Dirac-Fock model \cite{Swirles-35}, allowing to describe $N$ \emph{interacting} electrons in an external electrostatic field. This is a nonlinear model which is used in Quantum Chemistry to compute the state of such electrons in heavy atoms. The energy functional is strongly indefinite and therefore it is really not obvious how to find an adequate definition of the ground state, and to prove the existence of critical points in general. Explaining how this can be achieved is the main goal of the section. The  model consists of a system of $N$ coupled nonlinear equations, posed on $L^2(\R^3,\C^4)^N$:
$$D_{_{c,\Phi}}\,\phi_i=\varepsilon_i\phi_i,\qquad 0<\varepsilon_i<m c^2,$$
where 
$$D_{_{c,\Phi}}=D_c+V+\Gamma_{\Phi},$$
$\Gamma_{\Phi}$ being an operator depending nonlinearly on $\Phi:=(\phi_1,...,\phi_N)$ and which models the interactions between the electrons. The functions $\phi_i$'s, which are assumed to satisfy the constraints $\int_{\R^3}(\phi_i,\phi_j)=\delta_{ij}$, represent the states of the $N$ electrons. Being a system of nonlinear eigenvalue problems with eigenvalues in a spectral gap, the Dirac-Fock equations carry some similarity with the equations studied in Section 1. But there is a big difference: the $L^2$ norm of the solutions is now fixed \emph{a priori}, and the eigenvalues $\varepsilon_i$ are unknown Lagrange multipliers associated with these constraints. This makes the problem harder, the main difficulty being to keep the multipliers $\varepsilon_i$ in the interval $(0,mc^2)$. The positivity of $\varepsilon_i$ is obtained thanks to a new penalization method. See \cite{Buffoni-Esteban-Sere} for a generalization of this method, with applications to nonlinear periodic Schrödinger models for crystals. The inequality $\varepsilon_i<mc^2$ follows from Morse-type estimates, as in the existence proof of Lions for the nonrelativistic Hartree-Fock model \cite{Lions-87}. To obtain these Morse-type estimates, the easiest way is to use a general theorem of Fang-Ghoussoub \cite{Fang-Ghoussoub-92}. Note that, since the functional is strongly indefinite, one has to work in fact with a relative Morse index.

Finally, in the last section we present a more involved physical model in which the behavior of the electrons is coupled to that of the Dirac sea, in the presence of an external electrostatic field $V$. In this model, Dirac's interpretation of the negative energies is really taken into account: the vacuum is considered as being an unknown physical object which can react to an external stimulation. The important feature of the model will be that the energy functional is \emph{bounded from below}, as first proposed by Chaix and Iracane \cite{Chaix-Iracane-89}, showing the importance of the vacuum polarization effects. The main drawback will be that one necessarily has to deal with infinitely many interacting particles (the real ones and the virtual ones of the Dirac sea), which creates lots of mathematical difficulties. In particular, the main unknown of the model is, this time, an orthogonal projector $P$ of infinite rank. The optimal projector $P$ representing the ground state of the system is solution of a nonlinear equation of the form
\begin{equation}
P=\chi_{(-\infty, \mu]}(D_c+V+\Gamma'_{P})
\label{equation_model}
\end{equation}
where $\Gamma'_{P}$ is an operator depending on the projector $P$ and describing the interactions between all particles (the real and the virtual ones). We have used the standard notation $\chi_{I}(A)$ for the spectral projector of $A$ associated with the interval $I$.
Solutions of \eqref{equation_model} are obtained by a minimization principle, on a set of compact operators. One has to be very careful in the choice of this set, and in the definition of the energy. A serious difficulty is the presence of ultraviolet divergencies.

\medskip

\medskip

\noindent \textbf{Acknowledgment.}  First of all, M.J.E. and E.S. would like to thank T. Cazenave, I. Ekeland, P.-L. Lions and L. V\'azquez for having introduced them to this field.  We would like also to thank J. Dolbeault, V. Georgiev, C. Hainzl, M. Loss and J.-P. Solovej with whom we have collaborated on differents parts of the works described in this review. Some years ago P. Chaix was very patient in explaining to us some models of QED on which we have worked later; the last section of this paper has its origin in those conversations. Several discussions with B. Buffoni  were very inspiring to us when looking for new variational principles in the case of strongly indefinity functionals. We want also to acknowledge our participation in the european research projects ``Analysis and Quantum'' (HPRN-CT-2002-00277) and ``HYKE" (HPRN-CT-2002-00282). The interaction with our partners in those two projects has been a source of inspiration and motivation for us along the years and we thank them for that. 
We also acknowledge support from the ANR ``ACCQUAREL'' of the french ministry of research.
E.S. was partially supported by the ``Institut Universitaire de France".
Finally, we wish to thank P. Rabinowitz, who has been very patient encouraging us to write this review article.

\medskip

\subsection*{Notations and basic properties of the Dirac operator}\ \\
Before going further, let us fix some notations.  All throughout this review,
the conjugate of $ z \in \BbC\ $ will be denoted by $z^*$.  For $X=(z_1,...,z_4)^T$ a column vector in $\BbC^{\,4}\,$, we denote by $X^*$ the row 
co-vector $(z^\ast_1, ..., z^\ast_4)$. Similarly, if $ {A}= (a_{ij})$ is a $4 \times 4$ complex
matrix, we denote by ${A}^\ast$ its adjoint, $({A}^\ast)_{ij} = a^\ast_{ji}$. 

We denote by $( X , X' )$ the Hermitian product of two vectors $X, \ X'$ in 
$\BbC^{\,4}$, and  by $|X|$, the canonical hermitian norm of $X$ in
$\BbC^4$, i. e. $|X|^2={\sum_{i=1}^4 X_i^*X_i}\;.$
The usual  Hermitian product in
$L^2(\BbR^3,\BbC^{\,4})$ is denoted as
\begin{equation} (\psi,
\psi')_{_{L^2}}
\ = \ \int_{\BbR^3} \ \Bigl( \psi(x), \psi'(x) \Bigr) \ d^3x.\end{equation}

For the sake of simplicity, we shall use a system of units in which 
$$m=\hbar=1.$$
 Actually, by scaling one can also fix  the value of another physical constant, like for instance the speed of light $c$ or the charge of an electron $e$. We shall use both possibilities in this review (they are of course equivalent). 

\medskip

Let us now list some basic and important properties of the free Dirac
operator. We refer to the book of Thaller \cite{Thaller-92} for details.

\begin{proposition}[Basic properties of the free Dirac operator]\label{R1}
The free Dirac operator $D_c$ is a self-adjoint operator on $L^2 (\BbR^3, \BbC^{\,4})$,
with domain $ H^1 (\BbR^3, \BbC^{\,4})$ and form-domain $ H^{1/2} (\BbR^3, \BbC^{\,4})$. Its spectrum is purely continuous, $\sigma(D_c)=(- \infty, -c^2 \rbrack\cup \lbrack c^2, + \infty)$. Moreover, there are two orthogonal projectors (both having infinite rank) from  $L^2 (\BbR^3, \BbC^{\,4})$ into itself, $P^0_{+,c}$ and $P^0_{-,c} =\1_{_{L^2}} - P^0_{+,c}$, such that
\begin{equation}\label{proj} \left\lbrace \begin{array}{l} D_c P^0_{+,c} \ = \ P^0_{+,c}D_c \ = \ \sqrt{c^4-c^2\,\Delta} \ \
P^0_{+,c} \ = \ P^0_{+,c} \ \sqrt{c^4-c^2\,\Delta} \\ D_c P^0_{-,c} \ = \ P^0_{-,c}D_c \ = \ - \sqrt{c^4-c^2\,\Delta} \ \ P^0_{-,c} \ = \
-P^0_{-,c} \ \sqrt{c^4-c^2\,\Delta}. \\
\end{array} \right. \end{equation}
The projectors $P^0_{+,c}$ and $P^0_{-,c}$ are multiplication operators in the Fourier domain, given by 
\begin{equation}
{ P^0_{\pm,c}}(p)=\frac{\pm
{D_c}(p)+\sqrt{c^2\,|p|^2+c^4}}{2\sqrt{c^2\,|p|^2+c^4}}.
\end{equation}
\end{proposition}

Note that Proposition \ref{R1} enables us to split the space
$$\gH := L^2(\BbR^3,\BbC^4)$$
as the direct sum of two infinite dimensional Hilbert spaces $\gH_{\pm,c}^0=P_{\pm,c}^0\gH$. The restriction of $D_c$ to $\gH_{\pm,c}^0$ is a self-adjoint operator in this subspace, with domain
$\gH_{\pm,c}^0\cap H^1(\BbR^3,\BbC^4)$. Furthermore, it will be convenient to use the following norm in $\gH$, equivalent to the usual norm of $H^{1/2}(\R^3,\C^4)$,
\begin{equation}\norm{\psi}_\gH:= \Bigl( \psi, (D_c\,P^0_{+,c} - D_c\,P^0_{-,c})
\psi
\Bigr)^{1/2}_{\gH \times \gH'} \ = \  \Bigl(\psi, |D_c| \psi \Bigr) ^{1/2}
\end{equation}
The subspaces $\gH_{+,c}^0\cap H^{1/2}(\R^3,\C^4)$ 
and $\gH_{-,c}^0\cap H^{1/2}(\R^3,\C^4)$ are orthogonal for this norm as well as for the $L^2$ norm.

When $c=1$, one recovers the usual $H^{1/2}$ norm. In this case, we shall use the convenient notation $P^0_{\pm,1}=P^0_{\pm}$ and $\gH^0_{\pm,1}=\gH^0_{\pm}$.

\section{Nonlinear Dirac equations for a free particle}
In this section, we present some nonlinear Dirac equations for a free particle. We therefore do not consider any external potential (but possibly a self-generated one). Stationary solutions of such equations represent the state of a localized particle which can propagate without changing its shape. 
The first to propose and study models for the description of this phenomenon were Ivanenko \cite{
Ivanenko-38}, Weyl \cite{Weyl-50} and 
Heisenberg  \cite{Heisenberg-53}. We refer to Ra\~nada \cite{Ranada-82} for a very interesting review on the historical background of this kind of models.

In this section, we shall always assume that $c=1$. A general form for the equations that we want to present  is 
\bq i{\partial_t}\Psi -D_1\Psi+\nabla F(\Psi )\; =\; 0\;,
\label{(2.1)}\eq
where $\Psi(t,\cdot)\in L^2(\R^3,\C^4)$.
Throughout this chapter we  assume that 
$\,F\in C^2(\BbC^{\,4},\BbR)\, $ satisfies
\bq F(e^{i\theta }\Psi )\; =\; F(\Psi )\qquad \hbox{for
all}\; \theta\in \BbR \; .
\label{(2.2)}\eq
The charge of the particle $-e$ does not appear since it is incorporated into the nonlinear functional $F$.

The relativistic invariance requirement imposes severe restrictions on the possible nonlinearities $F$. In two very interesting papers \cite{Finkelstein-Lelevier-Ruderman-51, Finkelstein-Fronsdal-Kaus-56}, Finkelstein {\sl et al } proposed various models for extended particles corresponding to various fourth order self-couplings $F$. In those papers, they gave some numerical description of the structure of the set of solutions, for different values of the parameters. Among the considered functions $F$, in \cite{Finkelstein-Lelevier-Ruderman-51, Finkelstein-Fronsdal-Kaus-56} we find the family of  general self-couplings
\bq  F_b(\Psi):=\lambda\left\{(\bar\Psi\Psi)^2+b(\bar\Psi\gamma^5\Psi)^2\right\} \label{finkel}\eq
where $\,\gamma^5=-i\alpha_1\alpha_2\alpha_3$,  $\,b\,$ is a real constant, and where we have used the notation
 $$\bar{\Psi }\Psi :=(\beta\Psi ,\Psi ).$$
In the sequel, without any loss of generality, we will assume that $\lambda=1/2$.
 
 \medskip
 
Stationary solutions of (\ref{(2.1)}) are functions of the type
\bq \Psi (t,x)\; =\; e^{-i\omega t} \psi (x),
\label{(2.3)} \eq
such that $\psi $ is a non-zero localized solution of the
following stationary nonlinear Dirac equation
\bq D_1\psi -\omega \psi -\nabla F(\psi )\;
=\; 0\qquad \hbox{in}\quad \BbR^3. \label{(2.4)}\eq
It is interesting to note that the latter equation has a variational structure: it is indeed the Euler-Lagrange equation associated with the functional 
\bq I^\omega (\psi )\; =\; \int_{\BbR^3}^{}\biggl (\frac{1}{2}( D_1\psi(x),\psi(x) )
-\frac{\omega}{2}\vert \psi(x) \vert ^2-F(\psi(x) )\biggr )
dx\,. \label{(2.5)}\eq
Hence,  the solutions of (\ref{(2.4)}) are formally the critical points of the ``energy"  functional $\,I^\omega$. In this context, we say that $\,\psi\,$ is a {\sl localized}  solution if $\,I^\omega(\psi)\,$  is well-defined, that is, if $\psi\in H^{1/2}(\BbR^3,\BbC^4)$
and $F(\psi)\in L^1(\BbR^3,\BbR)\,.$ Due to the structure of the Dirac operator $D_1$, the functional $I^\omega$ is not bounded below and solutions of \eqref{(2.4)} cannot be obtained by a minimization method.

\medskip

\subsection{Soler models: existence by O.D.E. techniques}
The case $\,b=0\,$ in the definition \eqref{finkel} of $F_b$  was proposed by Soler in \cite{Soler-70} to describe elementary fermions. In this case,  \eqref{(2.5)} reduces to
\bq D_1\psi-\omega \psi -(\bar\psi\psi)\beta\psi
=\; 0\qquad \hbox{in}\quad \BbR^3 \label{soler}\eq
which is usually called the \emph{Soler model}. Its more general version
 \bq D_1\psi -\omega \psi -g(\bar\psi\psi)\beta\psi
=\; 0\qquad \hbox{in}\quad \BbR^3 \label{gensoler}\eq
is often called the \emph{generalized Soler equation}, and it is the Euler-Lagrange equation associated with $\,I^\omega\,$ for $\,F(\psi)=\frac12G(\bar\psi\psi)\,$, $\,G'=g$, $G(0)=0$.

The main advantage of (\ref{gensoler}) is the existence of a special ansatz compatible with the equation: 
\bq\label{(2.7)} \psi (x)\; =\; \left( \begin{matrix} v(r)
\left( \begin{matrix}1 \\ 0 \\ \end{matrix}\right) \\
iu(r) \left( \begin{matrix} \cos\,  \theta  \\
\sin\, \theta \; e^{i\xi } \end{matrix} \right) \end{matrix}\right) \,.
\eq
In this ansatz,  equation (\ref{gensoler}) reduces to the O.D.E. system
\bq \label{ODEs}\left\{\begin{array}{ll}
(u'+\frac{2u}{ r})\; & = \; v\big(g(v^2-u^2)-(1-\omega
)\big)\\
\quad  v'\; & =\; u\big(g(v^2-u^2)-(1+\omega
)\big)\end{array}\right.
\eq

The O.D.E. system (\ref{ODEs}) has been extensively studied. In \cite{Vazquez-77} V\'azquez proved some qualitative properties of the solutions in the case $\,0<\omega<1$, and showed the non-existence of localized solutions when $\,|\omega|>1$. Cazenave and V\'azquez obtained the first rigorous existence result for this problem in \cite{Cazenave-Vazquez-86}. More precisely, in \cite{Cazenave-Vazquez-86} they proved the existence of a solution $(u,v)$ to ({\ref{ODEs}) for nonlinearities $g$ of class  $\,C^1\,$ satisfying:
\bq g:[0,+\infty)\to [0,+\infty)\,,\; g(0)=0\,,\;g'>0\,,\; \lim_{s\to+\infty}g(s)=+\infty \,, \eq
whenever $\,0<\omega<1$.
Moreover, this solution $\,(u,v)\,$ is such that $u$ and $v$ are positive on $\BbR^+$, $\,u(0)=0,\,v(0)>0 $. Additionally, $u$ and $v$ (as well as their first derivatives) decay exponentially at infinity.

Later on, Merle \cite{Merle-88} improved the above result by extending it to a more general class of nonlinearities $g$. Then, Balabane {\sl et al} proved the following general multiplicity result:

\begin{theorem}\label{les4} {\rm (\cite{Balabane-Cazenave-Douady-Merle-88})}  Assume that $\, g: \BbR\to \BbR\,$ is a function of class $C^1$ such that $\,g(0)=0\,$, $\,g\,$ is increasing in $\,(0,+\infty)$, $g(s)>1+\omega\,$ for $s\,$ large, $\,g'(g^{-1}(1-\omega))>0\,$ and $\,g(s)\leq 0\,$ for $\,s\leq 0$. Then, for any $\omega\in (0,1)\,$,  there exists an increasing sequence of  positive numbers $\{x_n\}_{n\geq1}$ such that for every $n\geq 1$, there is a solution $(u_n,v_n)$ of (\ref{ODEs}) satisfying
\begin{itemize}
\item $u_n(0)=0 \,, v_n(0)=x_n\,,$
\item $u_n\,$ and $\, v_n\,$ have $\,n-1\,$ zeros in $\BbR^+$,
\item $u_n\,$ and $\, v_n\,$ decay exponentially at infinity, as well as their first derivatives.
\end{itemize}
Moreover, if  for all $\,s$, $\,g(s)=s$, then the sequence  $\{x_n\}$ is bounded.
\end{theorem}

In the case of singular nonlinearities, compactly supported solutions may exist. More concretely, the following result was proved in \cite{Balabane-Cazenave-Vazquez-90}:

\begin{theorem}\label{solsing} {\rm(\cite{Balabane-Cazenave-Vazquez-90})} Assume that $\, g: (0,+\infty) \to (-\infty, 0)\,$ is a function of class $C^1$ which  is nondecreasing and  integrable near the origin. Suppose also that there exists a number $a$ such that 
$\,g(a^2)=1-\omega\,$, while $\,g'(a^2)>0$. Then, for every $\,0<\omega<1\,$ there exists a positive solution $(u,v)$ of (\ref{ODEs}). Moreover, the support of  $(u,v)$ is bounded if and only if 
$$\int_0^{1} \frac{ds}{G(s)}< +\infty\,,\quad \mbox{where}\quad G(s):=-\int_0^s g(x)\,dx\,.$$
\end{theorem}

\medskip

\subsection{Soler models: existence by variational techniques}
  All the above results were obtained by  a dynamical systems approach. But it is also possible to exploit the variational structure of (\ref{gensoler}) (and also of the O.D.E. system (\ref{ODEs})) to prove existence of solutions. 
  
  In the case of (\ref{ODEs}), the use of variational methods does not radically improve the results that were obtained by O.D.E. methods (see above). The assumptions needed to use variational techniques are slightly different. In \cite{Esteban-Sere-95}, Esteban and Séré obtained the following result:

\begin{theorem}\label{TSoler1} {\rm(\cite{Esteban-Sere-95})}  Let $F:\BbC^{\,4}\to \BbR$
satisfy 
$$F(\psi )\; =\;\frac{1}{2}
G(\overline{\psi }\psi ),\quad G\in
C^2(\BbR,\BbR),\quad G(0)=0\;,
$$
 with $G\in C^2(\BbR,\BbR)$.
Denoting by $g$ the first derivative of $G$,
we make the following assumptions:
\bq \forall x\in \BbR,\quad x\,g(x)\; \geq \; \theta \, G(x) \quad \text{for some}\quad \theta >1,
\label{(H1)}\eq
\bq G(0)=g(0)=0,
\label{(H2)}\eq
\bq (\forall x\in\BbR,\quad G(x)\; \geq \; 0\;)\quad\hbox{and}\quad
G(A_0)\,>\,0\quad\hbox{for some $A_0\,>\,0$},
\label{(H3)} \eq
\bq 0<\omega <1\; .
\label{(H4)} \eq
Then there exist infinitely many solutions of Equation (\ref{(2.4)})
in \break $\bigcap_{\,2\leq q<+\infty
}^{}\!W^{1,q}(\BbR^3,\BbC^{\,4})$. $\!$Each
of them solves a min-max variational problem on the functional $I^\omega
$. They are of the form
(\ref{(2.7)}) and thus correspond to  classical solutions of (\ref{ODEs}) on
$\BbR_+$. Finally, they all decrease exponentially at infinity, together with their first derivatives. 
\end{theorem}

The interest of  using variational techniques appears much more clearly when one studies equations for which no simplifying ansatz is known, for instance in the case where $F=F_b$ with $b\ne 0$. In that case, Equation \eqref{(2.4)}  cannot be reduced to a system of O.D.Es similar to \eqref{ODEs}. A general result proved by Esteban and S\'er\'e in this context is  the following:

\begin{theorem}\label{TSoler2}   {\rm(\cite{Esteban-Sere-95}) }  Let be $F(\psi )=\lambda
\bigl (\vert \psi \bar{\psi }\vert^{\alpha
_1}+b\vert \bar{\psi }\gamma ^5\psi \vert
^\alpha_2 \bigr )$, with $1<\alpha_1 $, $\alpha_2 <\frac{3}{2};$
 $\lambda ,b>0$.
Then, for every $\omega \in
(0,1)$, there exists a non-zero solution of (\ref{(2.4)})  in
$\bigcap_{\,1\!\leq q<+\!\infty
}W^{1,q}(\BbR^3,\BbC^{\,4})$. \end{theorem}

In fact, Theorem \ref{TSoler2} is a consequence of the more general following result in \cite{Esteban-Sere-95}

\begin{theorem}\label{TSoler3}   {\rm (\cite{Esteban-Sere-95})}  Assume that
$F:\BbC^{\,2}\to \BbR$ satisfies :
 \bq \forall\psi \in \BbC^{\,4},\qquad 0\; \leq \; F(\psi )\; \leq \; a_1\left( \vert
\psi \vert ^{\alpha_1}+\vert \psi \vert ^{\alpha_2}\right)\,,
\label{(H5)}\eq
with  $\,a_1>0$ and $\;2<\alpha _1\leq \alpha _2<3$. Assume moreover that
\bq
\left\{\begin{array}{l}
F\in C^2(\BbC^{\,4},\BbR),\quad  F'(0)=F''(0)=0\;
,\\
\vert F''(\psi )\vert \; \leq \; a_2\vert
\psi \vert ^{\alpha _2-2},\quad a_2>0,\quad \hbox{for $\vert
\psi \vert $ large;}\end{array}\right.
\label{(H6)}\eq
 \bq\forall\; \psi
\in \BbC^{\,4},\quad\nabla F(\psi )\cdot \psi \; \geq \; a_3
F(\psi ),\quad a_3 >2\; ;
\label{(H7)}\eq
\begin{equation}
\exists\; a_4 >3,\; \forall\; \delta >0,\; \exists\; C_\delta >0,\; \forall\; \psi \in
\BbC^{\,4},\quad
\vert \nabla F(\psi )\vert \; \leq \; 
\left( \delta +C_\delta F(\psi )^{\frac{1}{a_4}}\right) \vert \psi \vert;
\label{(H8)} 
\end{equation}
 \bq \forall\; \psi \in \BbC^{\,4},\quad F(\psi )\; \geq \; a_5\vert \psi
\bar{\psi }\vert ^\nu -a_6,\quad \nu >1,\;  a_5,a_6>0.
\label{(H9)} \eq

Then, for every $\omega \in (0,1)$, there exists a
non-zero solution of (\ref{(2.4)})  in \break $\bigcap_{\,2\leq q<+\infty
}W^{1,q}(\BbR^3,\BbC^{\,4})$. \end{theorem} 

\medskip 
 
 \subsection{Existence of solutions by perturbation theory}
Another way of finding solutions to nonlinear Dirac equations is perturbation theory. In this approach, one uses  previously known information about the nonlinear Schrödinger equation \cite{Weinstein-85}, which is approached in the non-relativistic limit. Ounaies proved in \cite{Ounaies-00} that solutions of some nonlinear Dirac equations, when properly rescaled, are close to solutions of the nonlinear Schr\"odinger equation, with the same nonlinearity, when the phase $\omega$ approaches $\,1$. More precisely, assume for instance that 
$$F(\psi):= \frac12\left(G(\bar\psi\psi)+H(\bar\psi\gamma^5\psi)  \right)\,,$$
where $G,H$ are two functions of class $C^2$ such that $G(0)=H(0)=0$, and such that $g:=G'$ and $h:=H'$ are homogeneous of degree $\,\theta\in (0,1]$. Then, if we write any $4$-spinor $\psi$ as $\psi=\left(^\varphi_\chi\right)$, the main theorem in \cite{Ounaies-00} states the following

\begin{theorem}  {\rm({\cite{Ounaies-00}})}  Under the above assumptions, let $\,1-\omega= a^{2\theta}=\lambda^2=b^{\frac{2\theta}{\theta+1}}=\varepsilon$. If we rescale  the functions  $\,\varphi,\, \chi\,$ as follows
$$\varphi(x):= a\,\bar\varphi(\lambda x),\; \chi(x):= b\,\bar\chi(\lambda x)\,,$$
 then,  $\psi=\left(^\varphi_\chi\right)$ is a solution to (\ref{(2.4)}) if and only if $\,\bar\varphi,\, \bar\chi\,$ are solutions to the system
$$\left\{\begin{array}{rcl}
(-i\,\sigma\cdot\nabla)\bar\chi+\bar\varphi-g(|\bar\varphi|^2)\bar\varphi+K_1(\varepsilon,\bar\varphi,\bar\chi) & = & 0\\
 (-i\,\sigma\cdot\nabla)\bar\varphi -2\bar\chi+K_2(\varepsilon,\bar\varphi,\bar\chi)&=&0.
\end{array}\right.
$$
Here $K_1$ and $K_2$ are small functions for small $\varepsilon$, $\bar\varphi$ and $\bar\chi$ taking values in a bounded set of $\BbC^2$. Moreover, for $\,\varepsilon\,$ small enough, there exist solutions to the above equation. They are close to a solution of the nonlinear Schr\"odinger equation
\begin{equation}
-\frac{1}{2}\Delta \bar\varphi+\bar\varphi-g(|\bar\varphi|^2)\bar\varphi=0\,,\quad \bar\chi=-\frac i2(\sigma\cdot\nabla)\bar\varphi\,.\label{non_linear_Schr}
\end{equation}
\end{theorem}

\begin{remark} Note that the function $\,h= H'\,$ does not appear in the above limiting equation.
\end{remark}

The proof of this theorem makes use of the implicit function theorem in an appropriate manner. Important ingredients are the uniqueness (up to translation) of the solution to the elliptic equation \eqref{non_linear_Schr} for $\,\bar\varphi\,$ and its nondegeneracy \cite{Kwong-Li-92, Coffman-72, Weinstein-85}.

\medskip

\subsection{Nonlinear Dirac equations in the Schwarzschild metric}
All the above models correspond to the Dirac equation written in the Minkowski metric, this is, in flat space. But the space-time geometry plays an important role when one wants to take relativistic effects into account. For instance, when considering the Schwarzschild metric outside a massive star,  the nonlinear Dirac equation appears to be different. 

In \cite{Bachelot-Motet-98} A. Bachelot-Motet has studied  numerically this problem in the case of the symmetric solutions as above. One has to study a system of O.D.Es similar to (\ref{ODEs}) but with $r$-dependent coefficients instead of constant ones. More precisely, in the ansatz (\ref{(2.7)}) and for the case $\,F(s)=\lambda |s|^2\,$, system (\ref{ODEs}) becomes 
\bq \label{schwarzs}\begin{array}{ll}
fu'+{\frac{u}{r}}(f+f^{1/2})\; & = \; v\bigl [\lambda(v^2-u^2)-(f^{1/2}-\omega
)\bigr ]\\
\quad fv'+{\frac{v}{r}}(f-f^{1/2})\; & =\; u\bigl [\lambda(v^2-u^2)-(f^{1/2}+\omega
)\bigr ]\,,\end{array}
\eq
where $\,f(r)=1-\frac1r$. 

Notice that this problem is not to be considered in the whole space: since the physical situation corresponds to the outside of a massive star, the natural domain is the complement of a ball, or in radial coordinates, the interval $\, r>r_0\,$, for some $\,r_0>0\,$. In this case, the usual ``MIT-bag" boundary condition reads
$$u(r_0)=-v(r_0)\,.$$

The very interesting numerical results obtained by Bachelot-Motet suggested conditions for some existence and multiplicity results for (\ref{schwarzs}) that were later rigorously proved by Paturel in \cite{Paturel-00}. Note that in \cite{Paturel-00} the solutions are found as critical points of a reduced energy functional by a mountain-pass argument, while as we see below, we use a linking method to produce our solutions.

\subsection{Solutions of the Maxwell-Dirac equations}
The nonlinear terms appearing in all the above models are local, that is, are functions of the spinor field $\,\Psi$. But in some cases, one has to introduce nonlocal terms, like for instance when considering the interaction of the Dirac field with a self-generated field. In this case, the equations become integro-differential. 

Our first example is the Maxwell-Dirac system of classical field equations, describing the interaction
of a particle with its self-generated electromagnetic field. In order to write the equations in relativistically covariant form, we introduce the usual four-dimensional notations: let be $\gamma^0:=\beta$ and $\gamma^k:=\beta\alpha_k$. For any wavefunction $\Psi(x_0,x):\R\times\R^3\mapsto\C^4$ (note that $x_0$ plays the role of the time $t$), we denote $\bar\Psi:=\beta\Psi$. In the Lorentz gauge the Maxwell-Dirac equations can be written as follows 
\bq \left\lbrace
\begin{array}{cc}
 &(i\gamma^\mu \partial _\mu \!-\!\gamma ^\mu A_\mu
-1)\Psi \; =\; 0\qquad\qquad \hbox{in}\quad \BbR\times \BbR^3 \\
 &\partial
_\mu A^\mu =0,\quad 4\pi \,\partial _\mu \partial ^\mu \, A_\nu
= (\bar\Psi,\gamma^\nu\Psi) \qquad \;\, \hbox{in} \quad \BbR\times \BbR^3. \\
\end{array}
 \right.
\label{M-D}
\eq
Notice that we have used Einstein's convention for the summation over $\mu$. We also introduce the electromagnetic current $J^\nu:=(\bar{\Psi },\gamma^\nu \Psi )$.

Finite energy stationary solutions of classical nonlinear wave equations 
 have been sometimes used  to describe extended  particles. Of course the electromagnetic field should in principle be quantized like in Quantum Electrodynamics. In the Maxwell-Dirac model, the field is not quantized but it is believed that interesting qualitative results can be obtained by using classical fields (see, e.g. \cite[Chapter 7]{Grandy}). 

 \medskip
 
 Another example of a self-interaction is the Klein-Gordon-Dirac system  which arises in the
so-called Yukawa model  (see, e.g. \cite{Bjorken-Drell-65}). It can be written as
\bq \left\lbrace \begin{array}{cc}
&(i\gamma^\mu \partial _\mu -\chi-1)\Psi \; =\;
0\qquad\quad \hbox{in}\quad \BbR\times \BbR^3 \\
 &\partial ^\mu \partial _\mu
\chi +M^2\chi \; =\; {\frac{1}{4\pi} }\, (\bar{\Psi },\Psi )
\qquad\quad \hbox{in}\quad  \BbR\times \BbR^3 \; .\\
\end{array}
\right.
\label{KG-D}\eq
Other related models, that we will not discuss, include the Einstein-Dirac-Maxwell equations, which have been investigated by  F. Finster, J. Smoller and S.-T. Yau \cite{Finster-Smoller-Yau1}, \cite{Finster-Smoller-Yau2}.
The above systems \eqref{M-D} and \eqref{KG-D} have been extensively studied and many results are available concerning the Cauchy problem  (we refer to \cite{Esteban-Georgiev-Sere-96} and \cite{Grandy}, chapter 7, for detailed references).

\medskip

A stationary solution of the Maxwell-Dirac system \eqref{M-D} is a particular solution $\,(\Psi, A):\R\times\R^4 \mapsto \C^4\times\R^4$ of the form
\bq \left\lbrace\begin{array}{cc}
 &\Psi
(x_0,x)\; =\; e^{-i\omega x_0}\psi (x)\quad \text{with}\quad \psi :\BbR^3\to
\BbC\,^4,\\
 &A^\mu (x_0,x)\; =\; J^\mu *{\frac{1}{ \vert x\vert }}\; =\;
\int_{\BbR^3}^{} \, \frac {dy}{ \vert x\!-\!y\vert }\; J^\mu (y).
\end{array}
\right.
\label{S}\eq
The existence of such stationary solutions of \eqref{M-D} has been an open problem for a long time (see, e.g.
\cite[p. 235]{Grandy}).
Indeed, the interaction between the spinor and its own electromagnetic field makes
equations (\ref{M-D}) nonlinear. 

Concerning stationary solutions of (\ref{M-D}), let us mention the pioneering works of Finkelstein et al
\cite{Finkelstein-Lelevier-Ruderman-51} and Wakano \cite{Wakano-66}.  The latter considered this system in the  
approximation $A_0\not\equiv 0$, $A_1=A_2=A_3\equiv 0$, the so-called Dirac-Poisson system. This problem can be reduced to a system of three coupled differential equations by using the spherical spinors \eqref{(2.7)}. Wakano obtained numerical evidence for the existence of stationary solutions of the Dirac-Poisson equation. Further work in this direction (see \cite{Ranada-82}) yielded the same kind of numerical results for some modified Maxwell-Dirac equations which include some nonlinear self-coupling.

In \cite{Garrett Lisi-95} Lisi  found numerical solutions of the Dirac-Poisson and of the Maxwell-Dirac systems. The computation of the
magnetic part of the field $A$ for these solutions showed that Wakano's approximation was reasonable, since the field components $(A_1,A_2,A_3)$ stay small compared with $A_0$. 
See also \cite{Radford-96,Radford-03,Booth-00} where various kinds of stationary solutions are considered, like the so-called \emph{static solutions} which have no current ``flow''.

In the case  $-1 <\omega<0$, Esteban, Georgiev and S\'er\'e \cite{Esteban-Georgiev-Sere-96} used variational techniques to prove  the existence of stationary solutions of (\ref{M-D}). 
Any solution of (\ref{M-D}) taking the form (\ref{S}) corresponds (formally) to a critical point $\psi$ of the following functional:
\begin{multline*}
I_\omega (\psi )\; =\; \int_{\BbR^3}^{}
{\frac{1}{2}}\, (i\alpha ^k\partial _k\psi
,\psi ) -{\frac{1}{2}}\, (\bar{\psi },\psi )
-{\frac{\omega}{2}}\, \vert \psi \vert ^2
 -{\frac{1}{4}}\iint_{\BbR^3\times \BbR^3}^{}
\frac{J^\mu (x)\, J_\mu (y)}{ \vert x\!-\!y\vert }\; dx\,
dy. \\
\end{multline*} 
This remark was used in \cite{Esteban-Georgiev-Sere-96} to find a stationary solution of (\ref{M-D}) in the
appropriate space of functions.

\begin{theorem}\label{TTMD} {\rm(\cite{Esteban-Georgiev-Sere-96})}  For any $\omega$
 strictly between $-1$ and $0$, there exists a non-zero critical point $\psi
_\omega $ of $I_\omega $. This function $\psi _\omega $ is  smooth
in $x$ and   exponentially decreasing at infinity. Finally, the fields $\Psi
(x_0,x)=e^{-i\omega x_0}\, \psi _\omega $, $A^\mu
(x_0,x)=J_\omega ^\mu*\frac{1}{\vert x\vert }$ are
solutions of the Maxwell-Dirac system (\ref{M-D}).\end{theorem}

Later, using cylindrical coordinates,  S. Abenda \cite{Abenda-98} extended the above result to the
case $-1<\omega<1$. Indeed, in the class of cylindrically symmetric functions, the energy functional has better properties which allow to use the same variational procedure as in the work of Esteban-S\'er\'e, but in the more general case $\, \omega\in (-1, 1)$. 

Many questions are still open about the existence of stationary
solutions for \eqref{M-D}. It is easy to see that they have all a negative
``mass". Wakano already observed this phenomenon for the soliton-like solutions
of the Dirac-Poisson system. However, it was shown in \cite{Wakano-66} that a positive mass can be reached by
taking into account the vacuum polarization effect. 

\medskip

For the case of the Klein-Gordon-Dirac
equations the situation is slightly simpler because they are compatible with the ansatz  (\ref{(2.7)}) introduced above. So, in this case the authors of \cite{Esteban-Georgiev-Sere-96} did not only obtain existence of solutions, but also multiplicity:

\begin{theorem} {\rm(\cite{Esteban-Georgiev-Sere-96})}  For any $\omega$
 strictly between $-1$ and $0$, there exists infinitely many solutions to the Klein-Gordon-Dirac system  (\ref{KG-D}). These solutions are all smooth  and   exponentially decreasing at infinity in $x$.  \end{theorem}

\medskip

We finish this section by explaining the general ideas of the proof of Theorem \ref{TTMD}. The proofs of Theorems \ref{TSoler1},  \ref{TSoler2} and  \ref{TSoler3}  basically follow the same lines and we will skip them. 

\medskip

\noindent{\bf Sketch of the proof of Theorem \ref{TTMD}.}
As already mentioned in the introduction, the presence of the negative spectrum for the Dirac operator forbids the use of a minimization argument to construct critical points. Instead, the solution will be obtained by means of a min-max variational method based on complicated topological arguments. This kind of method used to treat problems with infinite negative and positive spectrum have been already used under the name of \emph{linking}. The linking method was introduced by V. Benci and P. Rabinowitz in a compact context \cite{Benci-Rabinowitz-79}.
The reasons  making the use of variational arguments nonstandard in our case are : (1) the equations
are translation invariant, which creates a lack of compactness; (2) the
interaction term $\,\,J^\mu A_\mu\,\,$ is not positive definite.  Note  that as we have alredy pointed out, in some cases one can perform a reduction procedure and obtain a reduced functional for which critical points can be found by a mountain-pass argument \cite{Paturel-00}.

\medskip

\noindent{\bf First  step: Estimates.}
Defining
$A^\mu =J^\mu \ast \frac{1}{|x|}$, then one deduces
$J^\mu A_\mu = J^0A^0-\sum_{k=1}^{3}J^kA^k$ and
$$ L(\psi):= \iint_{\BbR^3\times \BbR^3}^{}
\frac{J^\mu (x)\, J_\mu (y)}{ \vert x\!-\!y\vert }\; dx\,
dy = \int_{\BbR^3} J^\mu A_\mu\,dx\,.$$
Let us also introduce the functional
$$ Q(\psi)\,=\,\iint_{\BbR^3\times \BbR^3}^{}\, \frac{(\overline{\psi
},\psi )(x)\, (\overline{\psi },\psi )(y)}{\vert x\!-\!y\vert }\, dx\, dy\;.\leqno (2.2) $$
It is easy to prove  that $Q$ is non-negative, continuous and convex on 
$\gH=H^{1/2}(\R^3,\C^4)$, and vanishes only when
$(\overline{\psi},\psi )(x)\,=\,0$ a.e. in $\BbR^3\,.$ 

 Let us state a lemma giving some properties of the quadratic forms in $\,I_\omega$:
 \begin{lemma}\label{LL21}  For any $\psi \in
\gH$, the following inequalities hold:
 
 \smallskip
(i) $\qquad J^\mu A_\mu (x)\; \geq \; 0\;
,\quad \mbox{a.e. in}\; \;\BbR^3,$
 
  \smallskip
(ii) $\qquad \int_{\BbR^3}^{}J^\mu A_\mu \; \geq \;
Q(\psi),$
   
  \smallskip
(iii) $\qquad A_0\; \geq \; \biggl
(\sum_{k=1}^{3}\vert A_k\vert ^2\biggr )^{1/2},$

  \smallskip
 (iv) $\qquad \vert \gamma ^\mu A_\mu \psi \vert \; \leq \;
C\sqrt{A^0}\; \sqrt{A^\mu J_\mu} \quad \mbox{a.e.  in}\; \;\BbR^3.$
 
\end{lemma}

\begin{remark}  Note that when 
the function $\psi$ is cylindrically symmetric, the functional $L$ defined above is 
not only non-negative, but actually controls from below 
$\norm{\psi}_{\gH}^4$ (see Lemma 1 in \cite{Abenda-98}). This is the 
reason why Abenda has been able to treat the case $\,\omega\in (-1, 1)$, extending 
Theorem \ref{TTMD}.
\end{remark}

Another important information is given by the
\begin{lemma}Let be $\mu >0$. There is
a non-zero function $e_+:(0;\infty)\to \gH_1^+=\Lambda_1^+\gH$ such that, if
$\Lambda^+\psi =e_+(\mu)$, then
$$\frac12\int_{\BbR^3}^{}(\psi ,D_1\psi )-\frac{1}{4}Q(\psi)\; \leq \; \frac{\mu}{2}\; \Vert
\psi \Vert _{L^2}^2\; .$$
\end{lemma}\medskip

\noindent{\bf Second  step: Modified functional and variational argument.}
In order to obtain some coercivity, a modified functional $\,I_{\omega, \varepsilon}\,$ was considered in \cite{Esteban-Georgiev-Sere-96}. It reads
$$I_{\omega ,\varepsilon }(\psi )\; =\; I_\omega
(\psi )-\frac{2\varepsilon}{5}\, \Vert \psi \Vert
_{L^{5/2}}^{5/2}.\label{(2.16)}$$
where $ \varepsilon >0$. 
The critical points of $I_{\omega ,\varepsilon }(\psi
)$ satisfy
\begin{equation}\label{CEqeps} \left\lbrace
\begin{array}{l}
i\gamma ^k\partial _k\psi -\psi -\omega \gamma
^0\psi -\gamma ^\mu A_\mu \psi 
- \varepsilon \gamma ^0\vert \psi \vert
^{\frac12}\psi \; =\; 0\\
-4\pi \Delta A_0=J^0=\vert \psi \vert ^2,\quad -4\pi \Delta A_k=-J^k\,.\\
\end{array} \right. \end{equation} 

Let $\theta$ be a smooth function satisfying $\theta(s)=0\,$ for $\,s\leq -1$ and $\theta(s)=1\,$ for $\,s\geq 0\,$. The gradient being defined by $-\nabla I_{\omega,\varepsilon }=-\vert D_1\vert ^{-1}\, I'_{\omega ,\varepsilon }$, let us consider the flow for positive times $t$, $\eta _{\omega ,\varepsilon }^t$,  of a modified gradient :  

\begin{equation}\label{flot11} \left\lbrace
\begin{array}{l}
\eta ^0\; =\; \un_\gH\\
\partial _t\eta_{\omega ,\varepsilon }
 ^t\; =\; - \left(\theta(I_{\omega, \varepsilon})\nabla I_{\omega ,\varepsilon
}\right)\circ \eta_{\omega ,\varepsilon } ^t\; .
 \\
\end{array} \right. \end{equation} 

It can be seen that for $\,\varepsilon>0\,$ the functional $\,I_{\omega ,\varepsilon }\,$ enjoys the properties needed for the Benci-Rabinowitz linking argument \cite{Benci-Rabinowitz-79}, except that its gradient is not of the form $L+K$ with $L$ linear and $K$ compact. Due to this lack of compactness, one cannot use Leray-Schauder's degree. One can work instead with a generalized version of
the Leray-Schauder $\Z_2$-degree, due to Smale \cite{Smale-65}  to show the existence of a positive critical level of  $\,I_{\omega ,\varepsilon }\,$.  This idea was introduced by Hofer-Wysocki \cite{Hofer-Wysocki-90} in the study of homoclinic orbits of nonconvex Hamiltonian systems, where a similar lack of compactness occurs. Hofer and Wysocki worked with the unregularized $L^2$-gradient. This gradient does not have a well-defined flow, but for the linking argument it is sufficient to consider certain smooth gradient lines, which are pseudo-holomorphic curves satisfying boundary conditions. Later S\'er\'e \cite{Sere-95}, studying homoclinic orbits on singular energy hypersurfaces, worked with the $H^{1/2}$-regularized gradient, which has a well-defined flow leading to an easier and more flexible linking argument. This approach is adapted to nonlinear Dirac in \cite{Esteban-Sere-95} and to Dirac-Maxwell and Dirac-Klein-Gordon in \cite{Esteban-Georgiev-Sere-96}.
Consider the sets
$${\mathcal N}^-\; =\; \bigl \{\psi =\psi _- +\lambda
e_+(\mu)\; ,\; \psi _-\in \gH_1^-,\; \Vert \psi _-\Vert
_\gH\leq \Vert e_+(\mu)\Vert_\gH,\; \lambda \in [0,1]\bigr \}$$
and
$$\Sigma _+=\{\psi \in \gH_1^+\; /\;
\Vert \psi \Vert _\gH=\rho \}\,, \quad \rho>0\,.$$
Then one can prove the
\begin{proposition}\label{variat-arg} For any $-1 <\omega<-\mu$, $\varepsilon >0$ and $\Sigma _+$, ${\mathcal N}_-$
constructed as above, there exists a positive constant $\,c_\omega$, such that the set $\eta_{\omega,\varepsilon} ^t({\mathcal N}_-)\cap
\Sigma _+$ is non-empty, for all $t\geq 0$. Moreover, the
number
$$c_{\omega ,\varepsilon }\; =\; \inf_{t\geq 0}\;
I_{\omega ,\varepsilon} \circ \eta _{\omega ,\varepsilon
}^t({\mathcal N}_-) $$
is strictly positive, it is a critical level for $\,I_{\omega,\varepsilon}\,$ and $c_{\omega ,\varepsilon }\to
c_\omega >0$ as $\varepsilon \to 0$.

Additionally, for any $\omega ,\varepsilon $ fixed, there
is a sequence $\{\varphi _{\omega ,\varepsilon }^n\}_{n\geq
0}$ such that as $\,n\to +\infty$, 

\begin{equation} \left\lbrace
\begin{array}{l}
I_{\omega ,\varepsilon }(\varphi _{\omega ,\varepsilon }
^n)\; \mathop \to\; c_{\omega ,\varepsilon
}\; ,\\
\bigl (1+\Vert \varphi _{\omega ,\varepsilon }^n\Vert
\bigr )\nabla I_{\omega ,\varepsilon }
(\varphi _{\omega ,\varepsilon }^n)\; \mathop  \to
\; 0\; .\\
\end{array} \right. \end{equation} 
\end{proposition}

\medskip

\begin{remark}
In \cite{Esteban-Georgiev-Sere-96} and  \cite{Esteban-Sere-95}, an easy regularization step is missing. Indeed, Smale's degree theory requires $C^2$-regularity for the flow, which corresponds to $C^3$-regularity for the functional. In the case of the local nonlinear Dirac equation, such a regularity can be easily  achieved by a small perturbation of the function $\,F(\psi) + \varepsilon|\psi|^{\alpha_2-1}\psi$. Since all the estimates will be independent of the regularization parameter, the solutions of the non-regularized problem will be obtained by a limiting argument.
\end{remark}

\medskip

Note that the linking argument of   \cite{Sere-95}, \cite{Esteban-Sere-95} and  \cite{Esteban-Georgiev-Sere-96}
has inspired later work (see \cite{Troestler-Willem-96, Kryszewski-Szulkin-98}), where an abstract linking theorem in a noncompact setting is given, valid first for $\,C^2\,$ and then for $\,C^1$-functionals.

\medskip \noindent{\bf Third  step: Properties of the critical sequences.}
The concentration-compactness theory of P.-L. Lions \cite{Lions-84} allows us to analyze the behavior of critical sequences of 
$\,I_{\omega,\varepsilon}\,$ as follows:

\begin{proposition}\label{cccc} Let $\omega \in
(-1,0)$ and $\varepsilon \geq 0$ be fixed.
Let $(\psi _n)\subset \gH$ be
a sequence in $\gH$ such that 
\bq\label{AA21} 0\;<\;\inf_{n} \Vert \psi _n\Vert _\gH\; \leq \;
\sup_{n} \Vert \psi _n\Vert _\gH \;<\;+\infty \eq
and $I_{\omega
,\varepsilon }'(\psi _n)\displaystyle \mathop \to 0$ in $\gH' \,$ as  $\,n\,$ goes to $+\infty$.
 Then
we can find a finite integer
$p\geq 1\,,$ $p$ non-zero
solutions $\phi ^1,\ldots ,\phi ^p$ of (\ref{CEqeps})  in $\gH$ and $p$
sequences $(x_n^i)\subset \BbR^3$,
$i=1,\ldots ,p$ such that for $i\neq j$, $\vert
x_n^i-x_n^j\vert \displaystyle \mathop \to_{n\to +\infty
}+\infty\,,$ and, up to extraction of a subsequence,
$$\Bigl \Vert \psi _n-\sum_{i=1}^{p}\phi ^i(\cdot
-x_n^i)\Bigr \Vert _\gH\; \mathop \to_{n\to +\infty }\;
0\; .
$$
\end{proposition}

Obtaining estimates in $\,H^{1/2}(\R^3)$ for the sequence $\{\varphi _{\omega ,\varepsilon }^n\}_{n\geq
0}$ of Proposition \ref{variat-arg} is quite easy because of the coercivity introduced by the perturbation term in $\varepsilon$.  Moreover, $c_{\omega, \varepsilon}$ being strictly positive, the sequence
$\{\varphi _{\omega ,\varepsilon }^n\}_{n\geq
0}$ is also bounded from below away from $0$.  So, Proposition \ref{cccc} applies to prove the existence of a solution to (\ref{CEqeps}) for every $\varepsilon>0$. Next, we want to pass to the limit when $\varepsilon$ goes to $0$. Note that we are doing so along a sequence of functions which are exact solutions of the approximate problem (\ref{CEqeps}).
 This part of the proof is done by first proving the equivalent of the Pohozaev identity for equation (\ref{CEqeps}), $\varepsilon\geq 0$,  and then by introducing some special topologies in the spaces
$\,L^q\,$ which are related to the decomposition of $\,\BbR^3\,$ as the union of unit cubes. Analyzing the solutions to  (\ref{CEqeps}) in those topologies, we find the following

\begin{theorem}\label{presque} There is a constant $\,\kappa>0\,$ such that if
 $\,-1<\omega < 0$
and $0<\varepsilon\leq 1$,  there is  a function $\psi _\varepsilon
\in \gH$ such that $I_{\omega ,\varepsilon }'(\psi
_\varepsilon )=0$ and
$${\kappa}\, \leq\,  I_{\omega ,\varepsilon
}(\psi _\varepsilon )\, \leq \, c_{\omega ,\varepsilon}\;.
$$
\end{theorem}

\medskip

\noindent{\bf Last  step: Passing to the limit $\,\varepsilon\to 0$.}
Eventually, we use Proposition \ref{cccc} to pass to the limit $\,\varepsilon\to 0$.
When obtaining the estimates (\ref{AA21}) for the critical sequences of $\,I_{\omega ,\varepsilon }\,$, we observe that the lower estimate for the norm $\,||\cdot||_\gH\,$ is actually independent of $\,\varepsilon$. Assume, by contradiction that the upper estimates do not hold for the sequence $\,(\psi_\varepsilon)$. Then, we consider the normalized functions 
$$ \tilde \psi _\varepsilon=\Vert \psi
_{\varepsilon}\Vert _\gH^{-1}\, \psi _{\varepsilon}\;$$
and apply Proposition \ref{cccc} to the sequence $\,(\tilde\psi_\varepsilon)$. Under the assumption that $\,||\psi_\varepsilon||_\gH\to +\infty$, we use all the previous estimates to infer that for $\,j=1,\dots, p$,
$$\Sum_{j=1}^p\int_{\BbR^3}\bar\phi^j\phi^j+\omega|\phi|^2\,dx=0\,,\quad Q(\phi^{j})=0.$$
But the latter implies that for every $j$,  $\,\bar\phi^j\phi^j=0$ a.e. and so, from the r.h.s. identity we obtain that $\,\phi^j=0$ a.e. for all $j$. This contradicts Theorem \ref{presque}. \hfill$\square$

\bigskip

\subsection{Nonlinear Dirac evolution problems}\label{time_dep}
The results which we have mentioned so far are concerned with the existence of  stationary solutions of various nonlinear Dirac evolution equations. These particular solutions are global and do not change their shape with time. The study of the nonlinear Dirac evolution problem 
\begin{equation}
 \left\{\begin{array}{l}
i{\partial_t}\Psi -D_1\Psi+G(\Psi )\; =\; 0 \\
\Psi(0)=\Psi_0
 \end{array}\right.
\end{equation}
is also interesting in itself and, even if this is not the aim of the present paper, let us mention some references. 

For the case of local nonlinearities as the ones  considered in this section, several works have proved well-posedness for small initial data in well chosen Sobolev spaces. For nonlinearities containing only powers of $\,\Psi\,$ of order $p\geq 4$, Reed proved in \cite{Reed-76}  the global well-posedness for small initial data in $\,H^s\,$, $s>3$. A decay estimate at infinity was also obtained in this paper. Later,  Dias and Figueira  \cite{Dias-Figueira-87} improved this result to include powers of order $ p=3$  and for $s>2$. Najman \cite{Najman-92} took the necessary regularity of the initial data down to $H^2$.  In \cite{Escobedo-Vega-97} Escobedo and Vega proved an ``optimal result" which states that for the physically relevant nonlinearities of order $p\geq 3$ of the type
\bq  G(\Psi):=\lambda\left\{(\bar\Psi\Psi)^{\frac{p-1}{2}}\beta\Psi+b(\Psi,\gamma^5\Psi)^{\frac{p-1}{2}}\gamma_5\Psi\right\}, \label{finkel2}\eq
there is local well-posedness of the evolution equation in $H^s$, for $s>\frac32-\frac1{p-1}$, when $p$ is an odd integer, while $s$ has to be in the interval $\,(\frac32-\frac1{p-1}, \frac{p-1}{2})\,$ otherwise. Moreover, if $p>3$, then the problem is globally well-posed for small initial data in $\,H^{s(p)}$, with $\,s(p)= \frac32-\frac1{p-1}\,$.
For a more recent result, see for instance a paper of Machihara, Nakanishi and Ozawa \cite{Machihara-Nakanishi-Ozawa-03}, in which the existence of small global solutions is proved in $H^s$ for $s>1$, and the nonrelativistic limit is also considered.

An interesting question to ask is about the (linear or nonlinear) stability properties  of the stationary solutions with respect to the flow generated by the evolution equation. At present this seems to be a widely open problem (see \cite{Ranada-82} and \cite{Strauss-Vazquez} for a discussion). Recently, Boussaid \cite{Boussaid} has obtained the first stability results, for small stationary solutions of nonlinear Dirac equations with exterior potential.

\medskip

Concerning the Cauchy problem for the Maxwell-Dirac equations, the first result about the local existence and uniqueness of solutions was obtained by L. Gross in \cite{Gross-66}. Later developments were made by Chadam \cite{Chadam-73}  and Chadam and Glassey \cite{Chadam-Glassey-76} in $1+1$ and $2+1$ space-time dimensions and in $3+1$ dimensions when the magnetic field is $0$. Choquet-Bruhat studied in \cite{Choquet-Bruhat-81} the case of spinor fields of zero mass and Maxwell-Dirac equations in the Minskowski space were studied by Flato, Simon and Taflin in \cite{Flato-Simon-Taflin-87}. In \cite{Georgiev-91}, Georgiev obtained a class of initial values for which the Maxwell-Dirac equations have a global solution. This was performed by using a technique introduced by Klainerman (see \cite{Klainerman-85,Klainerman-86A, Klainerman-86B}) to obtain $L^\infty$ a priori estimates via the Lorentz invariance of the equations and a generalized version of the energy inequalities. The same method was used by Bachelot \cite{Bachelot-88} to obtain a similar result for the Klein-Gordon-Dirac equation. Finally, more recent efforts have been directed to proving existence of solutions for the time-dependent Klein-Gordon-Dirac and Maxwell-Dirac equations in the energy space, namely $C(-T,T ; H^{1/2}\times H^1)$. The existence and uniqueness of solutions to the Maxwell-Dirac system in the energy space has been proved by Masmoudi and Nakanishi in \cite{Masmoudi-Nakanishi-03A, Masmoudi-Nakanishi-03B}, improving Bournaveas' result in \cite{Bournaveas-96}, where the space considered was $C(-T, T; H^{1/2+\varepsilon} \times H^{1+\varepsilon})$.

Note that as mentioned above, the stationary states of the form (\ref{S}) are particular solutions of the Maxwell-Dirac equations.  Physically they correspond to bound states of the electron.

\section{Linear Dirac equations for an electron in an external field}

When looking for stationary states describing the dynamics of an electron moving in an external field generated by an electrostatic potential $V$, one is led to study the eigenvalues and eigenfunctions of the operator $\,D_c+V\,$. If the electron has to enjoy some stability, the eigenvalues should also be away from the essential spectrum. In the case of not very strong potentials $V$, the essential spectrum of $\,D_c+V\,$ is the same as that of $\,D_c\,$, that is, the set $\,(-\infty,-c^2] \cup[c^2,+\infty)$. So the eigenvalues which  are of interest to us are those lying in the gap of the essential spectrum, i.e. in the interval $\,(-c^2,c^2)$. More precisely, in general a state describing an electron is always assumed to correspond to a positive eigenvalue. It is therefore important to be able to determine whether there are positive eigenvalues or not, and what is the behaviour of the `first' eigenvalue when $V$ varies (whether it crosses $0$ or dives into the lower negative essential spectrum for instance). Note that one expects that for a reasonable potential there are no eigenvalues embedded in the essential spectrum. Very general conditions on $V$  which ensure nonexistence of embedded eigenvalues have been given by  \cite{Berthier-Georgescu-83, Berthier-Georgescu-87} .  Note  finally  that in this section $\,c\,$ is kept variable.

Formally, the eigenvalues of the operator $\,D_c+V\,$ are
critical values of the Rayleigh quotient
\begin{equation} Q_V(\psi):=\frac{((D_c+V)\psi,\psi)}{
(\psi,\psi)}\end{equation} 
in the domain of $\,D_c+V\,$.  Of course, one cannot use a minimizing argument to find such critical points since, due to the negative continuous spectrum of the free Dirac operator, $Q_V$ is not bounded-below. 
Many works have been devoted to finding non-minimization variational problems yielding the eigenvalues of $\,D_c+V\,$ in the interval $\,(-c^2, c^2)$.   Another important issue is to avoid the appearance of spurious states (some eigenvalues of the finite dimensional problem may not approach the eigenvalues of the Dirac operator $\,D_c+V\,$) as it has been the case in many proposed algorithms (see for instance \cite{Drake-Goldman-81}). W. Kutzelnigg has
written two excellent reviews \cite{Kutzelnigg-84, Kutzelnigg-97} on this  subject, where many relevant references
can be found. The main techniques which have been developed so far and used in practice can  be divided  in three groups:

\begin{enumerate}
\item Use of effective Hamiltonians whose point spectra  are close to the spectrum of the Dirac operator  in the limit $\,c\to +\infty$. For instance, one can cut at a  finite level some infinite asymptotic formal expansion in negative powers of $c$. To this category of works belong for instance
\cite{Durand-86, Durand-Malrieu-87,  vanLenthe-vanLeeuwen-Baerends-Snijders-94, vanLeeuwen-vanLenthe-Baerends-Snijders-94, LeYaouanc-Oliver-Raynal-97, Kutzelnigg-97}. 
\item Use of a Galerkin approximation technique to approach the eigenvalues, and this
without falling into the negative continuum
$\,(-\infty, -c^2)$. This is equivalent to  projecting the equation onto a well-chosen finite dimensional space.  This procedure has been well explained for instance  in~\cite{Drake-Goldman-88A, Drake-Goldman-88B, Kutzelnigg-84}.
\item Replacement of the problematic minimization of $Q_V$ by another one. For instance, it was proposed to minimize the Rayleigh
quotient for the squared Hamiltonian $\,(D_c+V)^2\,$ (see, e.g.
\cite{Wallmeier-Kutzelnigg-81, Bayliss-Peel-83}) or later on, to maximize the Rayleigh quotients
for the ``inverse Hamiltonian" $\,\frac{D_c+V}{|D_c+V|^2}\,$
(see \cite{Hill-Krauthauser-94}). 
\end{enumerate}

\medskip

Before we go further, let us recall some useful inequalities which are usually used to control the external field $V$  and show that $D_c+V$ is essentially self-adjoint. We recall that $\gH=H^{1/2}(\R^3,\C^4)$ and that $\gH_\pm^0$ are the positive and negative spectral subspaces of $D_1$.

\begin{proposition}[Hardy-like inequalities]\label{R2}
The Coulomb potential $W(x) = { \frac{1}{ \vert x \vert}}$ satisfies the following Hardy-type inequalities:
\begin{equation}\label{kato-ineg}W\leq \frac{\pi}{2}\sqrt{-\Delta}\leq \frac{\pi}{2c}\, \vert D_c \vert,
\end{equation}
\begin{equation}\label{tix-ineg} 
\forall\psi\in \gH^0_+\cup \gH^0_-,\quad 
\Bigl( \psi, W (x) \psi \Bigr)_{L^2} \leq
\frac{1}{2}\left(\frac{\pi}{2}+\frac{2}{\pi}\right) \left(
\psi,\vert D_1 \vert \psi \right)_{L^2},
\end{equation} 
\begin{equation}  W^2  \leq -4\Delta \leq 4 |D_1|^2.
\end{equation}
\end{proposition}

The inequalities of Proposition \ref{R2} are classical (see, e.g. \cite{Herbst-77, Kato-66} for \eqref{kato-ineg}), except for (\ref{tix-ineg}). The proof of the latter is based on a method of
Evans, Perry and Siedentop \cite{Evans-Perry-Siedentop-96} and is contained in the recent papers 
\cite{Burenkov-Evans-98, Tix-97, Tix-98}.

\subsection{A variational characterization of the eigenvalues of $D_c+V$}
Formally, the eigenvalues of $\,D_c+V\,$ lying in the gap of the
essential spectrum should be described by some kind of {\em min-max} argument. This was mentioned in several papers dealing with
numerical computations of Dirac eigenvalues before it was formally addressed in different contexts in a
series of papers \cite{Esteban-Sere-97,Griesemer-Siedentop-99,Griesemer-Lewis-Siedentop-99, Dolbeault-Esteban-Sere-00A,  Dolbeault-Esteban-Sere-00B}. 

For the sake of clarity, we are going to present only a particular version of those min-max arguments allowing to characterize eigenvalues of the operator $\,D_c+V\,$ for appropriate potentials $\,V$. This method derives from a proposition 
 made by Talman \cite{Talman-86} and Datta, Deviah \cite{Datta-Deviah-88} and based on the decomposition of any spinor  $\,\psi=\binom{\varphi}{ \chi}\,$ as the sum of its upper and its lower components:
\bq\,\psi=\binom{\varphi}{ 0}+\binom{0}{ \chi}\,.\eq

This proposal consisted in saying that the first eigenvalue of $D_c+V$ could be obtained by solving the min-max problem
\bq\label{Tal} \min_{\varphi\neq 0}\;\max_\chi \frac{((D_c+V)\psi,\psi)}{(\psi,\psi)}\,.\eq

The first rigorous result on this min-max principle was obtained by Griesemer and Siedentop
\cite{Griesemer-Siedentop-99}, who proved that (\ref{Tal}) yields indeed the first positive eigenvalue of $\,D_c+V\,$ for potentials $V$ which are in $L^\infty$  and not too large.
In \cite{Dolbeault-Esteban-Sere-00B}, Dolbeault, Esteban and Séré proved that if $\,V\,$ satisfies the assumptions
\bq\label{V1} V(x) \quad\;\longrightarrow_{_{{\hspace{-8mm}}|x|\to +\infty}} \; 0 \,, \eq
\bq\label{V2} -\frac{\nu}{|x|}-K_1\leq V\leq K_2 =\Sup_{x\in \BbR^3} V(x)\,,
 \eq 
\bq\label{V3}\quad K_1, K_2\geq 0,\;\, K_1+K_2-c^2< \sqrt{c^4-\nu^2\,c^2}
\eq 
with $\,\nu\in (0,c)$, $K_1,K_2\in\BbR$, then the first eigenvalue $\lambda_1(V)$ of $\,D_c+V\,$ in the interval $\,(-c^2, c^2)\,$ is given by the formula
\bq\label{Talm}\lambda_1(V)=\inf_{\varphi\neq 0}\sup_{\chi}\frac{(\psi,(D_c+V)\psi)}{(\psi,\psi)}\;,\quad \psi=\binom{\varphi}{ \chi}\,.\eq
Actually, under the conditions \eqref{V1}-\eqref{V2}-\eqref{V3}, it can be seen that $D_c+V$ has an infinite sequence of eigenvalues $\{\lambda_k(V)\}_{k\geq1}$ converging to 1, and it was proved in \cite{Dolbeault-Esteban-Sere-00B} that each of them can be obtained by a min-max procedure:

\begin{theorem}\label{TT1} {\rm (Min-max characterization of the eigenvalues of $D_c+V$ \cite{Dolbeault-Esteban-Sere-00B})}. Let $V$ be  a scalar potential satisfying
(\ref{V1})-(\ref{V2})-(\ref{V3}).
Then,  for all $\,k\geq 1$,  the $k$-th eigenvalue $\lambda_k(V)$ of the operator $\,D_c+V\,$ is given by the following min-max formula
\bq\label{L11} \lambda_k (V)= 
 \inf_{\substack{Y  \ {\rm subspace \ of \ } C^\infty_{o} (\BbR^3,
\BbC^{\,2})\\ {\rm dim } Y=k}} \ \sup_{\varphi\in
Y\setminus\{0\}}\lambda^{T}\!(V,\varphi)\,,\eq where

\bq \lambda^{T}\!(V,\varphi):=\sup_{\substack{\psi=\left(^\varphi_\chi\right)\\ \chi\in
{C^\infty_0} (\BbR^3, \BbC^{\,2})}}\frac{((D_c+V)\psi, \psi)}{(\psi,\psi)}\label{XX1}\eq
is the unique number in $(K_2-c^2,+\infty)$ such that 
\bq \lambda^T(V,\varphi) \Int_{\BbR^3} |\varphi|^2 dx \! = \! \Int_{\BbR^3} \Bigl(
\Frac{c^2\,|
(\sigma\cdot\nabla) \varphi|^2}{c^2-V+\lambda^T(V,\varphi)} + (c^2+V) |\varphi|^2
\Bigr) dx.
\label{LA} \eq
\end{theorem}

The above result is optimal for Coulomb potentials for which all the  
cases $\nu\in (0,c)$ are included. But note that assumptions (\ref 
{V2})-(\ref{V3}) can be replaced by weaker ones allowing in  
particular to treat potentials which have a finite number of isolated  
singularities, even of different signs. We describe some of these  
extensions at the end of this subsection.

Theorem \ref{TT1} is a useful tool from a practical point of view in the sense that the first eigenvalue (case $k=1$) of $D_c+V$ can be obtained by a \emph{minimization procedure} over the (bounded-below) nonlinear functional $\phi\mapsto\lambda^T(V,\phi)$. Higher eigenvalues are obtained by the usual Rayleigh-Ritz minimax principle on the same nonlinear functional. As we shall see below, this has important consequences from a numerical point of view.
 
 \medskip

 Theorem \ref{TT1} is a direct consequence of an abstract theorem proved by Dolbeault, Esteban and S\'er\'e \cite{Dolbeault-Esteban-Sere-00B}, providing variational characterizations for the eigenvalues of self-adjoint operators in the gaps of their essential spectrum.
 
 \begin{theorem}\label{gaps} {\rm (Min-max principle for eigenvalues of operators with gaps \cite{Dolbeault-Esteban-Sere-00B})} 
 Let ${\mathcal H}$ be a Hilbert space 
and $A:D(A) \subset {\mathcal H}
 \rightarrow {\mathcal H}$ a
self-adjoint operator. We denote by  ${\mathcal F} (A)$ the form-domain of $A$. 
Let
${\mathcal H}_+$, ${\mathcal H}_-$ be two orthogonal Hilbert subspaces of 
${\mathcal H}$ such that ${\mathcal H}={\mathcal H}_+
{\displaystyle \oplus} {\mathcal H}_-$
and let $\Lambda_\pm\,$ be the projectors associated with  ${\mathcal H}_\pm$.
We assume the
existence of a subspace of  $D(A) $, $F$,  which is dense in $D(A) $ and  such that :
\begin{itemize}
\item[$(i)$] $F_+ = \Lambda_+ F$ and $F_- = \Lambda_- F$ are two subspaces
of ${\mathcal F}(A)$.
\item[$(ii)$] $a_-=\sup_{x_- \in F_-\setminus \{ 0\}} \frac{(x_-, Ax_-)}{
\Vert x_- \Vert^2_{_{\mathcal H}}} <+\infty $ .\medskip
\end{itemize}
Moreover we define the sequence of min-max levels
\bq c_k(A) = \  \inf_{\substack{V \ {\rm subspace \ of \ } F_+ \\ {\rm dim}\ V =k  }} \  
\Sup_{  \scriptstyle x \in ( V \oplus F_- ) \setminus \{ 0 \} } \
\Frac{(x, Ax)}{||x||^2_{_{\mathcal H}}} \ ,
\qquad k \geq 1,
\label{min-maxesA} \eq
and assume that
\begin{itemize}
\item[$(iii)$] $\qquad\qquad c_1(A) > a_- \ .$
\end{itemize}

\medskip

Then
$$\forall k \geq 1,\quad  c_k(A) \ = \ \mu_k, $$
where, if $\,b = \inf \ (\sigma_{\rm ess} (A) \cap (a_-, + \infty)) \in (a_-, +
\infty]$,   $\mu_k\,$ denotes  the $k^{\rm th}$ eigenvalue of $A$ (counted with multiplicity) in
the interval
$(a_-, b)$ if it exists, or
 $\mu_k = b$ if there
is no $k^{\rm th}$ eigenvalue. 

As a consequence, $\displaystyle{b\,=\lim_{k\to\infty} c_k(A)\,=\sup_k
c_k(A)\,>\,a_-\;.}$ 
\end{theorem}

An important feature of this min-max principle is that the min-max levels do not depend on the splitting ${\mathcal H}={\mathcal H}_+
{\displaystyle \oplus} {\mathcal H}_-$ provided assumptions $(i)$, $(ii)$ and $(iii)$ hold true. In practice, one can find many different splittings satisfying these assumptions and choose the most convenient one for a given application.

In order to treat families of operators without checking the assumptions of the above theorem for every case, there is  a ``continuous" version of Theorem \ref{gaps} in 
\cite{Dolbeault-Esteban-Sere-00B} which we shall present now. 

Let us  start with a self-adjoint operator  $A_0:D(A_0) \subset {\mathcal H}
 \rightarrow {\mathcal H}$. and denote by  ${\mathcal F} (A_0)$ the form-domain of
$A_0$. Now, for $\nu$ in an interval $[0,\bar\nu)$ we define $\,A_\nu=A_0+\nu W\,$ where $W$ is a bounded operator.
The operator 
$\,A_\nu\,$ is self-adjoint with $\,{\mathcal D}(A_\nu)=
{\mathcal D}(A_0)$,   $\,{\mathcal F}(A_\nu)=
{\mathcal F}(A_0)$.
Let
${\mathcal H}={\mathcal H}_+\oplus {\mathcal H}_-$ be an orthogonal splitting
of ${\mathcal H}$, and $P_+\;,\;P_-$ the associated projectors, as
in Section 1.  We  assume the existence of a subspace of $D(A_0) $, $F$,  dense in $D(A_0) $ and
  such that:
\begin{itemize}
\item[$(j)$] $F_+ = P_+ F$ and $F_- = P_- F$ are two subspaces
of ${\mathcal F}(A_0)$;\medskip

\item[$(jj)$] there is $\,a_-\in\BbR$ such that for all $\,\nu\in
(0,\bar\nu)$, $$a_\nu:={\sup_{x_-
\in F_-\setminus
\{ 0\}} \frac{(x_-, A_\nu x_-)}{
\Vert x_- \Vert^2_{_{\mathcal H}}} \leq a_-}. $$
\end{itemize}

For $\nu \in (0,\bar\nu)$, let $b_\nu:=\inf (\sigma_{\rm ess}(A_\nu)\cap (a_\nu,+\infty))\,,$
and for $k\geq 1$, let $\mu_{k,\nu}$ be the $k$-th eigenvalue of $A_\nu$
in the interval $(a_\nu,b_\nu)$, counted with multiplicity, if it exists.
If it does not exist, we simply let $\mu_{k,\nu}:=b_\nu\;.$ Our next assumption is
\begin{itemize}
\item[$(jjj)$] there is $\,a_+>a_-\,$ such that for all 
$\,\nu\in
(0,\bar\nu)$, $\mu_{1,\nu}\geq a_+\,.$
\end{itemize}

Finally, we define the min-max levels
\bq c_{k,\nu} := \  \inf_{\substack{V \ {\rm subspace \ of \ } F_+  \\ {\rm dim}
\ V =
k }} \  \Sup_{ x \in ( V \oplus F_- ) \setminus \{ 0 \} } \
\Frac{(x, A_\nu x)}{||x||^2_{_{\mathcal H}}} \ ,
\qquad k \geq 1\, ,
\label{min-max-nu} \eq
and assume that

\begin{itemize}
\item[$(jv)$]  $\quad c_{1,0}>a_-\;$.
\end{itemize}

\smallskip

Then, we have the 
\begin{theorem}\label{gapscontinu}  {\rm(\cite{Dolbeault-Esteban-Sere-00B})}  Under conditions $(j)$ to $(jv)$, $A_\nu$ satisfies the
assumptions $(i)$ to $(iii)$ of Theorem \ref{gaps} for all $\,\nu\in [0,\bar\nu)$, and 
$\,c_{k,\nu}= \mu_{k,\nu}\geq a_+$, for all $\,k\geq 1$.
\end{theorem}

Theorems \ref{gaps} and \ref{gapscontinu} are very good tools in the study of the point-spectrum of Dirac operators $\,D_c+\nu V\,$, where $V$ is a potential which has singularities not stronger than $c/|x-x_0|\;(0<c<1)$. Of course, Theorem \ref{gapscontinu} cannot be directly applied to the case of unbounded potentials, but this can actually be done by first truncating the potential and then passing to the limit in the truncation parameter, as  we did in the proof of  Theorem \ref{TT1}.

Theorem \ref{gapscontinu} is an easy consequence of the proof of Theorem \ref{gaps}. In contrast, the proof of Theorem \ref{gaps} is more involved. We sketch now its main steps.

\medskip

\noindent{\bf Sketch of the proof of Theorem \ref{gaps}.}
For $E > a$ and $x_+ \in F_+$, let us define
\begin{eqnarray*}
\varphi_{E, x_+} : && F_- \rightarrow \BbR \\
&& y_- \mapsto \varphi_{E, x_+} (y_-) = \Bigl( (x_+ + y_-), A(x_+ + y_-)
\Bigr) - E
||x_+ + y_- ||^2_{_{\mathcal H}} . \end{eqnarray*}

From assumptions $(i)-(ii)$, $N (y_-) = \sqrt{(a+1)||y_-||^2_{_{\mathcal H}} -
(y_-, Ay_-)}$ is a norm on $F_-$. Let
$\overline{F}_-^{^N}$ be the completion of $F_-$ for this norm. Since
$||.||_{_{\mathcal H}}
\leq N$ on $F_-$, we have $\overline{F}_-^{^N} \subset {\mathcal H}_-$. For all $x_+\in F_+$,
there is an $x\in F$ such that
$\Lambda_+ x = x_+\,$. If we consider the new variable $z_-=y_--\Lambda_-x$, we can
define 
$$\psi_{E, x}(z_-):= \varphi_{E, \Lambda_+x}(z_-+\Lambda_-x)=(A(x+z_-), x+z_-)-E(x+z_-, x+z_-)\,.$$
Since $F$ is a subspace of $D(A)$, $\psi_{E, x}$ (hence $\varphi_{E, x_+}$) is
well-defined
and continuous for $N$, uniformly on bounded sets. So, $\varphi_{E, x_+}$ has a unique
continuous extension
$\overline \varphi_{E, x_+}$ on $\overline{F}_-^{^N}$, which is continuous for
the extended norm $\overline N$. It is well-known (see e.g. \cite{Reed-Simon-78}) that
there is a unique self-adjoint operator
$B : D(B) \subset {\mathcal H}_- \rightarrow {\mathcal H}_-$ such that $D(B)$ is a subspace of
$\overline{F}_-^{^N}$, and
$$\forall x_- \in D(B),\quad \overline {N} (x_-)^2 \ = \ (a+1)|| x_-||^2_{_{\mathcal H}} + (x_-, B x_-).$$
Now, $\overline \varphi_{E, x_+}$ is of class $C^2$ on
$\overline{F}_-^{^N}$ and
\begin{eqnarray}\label{coer} D^2 \overline\varphi_{E, x_+} (x_-) \cdot (y_-,y_-) & = & -
2 (y_-, By_-) - 2E  ||
y_- ||_{\mathcal H}^2\nonumber \\
& \leq & - 2 \min (1,E)\, \overline N (y_-)^2 \ . 
\end{eqnarray}
So $\overline \varphi_{E, x_+}$ has a unique maximum, at
the point $y_- = L_E (x_+)$. The Euler-Lagrange equations
associated to this maximization problem are :
\bq\label{N7} \Lambda_-Ax_+-(B+E)y_-=0\,.\eq

The above arguments allow us, for any $E > a$, to define a map
\begin{eqnarray}\label{N5} Q_E : & F_+ \rightarrow & \BbR \nonumber\\
& x_+  \mapsto & Q_E (x_+)  = \sup_{x_- \in F_-} \ \varphi_{_{E, x_+}} (x_-) =
\overline{\varphi}_{E, x_+} (L_E x_+) \\
&&\hspace{3mm} =\left(x_+, (A-E)x_+\right)+\left(\Lambda_-Ax_+,(B+E)^{-1}\Lambda_-Ax_+\right)
\;.\nonumber
\end{eqnarray}

It is easy to see that $Q_E$ is a quadratic form with domain $F_+ \subset
{\mathcal H}_+$ and it is monotone nonincreasing in $E>a$.

We may also, for $E > a$ given, define  the norm $n_E (x_+) =
||x_+ + L_E x_+ ||_{_{\mathcal H}}$.  We consider the completion
$X$ of $F_+$ for the norm $n_E$ and denote by $\overline n _E$
the extended norm.  Then, we define another
norm on $\,F_+$ by
$${\mathcal N}_E (x_+) = \sqrt{Q_E (x_+) + (K_E+1)
(n_E (x_+))^2}\, $$
with $K_E = \max (0, \frac{E^2 (E-\lambda_1)}{\lambda^2_1})\,$
and consider the
completion $\,G\,$ of $\,F_+\,$ for the norm $\,{\mathcal N}_{E}$. 
Finally, we use the monotonicity of the map $\,E\mapsto Q_E\,$ and classical tools of spectral theory to prove that the $k$-th eigenvalue of $A$ in the interval $\,(0, \inf \ (\sigma_{\rm ess} (A) \cap (a, + \infty)))\,$
 is the unique $E_k>a$ such that 
\bq\label{rayl} \inf_{\begin{array}{c}
 \scriptstyle V \ {\rm subspace \ of \ } G  \\ \scriptstyle
{\rm dim} \
V = k  \end{array}} \  \Sup_{{  \scriptstyle x_+ \in V\setminus \lbrace 0 \rbrace }}
\ \Frac{\bar Q_{E_k} (x_+)}{(\bar n_{E_k}(x_+))^2}\ =\ 0\;  . \eq
Note that since the above min-maxes correspond to the eigenvalues of the operator $\,T^E\,$ associated to the
quadratic form $\,Q_E$, (\ref{rayl}) is actually equivalent to 
$$\lambda_k(T^{E_k})=0\,.$$
and the fact that this inequality defines a unique $E_k$ relies on the monotonicity of $\,Q_E\,$ w.r.t. $\,E$.
\qed

\bigskip

In the application of Theorem \ref{gaps} to prove Theorems \ref{TT1} and \ref{gapscontinu}, various decompositions $\,{\mathcal H}=\tilde{\mathcal H}^+{\displaystyle \oplus} \tilde{\mathcal H}^-$ could be considered. One that gives excellent  results is  defined by
\bq\,\psi=\binom{\varphi}{ 0}+\binom{0}{ \chi}\,.\eq
This decomposition yields optimal results about the point spectrum for some potentials $V$. 
There are cases for which this is not anymore true. For instance, this happens when the potential $V$ has ``large" positive and negative parts, case which is not dealt with in the previous results.

Recently, Dolbeault, Esteban and S\'er\'e \cite{Dolbeault-Esteban-Sere-05} have considered the case where a potential can give rise to two different types of eigenvalues, not only those appearing in Theorems \ref{TT1} and \ref{gapscontinu}. More precisely, if $V$ satisfies (\ref{V1}), assume that it is continuous everywhere except at two finite sets of isolated points, $ \,\{x^+_i\}\,$, $\,\{x^-_j \}, \;i=1,\dots I,\; j=1,\dots, J, \,$ where
\bq\label{V22} \begin{array}{ll}
\displaystyle\lim_{x\to x^+_i} V(x) = +\infty\,, \quad &\displaystyle\lim_{x\to x^+_i} V(x)\, |x-x^+_i|\leq \nu_i,\\ &\\
\displaystyle\lim_{x\to x^-_j} V(x)=-\infty\,,\quad &\displaystyle\lim_{x\to x^-_j} V(x)\,|x-x^- 
_j|\geq -\nu_j,
\end{array}\eq
with $\,\nu_i, \nu_j\in (0,c)$ for all $\, i,\, j$. Under the above assumptions, as above, $D_c + V$ has a distinguished self-adjoint extension $A$ with domain ${\mathcal D}(A)$ such that 
$$H^1 (\BbR^3, \BbC^4) \subset {\mathcal D} (A) \subset H^{1/2}(\BbR^3, \BbC^4).$$
The essential spectrum of $A$ is the same as that of $D_c$ :
$$\sigma_{\rm ess} (A) \ = \ (- \infty, - c^2] \cup [c^2, + \infty),$$
see \cite{Thaller-92,Schmincke-72,Nenciu-76,Klaus-Wust-78}. Finally, $\,V\,$ maps $\, \mathcal{D}(A)\,$ into its dual, since (\ref{V1}) and (\ref{V22}) imply that for all $\,\phi\in H^{1/2}(\BbR^3)$, $V\phi\in H^{-1/2}(\BbR^3)$. The decomposition of $\gH$ considered here is related to the positive/negative spectral decomposition of the free Dirac operator $D_c$ :
$$\gH=\gH^0_+\oplus\gH_-^0\,,$$
with $\gH_\pm^0= P^0_\pm\gH$, where $P^0_\pm$ are the positive/negative spectral projectors of the free Dirac operator $D_c$.

As above, we assume the existence of a core $F$ ({\sl i.e.} a subspace of $\mathcal{D}(A) $ which is dense for the norm $\norm{\cdot}_{\mathcal{D}(A)}$), such that : 
\begin{itemize}
\item[(i)] $F_+ = P^0_+ F$ and $F_- = P^0_- F$ are two subspaces of ${\mathcal D}(A)$;
\item[(ii$^-$)] $a^-:=\sup_{x_- \in F_-\setminus \{ 0\}} \frac{(x_-,  
Ax_-)}{\Vert x_- \Vert^2_{\gH}} <+\infty $;
\item[(ii$^+$)]$a^+:=\inf_{x_+ \in F_+\setminus \{ 0\}} \frac{(x_+,  
Ax_+)}{\Vert x_+ \Vert^2_{\gH}} >-\infty $.
\end{itemize}
We consider the two sequences of min-max and max-min levels $\{\lambda_k^+\}_{k\geq 1}$ and $\{\lambda_k^-\}_{k\geq 1}$ defined by
\bq \lambda^+_k := \ \inf_{\substack{V \ \text{subspace of }\  F_+ \\ {\rm dim} \ V = k }} \ \Sup_{ \scriptstyle x \in  
( V \oplus F_- ) \setminus \{ 0 \} } \ \Frac{(x, Ax)}{\|x\|^2_{_ 
\gH}}\;,\label{min-max} \eq
\bq \lambda^-_k := \ \sup_{\substack{V \ {\rm subspace \ of \ } F_-\\ {\rm dim} \ V = k }} \ \inf_{ \scriptstyle x  
\in ( V \oplus F_+ ) \setminus \{ 0 \} } \ \Frac{(x, Ax)}{\|x\|^2_{_ 
\gH}}\;.\label{max-min} \eq

\begin{theorem}\label{TT67}  {Take a positive integer $\,k_0\,$ and any $k \geq {k_0}\,$ and let $A$ be the self-adjoint extension of $D_c+V$ defined above, where $V$ is a scalar potential satisfying~(\ref{V1}) and (\ref{V22}).}

{If $\,a^-<\lambda^+_{k_0}<c^2$, then $\,\lambda^+_k \,$ is either an eigenvalue of $D_c+V$ in the interval $\,(a^-, c^2)$, or $\,\lambda^+_k=c^2$. If additionally $\,V\geq 0$, then $\,a^-=c^2\,$ and $ \,\lambda^+_k = c^2$.}

{If $\, -c^2<\lambda^-_{k_0}<a^+$, then $\,\lambda^-_k\,$ is either an eigenvalue of $D_c+V$ in the interval $\,(-c^2, a^+)\,$ or $ \,\lambda^-_k=-c^2$. If additionally $\,V\leq 0$, then $\,a^+=-c^2\,$ and $\, \lambda^-_k = -c^2$.}\end{theorem}

The sequences $\{\lambda_k^+\}_{k\geq 1}$ and $\{\lambda_k^-\}_{k\geq 1}$ are respectively nondecreasing and nonincreasing.  {As a consequence of their definitions} we have: 
\bq\label{always} \mbox{for all }\; k\geq 1, \quad \lambda^+_k\geq  
\max\,\{a^-, a^+\} \;\mbox{ and }\; \lambda^-_k\leq \min\,\{a^-, a^+\} 
\,,
\eq
and if $\,a^-\geq a^+$, we do not state anything about the possible eigenvalues in the interval $\,[a^+, a^-]$. Note that, as it is showed in \cite{Dolbeault-Esteban-Sere-05}, there are operators for which {\sl all} or {\sl almost all }the eigenvalues lie in the interval $\,[a^+, a^-]$ and thus, they are not given by the variational procedures defining the $\,\lambda^\pm_k$'s.

Finally, let us remark that if we apply Theorem \ref{TT67} to deal with a family of operators  $D_c+\tau\, V$, with $V$ satisfying (\ref{V1})-(\ref{V22}), then we see that the eigenvalues $\lambda^+_k$'s and $\lambda^-_k$'s are of a ``different" kind , since
$$\lim_{\tau \to 0^+}\lambda_k^{\tau, \pm}= \pm c^2\,,\quad\mbox{ for  
all }\; k\geq 1\,.$$
In physical words, we could say that the $\lambda_k^+$'s correspond to electronic states, and the $\lambda_k^-$'s to positronic ones.

\subsection{Numerical method based on the min-max formula}
Let us now come back to the case $\,k=1\,$ of Theorem \ref{TT1}. Note that from (\ref{Talm}) and (\ref{XX1}) we see that (under the right assumptions on $\,V$) the first eigenvalue $\lambda_1(V)$ of $\,D_c+V\,$ in the gap
$\,(-c^2, c^2)\,$  can be seen as the solution of a minimization problem, that is,
\bq\label{taltal}\lambda_1(V)= \min_{\substack{\varphi\in C^\infty(\BbR^3, \BbC^2)\\ 
\varphi\ne 0}}\, \lambda^T(V,\varphi)\,,\eq
where $\phi\mapsto\lambda^T(V,\varphi)\,$ is a nonlinear functional implicitly defined by 
\bq \lambda^T(V,\varphi) \Int_{\BbR^3} |\varphi|^2 dx \! = \! \Int_{\BbR^3} \Bigl(
\Frac{c^2\,|
(\sigma\cdot\nabla) \varphi|^2}{c^2-V+\lambda^T(V,\varphi)} + (c^2+V) |\varphi|^2
\Bigr) dx.
\label{LAbis} \eq
The idea of characterizing  the first eigenvalue in a gap of the essential spectrum as the solution of a minimization problem is
not completely new. It has for instance already been used by Bayliss and Peel
\cite{Bayliss-Peel-83} in another context. It is also close to the Fesbach method and to some techniques used in Pencil Theories. 

The fact that one can reduce the computation of $\lambda_1(V)\,$ to that of a minimization problem (\ref{taltal})-(\ref{LAbis}) has an important practical consequence: these problems 
(\ref{taltal})-(\ref{LAbis}) can now be easily discretized to construct an algorithm, allowing us to approximate 
$\lambda_1(V)\,$ in an efficient manner. Indeed, the functional $\lambda^T(V, \cdot)$ to be minimized is \emph{bounded from below} in the whole space $\,H^{1/2}(\BbR^3,\BbC^2)\,$. This is a huge advantage compared to other methods in which the total Rayleigh quotient $\,Q_v\,$ is minimized on particular finite dimensional  subspaces of  $\,H^{1/2}(\BbR^3,\BbC^4)\,$: the latter are prone to  variational collapse (entering into the negative continuum) and can even furnish spurious solutions (see, e.g. \cite{Chaix-Iracane-89}). 

The discretization method based on \eqref{taltal} is completely free of all these complications and  satisfactory numerical tests for atomic and molecular models have been presented in \cite{Dolbeault-Esteban-Sere-Vanbreugel-00, Dolbeault-Esteban-Sere-02}. Notice that molecular simulations are more complicated to carry on because one cannot use the rotational symmetry like in the atomic case. In contrast  to the one-dimensional radially symmetric problem, the discretization has to be made in $\,\BbR^2\,$ when axial symmetry is present, or in $\,\BbR^3$ in the general case. Below we describe the algorithm that was used in 
\cite{Dolbeault-Esteban-Sere-Vanbreugel-00, Dolbeault-Esteban-Sere-02} to find eigenvalues of $\,D_c+V$ by the minimization method presented above.

Consider the following approximation procedure for
$\lambda_k(V)$, $k\geq 1$. Take any complete countable basis set ${\mathcal B}$ in the space
of admissible $2$-spinors $X$ and let ${\mathcal B}_n$ be an $n$-dimensional
subset of ${\mathcal B}$ generating the space $X_n$. We assume
that ${\mathcal B}_n$ is monotone increasing in the sense that if $n<n'$, then
${\mathcal B}_n$ is contained in~${\mathcal B}_{n'}$. Denote by
$\varphi_1,\varphi_2,\dots,\varphi_n$ the elements of ${\mathcal B}_n$.
For all $1\leq i, j\leq n$, we define the $n\times n$ matrix
$A_n(\lambda)$ whose entries are
\begin{equation}\label{145} A_n^{i,j}(\lambda)=\int_{\BbR^3} \Big(\;
\frac{(c\,(\sigma\cdot\nabla)\,\varphi_i, c\,(\sigma\cdot\nabla)\, \varphi_j)}{\lambda+c^2-V}
+(V+c^2-\lambda)\,(\varphi_i, \varphi_j)\;\Big)\; dx.\end{equation}
The
matrix $A_n(\lambda)$ is selfadjoint and has therefore $n$ real
eigenvalues. For $\,1\leq k\leq n$, we compute $\lambda_{k,n}$ as the solution of the equation
\begin{equation}\label{34}\mu_{k,n}(\lambda)=0\;,\end{equation}
where $\mu_{k,n}(\lambda)$ is the $k$-th eigenvalue of $A_n(\lambda)$.
Note that the uniqueness of such a $\lambda$ comes from the monotonicity
of the r.h.s. of equation (\ref{145}) with respect to $\,\lambda$. Moreover, since for a fixed
$\lambda$
\begin{equation}\label{35}\mu_{k,n}(\lambda)\searrow \mu_k(\lambda)
\quad{\hbox{ as }\quad n \to + \infty}\;,\end{equation}
we also have
\begin{equation}\label{36} \lambda_{k,n}\searrow \lambda_k(V) \quad{\hbox{ as
}\quad n \to+\infty}\;.\end{equation}

The elements of the
basis set used in \cite{Dolbeault-Esteban-Sere-Vanbreugel-00} were Hermite functions. 
In \cite{Dolbeault-Esteban-Sere-02} more efficient numerical results have been obtained by means of $B$-spline
functions. The interest of using well-localized basis set functions
is the sparseness and the nice structure of the corresponding
discretized matrix $A_n(\lambda)$. If the degree of the basis of B-splines
increases, the number of filled diagonals will also increase. So, a good
balance has to be found between the smoothness of elements of the
approximating basis set and the speed of the corresponding numerical
computations. In \cite{Dolbeault-Esteban-Sere-02} the simple choice
of considering second order spline functions on a variable length grid was made. In the atomic case, when $1$-dimensional
$B$-splines are used, very quick and accurate results can be obtained. In \cite{Dolbeault-Esteban-Sere-02} numerical tests were provided for some axially symmetric diatomic molecules. 

\medskip

In \cite{van_Lenthe-Baerends-Snijders-95} we can find an algorithm which has some analogy with the algorithm described above.

\subsection{New Hardy-like inequalities}
Another byproduct of the minimization characterization of the first eigenvalue of $\,D_c+V\,$ given in (\ref{taltal}) and of (\ref{LA}) is the following: for all $\, \varphi\in D(D_c+V)$,
\begin{equation}\Int_{\BbR^3}\left(
\Frac{c^2|(\sigma\cdot\nabla)\,\varphi|^2}{c^2-V+\lambda_1(V)}+(c^2+V-\lambda_1(V))\,|\varphi|^2\right)
\,dx\geq 0.\label{HardyV}\end{equation} 
In the particular case $\,V=-\nu/|x|$, $\,\nu\in (0,c)\,$, (\ref{HardyV}) means that for all $\,\varphi\in H^1(\BbR^3, \BbC^2)$, 
\begin{equation}\Int_{\BbR^3}\left(
\Frac{c^2|(\sigma\cdot\nabla)\,\varphi|^2}{c^2+\nu/|x|+\sqrt{c^4-\nu^2\,c^2}}+(c^2-\,\sqrt{c^4-\nu^2\,c^2})\,|\varphi|^2\right)
\,dx\geq   \nu\Int_{\BbR^3}\frac{|\varphi|^2}{|x|}\,dx.    \label{Hardynu}\end{equation} 
By scaling,  one finds that for all  $\,\varphi\in H^1(\BbR^3, \BbC^2)$, and for all
$\,\nu\in (0, 1)$, 
\begin{equation}\Int_{\BbR^3}\left(
\Frac{|(\sigma\cdot\nabla)\,\varphi|^2}{1+\nu/|x|+\sqrt{1-\nu^2}}+(1-\sqrt{1-\nu^2})\,|\varphi|^2\right)
\,dx\geq   \nu\Int_{\BbR^3}\frac{|\varphi|^2}{|x|}\,dx \,,     \label{Hardynubis}\end{equation} 
\bigskip
and, passing to the  limit when $\,\nu\,$ tends to $1$, we get:
\begin{equation}\Int_{\BbR^3}\left(
\Frac{|(\sigma\cdot\nabla)\,\varphi|^2}{1+1/|x|}+|\varphi|^2\right)
\,dx\geq \Int_{\BbR^3}\frac{|\varphi|^2}{|x|}\,dx \,.\label{Hardynu1}\end{equation} 

This inequality is a Hardy-like inequality related to the Dirac operator. It is not invariant under dilation, which corresponds to the fact that the Dirac operator $\,D_c\,$ is not homogeneous. But by another scaling argument, (\ref{Hardynu1}) yields, as a corollary, an inequality which is invariant by dilation,
\be
\Int_{\BbR^3}
{|x|\,|(\sigma\cdot\nabla)\,\varphi|^2}
\,dx\geq \Int_{\BbR^3}\frac{|\varphi|^2}{|x|}\,dx \label{Hardyhomog}\,,
\eq
which is actually equivalent to the ``classical" Hardy inequality
\begin{equation}\Int_{\BbR^3}
{|\nabla\varphi|^2}
\,dx\geq \frac 1{4}\Int_{\BbR^3}\frac{|\varphi|^2}{|x|^2}\,dx \,.\label{Hardyclass}\end{equation} 
For a $4$-dimensional version of \eqref{Hardyhomog}, see \cite{Wust-77,Kato-83}.

Finally, note that in  \cite{Dolbeault-Esteban-Loss-Vega-04} the Hardy-like inequality (\ref{Hardynu1}) (and slightly more general ones also) has been proved by analytical means, without using any previous knowledge about the Coulomb-Dirac operator's spectrum.

\subsection{The nonrelativistic limit}
Let us now indicate how we can relate the eigenvalues of the Dirac operator to those of the Schr\"odinger operator. This relation is established by taking the limit $\,c\to +\infty\,$, so by passing to the nonrelativistic limit.

A $\psi$
with values in $\BbC^{\,4}$ satisfies the eigenvalue equation
 \begin{equation}\label{galeq}
(D_c+V)\,\psi =\lambda\,\psi \end{equation} 
if and only if, writing $\psi=\binom{\varphi}{\chi}\,$ with $\varphi,\,\chi$ taking values
in $\BbC^{\,2}$, 
\begin{equation}\label{System2spinors}
\left\lbrace \begin{array}{l}R_c\,\chi \ =\ (\lambda-c^2- V)\
\varphi\,,\\ R_c\,\varphi\ =\ (\lambda+c^2- V)\ \chi\,,\\
\end{array}\right.
\end{equation}
with 
$$R_c= -i\,c\,(\vec{\sigma}.\vec{\nabla}) = \sum_{j=1}^3
\,-i\,c\,\sigma_j\,\frac\partial{\partial x_j}.$$
Recall that $\sigma_j$,
$j=1,2,3$, are the Pauli matrices. As long as ${\lambda\! +\! c^2\! -\!
V\! \neq\! 0}$, the system (\ref{System2spinors}) can be written as
\begin{equation}\label{elim} R_c\left(\frac{R_c\,\varphi}{g_{_\mu}}\right)+
V\varphi=\mu\,\varphi\,,\quad\chi=\frac{R_c\,\varphi}{g_{_\mu}}\end{equation}
where $g_{_\mu}=\mu+2c^2-V$ and $\mu=\lambda-c^2$. 

\medskip

Assume now that $\psi_c=\binom{\varphi_c}{\chi_c}\,$ is an eigenfunction of the operator $\,(D_c+V)\,$ associated with the eigenvalue $\,\lambda_c$ which satisfies
$$\liminf_{c\to\infty}(\lambda_c-c^2)>-\infty,\qquad \limsup_{c\to\infty}(\lambda_c-c^2)<0.$$
The system (\ref{elim}) can be written as
\bq\label{elimc} \frac{-c^2\Delta \varphi_c}{\mu_c+2c^2-V}+\frac{c^2(\vec{\sigma}.\vec{\nabla})\varphi_c\cdot(\vec{\sigma}.\vec{\nabla})V}{(\mu_c+2c^2-V)^2}+V\varphi_c=\mu_c\,\varphi_c\,,\quad \chi_c=\frac{-i\,c\,(\vec{\sigma}.\vec{\nabla})\varphi_c}{\mu_c+2c^2-V}\,,
\eq
with $\,\mu_c=\lambda_c-c^2$. It is then easy to prove (see \cite{Esteban-Sere-01}) that for $c$ large,  the functions $\,\varphi_c\,$, which in principle are only in $\,H^{1/2}(\BbR^3, \BbC^2)\,$, actually belong to the space  $\,H^1(\BbR^3, \BbC^2)\,$ 
and are uniformly bounded for the $\,H^1(\BbR^3, \BbC^2)\,$ norm. Moreover, after taking subsequences, we can find
$\,\bar\varphi\in H^1(\BbR^3, \BbC^2)\,$ and $\bar\mu<0\,$ such that
$$\lim_{c\to+\infty}\|\varphi_c-\bar\varphi\|_{H^1(\R^3,\C^2)}=0\,,\quad \lim_{c\to+\infty} \mu_c=\bar\mu\,,$$ and
$$ -\frac{\Delta \bar\varphi}{2}+V\bar\varphi=\bar\mu\,\bar\varphi\,.$$
Note that $\,\chi_c$, the lower component of the eigenfunction $\,\psi_c$, converges to $0$ in the nonrelativistic limit.

\medskip

It can be proved that for all the potentials $\,V\,$ considered in the theorems of this section, all the eigenvalues in the gap $\,(-c^2, c^2)\,$ satisfy the above conditions, and converge, when shifted by the quantity $\,-c^2$, to the associated eigenvalues of the Schr\"odinger operator $-\frac{\Delta}2$ perturbed by the same potential $V$.

\subsection{Introduction of a constant external magnetic field}
The previous results were devoted to the case of a scalar electrostatic field $V$. 
The Dirac operator for a hydrogenic atom interacting with a constant magnetic field in the $x_3$-direction is given by
\begin{equation}
D^B:= \alp \cdot \left[-i\nabla + \frac{B}{2} (-x_2,x_1,  
0)\right] +\beta - \frac{\nu}{|x|} \ , \label{one}
\end{equation}
where $\nu = Z \alpha>0$, $Z$ being the nuclear charge number (we fix the speed of light $c=1$ in this subsection) and $B$ is a constant.

The magnetic Dirac operator without the Coulomb potential $\nu/|x|$ has essential spectrum $(-\infty, -1] \cup [1, \infty)$ and no eigenvalue in the gap $(-1,1)$ for any $B\in\R$. The operator $D^B$ has the same essential spectrum
and possibly some eigenvalues in the gap. The ground state energy $\lambda_1(\nu, B)$ is the smallest among these. As the field gets large enough, one expects that the ground state energy of the Dirac operator decreases and eventually penetrates the lower continuum. The implication of this for a second quantized model is that electron--positron pair creation comes into the picture \cite{Nenciu, Pickl}.
The intuition comes from the Pauli equation, where the magnetic field tends to lower the energy because of the spin.
It is therefore reasonable to define the \textit{critical field strength} $B(\nu)$ as the supremum of the $B$'s for which $\lambda_1(\nu, b)$ is in the gap $(-1,1)$ for all $b < B$. As a function of $\nu$, $\lambda_1(\nu, B)$ is non-increasing, and as a result the function $B(\nu)$ is also non-increasing. Estimates on this critical field as a function of the nuclear charge $\nu$ can be found in \cite{Dolbeault-Esteban-Loss-06}. They have been obtained by adapting to this case the variational arguments of Theorem  \ref{TT1}. One of the first results in this paper states that for all $\nu\in (0, 1)$,
\begin{equation}
\frac{0.75}{\nu^2} \leq\; B(\nu)\;\leq \;\min \left(\frac{18\pi \nu^2} 
{[3\nu^2-2]_+^2}\ ,\
e^{\,C/\nu^2} \right)\,.
\end{equation}

As a corollary we see that as $\nu \to 1$ the critical field $B(\nu)$ stays strictly positive. This is somewhat remarkable, since in the case without magnetic field the ground state energy $\lambda_1(\nu,0)$, as a function of $\nu$ tends to $0$ as
$\nu \to 1$ but {\it with an infinite slope}. Thus, one might expect very large variations of the eigenvalue at $\nu=1$ as the magnetic field is turned on, in particular one might naively expect that the ground state energy leaves the gap for small fields $B$. This is not the case.

Next, again by using the min-max characterization of $\lambda_1(\nu,B)$, it is shown in \cite{Dolbeault-Esteban-Loss-06} that for $\nu>0$ small enough, and $B$ not too large $\,\lambda_1(\nu, B)$ is asymptotically close to the ground state energy of the Coulomb-Dirac magnetic operator in the lowest relativistic Landau level $c_0(\nu,B)$. This constant is proved to be given by
\begin{equation}
c_0(\nu, B)= \inf_{f\in C^\infty_0(\R, \C)\setminus\{0\}}\quad  
\lambda^B_0(f)\,,
\end{equation}
where $\, \lambda^B_0(f)\,$ is implicitly defined by
\begin{equation}
\lambda^B_0(f)\int_{-\infty}^{+\infty}  |f(z)|^2\,dz= \int_{-\infty}^ 
{+\infty}\left(\frac{|f'(z)|^2}{1+ \lambda^B_0(f)+\nu\,a^B_0(z)}+(1- 
\nu\,a^B_0(z))\,|f(z)|^2\right)dz,
\end{equation}
and
$$
a^B_0(z)=B\,\int_0^{+\infty}\frac{s\,e^{-\frac{Bs^2}{2}}}{\sqrt{s^2 
+z^2}}\, ds\,.
$$

In \cite{Dolbeault-Esteban-Loss-06} it is proved that for $B$ not too small and $\nu$ small enough,
\begin{equation}
c_0(\nu+\nu^{3/2}, B) \le \lambda_1(\nu, B) \le c_0(\nu, B).
\end{equation}
 and that since for $\nu$ small,  $\,\nu^{3/2}<<\nu$,
 $$ c_0(\nu+\nu^{3/2}, B)\sim  c_0(\nu, B)\sim   \lambda_1(\nu, B) \quad\mbox{as}\quad \nu\to 0\,.$$
 
The one dimensional $c_0(\nu, B)$ problem, while not trivial, is simpler to calculate than the $\lambda_1(\nu, B)$ problem. As a result, in the limit as $\nu \to 0$, this new theory yields the first term in the asymptotics of the logarithm of the critical field. In particular we have the following result,
$$
\lim_{\nu \to 0} \nu \log(B(\nu))\,=\, \pi
$$

\section{The Dirac-Fock equations for atoms and molecules}

In the previous two sections we described some results concerning the solutions of nonlinear or linear Dirac equations in $\R^3$, which represent the state of one electron only (or possibly many \emph{non-interacting} electrons). We now want to present the \emph{Dirac-Fock (DF) model} which allows to describe the state of \emph{interacting} electrons, like for instance $N$ electrons in a molecule. The DF model  is very often used in quantum chemistry computations and usually gives very good numbers when the correlation between the electrons is negligible. It is the relativistic counterpart of the better known non-relativistic Hartree-Fock equations, which can indeed be seen as the non-relativistic limit ($c\to\infty$) of the Dirac-Fock model as explained below. For this reason, we start by recalling briefly the Hartree-Fock model. 

\subsection{The (non-relativistic) Hartree-Fock equations} The Hartree-Fock equations are easily derived from the linear Schr\"odinger model in which one considers the following operator
\begin{equation}
H:=\sum_{i=1}^N\left(-\frac{\Delta_{x_i}}2 + V(x_i)\right) +\sum_{1\leq i<j\leq N}\frac{1}{|x_i-x_j|}
\label{Hamiltonien_Mol}
\end{equation}
whose associated quadratic form describes the energy of $N$ interacting electrons in the potential field $V$. Most often, $V$ is the Coulomb electrostatic potential created by a positive distribution of charge $\nu\geq0$, of total charge $Z$: 
$$V=-\nu\ast\frac{1}{|x|},\qquad \int_{\R^3}\nu=Z.$$
In the case of $M$ pointwise nuclei of charges $z_1,...,z_M$ and located at $\bar x_1,...,\bar x_M$, 
 one takes $\nu=\sum_{m=1}^Mz_m\delta_{\bar x_m}$ and $Z=\sum_{m=1}^Mz_m$. But extended nuclei can also be considered in which case $\nu$ is assumed to be a smooth $L^1$ non-negative function.
 
Due to the Pauli principle, the operator $H$ acts on $\bigwedge_{i=1}^NL^2(\R^3\times\{\pm\},\C)$, that is to say the space of $L^2$ functions $\Psi(x_1,\sigma_1...,x_N,\sigma_N)$ which are antisymmetric with respect to the permutations of the $(x_i,\sigma_i)$'s. When $Z> N-1$, it is known \cite{Zhislin-60,Zhislin-Sigalov-65} that the spectrum of $H$ has the form $\sigma(H)=\{\lambda_i\}\cup[\Sigma, \infty)$ where $\{\lambda_i\}$ is an increasing sequence of eigenvalues with finite multiplicity converging to the bottom of the essential spectrum $\Sigma$. We notice that the condition $Z> N-1$ plays a special role even for the linear theory based on the operator \eqref{Hamiltonien_Mol}, as one knows \cite{Vugalter-Zhislin-77,Hunziker-Sigal-00} that only \emph{finitely many} eigenvalues exist below $\Sigma$ when $N\geq Z+1$, and that there is \emph{no} eigenvalue below $\Sigma$ when $N\gg Z$ \cite{Ruskai-82,Sigal-82,Sigal-84,Lieb-84}. In the following, we shall always assume that $Z> N-1$.

In the Hartree-Fock approximation, one computes an approximation of the first eigenvalue $\lambda_1$ of $H$ by restricting the quadratic form $\Psi\mapsto \langle\Psi,H\Psi\rangle$ to the class of the functions $\Psi$ which are a simple (Slater) determinant:
\begin{equation}
\Psi=\phi_1\wedge\cdots\wedge\phi_N
\label{Slater}
\end{equation}
where $(\phi_1,...,\phi_N)$ is an orthonormal system of $L^2(\R^3\times\{\pm\},\C)=L^2(\R^3,\C^2)$, $\int_{\R^3}(\phi_i,\phi_j)_{\C^2}=\delta_{ij}$. Denoting $\phi_i=\left(^{\phi_i^+}_{\phi_i^-}\right)$, \eqref{Slater} means more precisely
$$\Psi(x_1,\sigma_1,...,x_N,\sigma_N)=\frac{1}{\sqrt{N!}}\det(\phi_i^{\sigma_j}(x_j)).$$
Since the set of all the $\Psi$'s having the form \eqref{Slater} is not a vector subspace of $\bigwedge_{i=1}^NL^2(\R^3\times\{\pm\},\C)$, one then obtains an energy functional which is nonlinear in terms of $\phi_1,...,\phi_N$.
The associated Euler-Lagrange equations form a system of $N$ coupled nonlinear PDEs:
\begin{equation}
H_\Phi\,\phi_k = \lambda_k \phi_k,\quad k=1,...,N
\label{Hartree-Fock}
\end{equation}
where $H_\Phi$ is the so-called \emph{mean-field operator} seen by each of the $N$ electrons
\begin{equation}
H_\Phi=-\frac\Delta2 +(\rho_\Phi-\nu)\ast\frac{1}{|\cdot|}-\frac{\gamma_\Phi(x,y)}{|x-y|},
\label{MF_op_HF}
\end{equation}
with $\rho_\Phi$ being the (scalar) \emph{electronic density}
and $\gamma_\Phi$ the so-called \emph{density matrix} of the $N$ electrons (this is a $2\times2$ matrix for any $(x,y)\in\R^3\times\R^3$):
\begin{equation}
\label{def_density_matrix}
\rho_\Phi:=\sum_{i=1}^N|\phi_i|^2\quad \text{and}\quad \gamma_\Phi(x,y):=\sum_{i=1}^N\phi_i(x)\otimes\phi_i(y)^*.
\end{equation}
We notice that \eqref{MF_op_HF} means
$$(H_\Phi\,\psi)(x) = -\frac{\Delta\psi(x)}2 +\left((\rho_\Phi-\nu)\ast\frac{1}{|\cdot|}\right)(x)\psi(x) - \int\frac{\gamma_\Phi(x,y)\psi(y)}{|x-y|}dy$$
for any $\psi\in H^2(\R^3,\C^2)$.
The existence of solutions to \eqref{Hartree-Fock} when $\int_{\R^3}\nu=Z> N-1$ has been proved first by Lieb and Simon \cite{Lieb-Simon-77} by a minimization method, and then by Lions \cite{Lions-87} by general min-max arguments. See also \cite{LeBris-Lions-05} for a recent survey.

\subsection{Existence of solutions to the Dirac-Fock equations} 
The relativistic Dirac-Fock equations were first introduced by Swirles in \cite{Swirles-35}. They take the same form as the Hartree-Fock equations \eqref{Hartree-Fock}, with $-\Delta/2$ replaced by the Dirac operator $D_c$. They are of course posed for functions taking values in $\C^4$ instead of $\C^2$. Note however that when $-\Delta/2$ is replaced by $D_c$ in the formula of the $N$-body Hamiltonian \eqref{Hamiltonien_Mol}, one obtains an operator whose spectrum is the whole line $\R$ as soon as $N\geq2$. To our knowledge, it is not known whether there exist or not eigenvalues which are embedded in the essential spectrum. In any case, the relativistic $N$-body problem is not well-posed. This somehow restricts the physical interpretation of the Dirac-Fock model, compared to its non-relativistic counterpart. We refer to the next section in which a better model deduced from Quantum Electrodynamics is presented.
Despite this issue, the Dirac-Fock equations have been widely used in computational atomic physics and quantum chemistry to study atoms and molecules involving heavy nuclei, and they seem to provide very good results when the correlation between the electrons is negligible.

In the case of $N$ electrons, the Dirac-Fock equations read
\begin{equation}\label{D-F}
D_{{c,\Phi}}\,  \phi_{_k} \ = \varepsilon_k \phi_{k},\quad  \ k = 1, ..., N, \end{equation} 
where  $\,\Phi=(\phi_1,\dots,\phi_N)\,$ satisfies $\int_{\R^3}(\phi_i(x),\phi_j(x))\,dx=\delta_{ij}$, i.e.
 \begin{equation}\label{normaliz}
\text{Gram}(\Phi)=\un, \end{equation} 
and 
 \begin{equation} \label{def_DF_mean_field_op}
D_{{c,\Phi}} = D_c  + (\rho_\Phi-\nu)*\frac{1}{|x|}
 -\frac{\gamma_\Phi(x,y)}{|x-y|}, \end{equation}
 \begin{equation}
\gamma_\Phi(x,y)  = \displaystyle{\sum^N_{\ell = 1 }} \phi_{_\ell} (x) \otimes 
\phi_{_\ell} (y)^\ast,\quad
 \rho_\Phi (x) =\tr_{\C^4}(\gamma_\Phi(x,x))= \displaystyle{\sum^N_{\ell = 1}} |\phi_{_\ell} (x)|^2.
 \label{def_gamma}
 \end{equation}
Notice that $\gamma_\Phi(x,y)$ is a $4\times4$ complex matrix, and that the operator whose kernel is $\gamma_\Phi(x,y)$, is nothing but the orthogonal projector onto the space spanned by $\phi_1,...,\phi_N$. We also denote it by $\gamma_\Phi$.
 
Indeed, like for the Hartree-Fock case, equations (\ref{D-F}) are the Euler-Lagrange equations of the Dirac-Fock functional 
\begin{multline}\label{DF_energy}
{\mathcal E^{\nu,c}_{\rm DF}} (\Phi) = \displaystyle{\sum^N_{\ell = 1}} 
\Bigl(\phi_{_\ell}, D_c
\phi_{_\ell} \Bigr)_{_{L^2}} - {\sum^N_{\ell = 1}}
 \Bigl( \phi_{_\ell}, \left(\nu*\frac{1}{|x|}\right)\phi_{_\ell}\Bigr)_{_{L^2}} \\
   + \frac{1}{2} \ \displaystyle  {\iint\limits_{\BbR^3 
\times \BbR^3}}\frac{\rho_\Phi(x) \rho_\Phi(y) -  {\rm tr } \Bigl( \gamma_\Phi(x,y) \gamma_\Phi (y,x) \Bigr)}{|x-y|} dx \,dy
\end{multline}
on the manifold
$$\mathcal{M}:=\left\{\Phi=(\phi_1,...,\phi_N)\in (H^{1/2}(\R^3,\C^4)^N,\quad \text{Gram}(\Phi)=\un\right\}.$$
It will be important to notice that the functional $\mathcal{E}^{\nu,c}_{\rm DF}$ only depends on the projector $\gamma_\Phi$ defined in \eqref{def_gamma}:
\begin{equation}
{\mathcal E^{\nu,c}_{\rm DF}} (\Phi) = ``\tr\left((D_c+V)\gamma_\Phi\right)"  + \frac{1}{2} \ \displaystyle  {\iint\limits_{\BbR^3 
\times \BbR^3}}\frac{ \rho_\Phi(x) \rho_\Phi(y) -  |\gamma_\Phi(x,y)|^2}{|x-y|} dx \,dy\,,
 \label{DF_gamma}
\end{equation}
where by $``\tr\left((D_c+V)\gamma_\Phi\right)"$ we denote $\Sum_{i=1}^N((D_c+V)\varphi_i,\varphi_i)$. Note that this expression is really a trace if the $\varphi_i$'s are in $H^1(\R^3, \C^4)$.

As a matter of fact, the Euler-Lagrange equations of $\mathcal E^{\nu,c}_{\rm DF}$ on $\mathcal M$ only depend on the space spanned by  $(\phi_1,...,\phi_N)$. This explains why, up to a rotation of the $\phi_i$'s, one can always assume that the Lagrange multiplier matrix associated with the constraint \eqref{normaliz} is diagonal, with diagonal elements $(\varepsilon_1,...,\varepsilon_N)$ appearing in \eqref{D-F}.

\medskip

Finding solutions of (\ref{D-F}) is then reduced to finding critical points of the functional ${\mathcal E^{\nu,c}_{\rm DF}}$ on the manifold $\mathcal M$. Once again, the unboundedness (from above and below) of the spectrum of the free Dirac operator makes the functional ${\mathcal E^{\nu,c}_{\rm DF}}$ totally indefinite. This together with the a priori lack of compactness of the problem posed in $\BbR^3$ and the fact that we have to work on a manifold $\mathcal{M}$ and not in the whole functional space, makes the variational problem difficult. A minimization procedure is once again impossible and another method has to be found. In \cite{Esteban-Sere-99}, Esteban and Séré defined a penalized variational problem (see below for details) which can be solved by first maximizing on some part of the spinor functions $\phi_i$ and then defining a more standard min-max argument for the remaining functional, together with Morse index considerations. 

The theorem proved in \cite{Esteban-Sere-99} and improved later by Paturel in \cite{Paturel-00} states the following:
\begin{theorem}\label{thmDF} {\rm(Existence of solutions to the Dirac-Fock equations \cite{Esteban-Sere-99, Paturel-00}) } With the above notations, assume that $N$ and $Z=\int_{\R^3}\nu$ are two positive integers satisfying $\, 
 \max (Z, N) < \frac{2c}{\pi/2+2/\pi}$ and $N-1< Z$. Then, there exists an infinite  sequence $(\Phi^{c,j})_{_{j \geq 0}}$ of critical points of the
Dirac-Fock functional ${\mathcal E^{\nu,c}_{\rm DF}}$ on $\mathcal M$.
 The functions $\phi^{c,j}_1, \dots \phi^{c,j}_N\, $  satisfy the normalization constraints (\ref{normaliz}) 
and they are strong solutions, in
$H^{1/2}(\BbR^3, \BbC^{\,4}) \ \cap \ \bigcap_{1\leq q < 3/2} W^{1,q}(\BbR^3,
\BbC^{\,4})$, of the Dirac-Fock equations
\begin{equation}\label{DFF4} D_{c,{\Phi^{c,j}}}\, \ \phi^{c,j}_{_k} \ = \ \varepsilon^{c,j}_{_k} \
\phi^{c,j}_{_k} \ , \quad 1 \leq k \leq N\ ,\end{equation}
  \begin{equation} 0 \ < \
\varepsilon^{c,j}_{_1} \ \leq ... \leq \ \varepsilon^{c,j}_{_N} \ < \ c^2 \ . \end{equation} Moreover,
\begin{equation} 0 \ < \ {\mathcal E^{\nu,c}_{\rm DF}} (\Phi^{c,j}) \ < \ Nc^2 \ , \end{equation}
\begin{equation} \lim_{j \rightarrow \infty}  {\mathcal E^{\nu,c}_{\rm DF}} (\Phi^{c,j}) \ = \ Nc^2 \ . \end{equation} 
\end{theorem}

\medskip

\begin{remark}  In our units, taking into account the physical value of the speed of light $c$, the above conditions become $$N\leq Z \leq 124.$$ 
\end{remark}

The proof of the above theorem is done by defining a sequence of min-max principles providing critical points of increasing  Morse index. We notice that the solution $\Phi^{c,0}$ obtained by Theorem \ref{thmDF} when $j=0$ will play an important role, since for $c$ large it will be actually interpreted as an ``electronic ground state" (see below). 

The condition $Z, N < 2c/(2\pi+\pi/2)$ is not that unnatural, since already in the linear case such a condition was necessary to use Hardy-like inequalities ensuring the existence of a gap of the spectrum of $D_c+V$ around $0$, see \eqref{tix-ineg}. Notice however that in \cite{Esteban-Sere-99} the following additional technical assumption was used:
\bq\label{unnatural} 3N-1 < \frac{2c}{\pi/2+2/\pi}.\eq
This assumption was removed by Paturel shortly afterwards in \cite{Paturel-00}. He did so by studying a finite dimensional
reduction of the problem and then passing to the limit, after having obtained the necessary bounds. 
This proof was done in the spirit of the Conley-Zehnder proof of the Arnold's conjecture \cite{Conley-Zehnder-83}.
The sketch of the proof that we give below is that of \cite{Esteban-Sere-99} because the variational arguments are easier to explain in that case.

\medskip

\noindent{\bf Sketch of the proof of Theorem \ref {thmDF} when (\ref{unnatural}) holds.} The first (and smallest) difficulty here  is that $\,\nu*\frac{1}{|x|}\,$ is not a compact perturbation of $D_c$ when $\nu=Z\delta_0$. This creates some technical problems. They are easily solved, replacing the Coulomb potential $\,\frac{1}{\vert x\vert}\,$ by a regularized potential $\,V_\delta$. The modified energy functional is denoted now by $\,{\mathcal E}^{\nu,c, \eta}_{\rm DF}$. At the end of the proof, we shall be able to pass to the limit $\eta\to 0$.

The second difficulty is that the Morse index estimates can
only give upper bounds on the multipliers $\,\varepsilon^{c,j}_k$ in (\ref{DFF4}). But  we also want to
ensure that $\,\varepsilon^{c,j}_k>0$, since these multipliers are interpreted as the energies of the different electrons.  
To overcome this problem, we replace the constraint
Gram $(\Phi)=\un\,$ by a penalization term $\pi_p(\Phi)$,
subtracted from the energy functional: 
 \begin{equation}
\pi_p (\Phi) \ = \ {\rm tr} \ \Bigl[ \Bigl( {\rm Gram} \ \Phi \Bigr)^p \ \Bigl( \un -{\rm Gram}
\ \Phi \Bigr)^{-1} \Bigr] \ . \end{equation} 
In this way we obtain a new functional $\,{\mathcal F}_{c, \eta, p}= {\mathcal E}^{\nu,c, \eta}_{\rm DF}-\pi_p\,$, defined now on the set of $\Phi$'s satisfying
$${\mathbf 0}<\mbox{Gram}\;\Phi<\un\,.$$  

Since in the basis in which $$\mbox{Gram}\;\Phi =\mbox{Diag}(\sigma_1,\dots,\sigma_N)\,, \; 0<\sigma_1\leq \cdots\leq \sigma_N<c^2\,,$$
 the matrix 
$\,\Bigl( {\rm Gram} \ \Phi \Bigr)^p \ \Bigl( \un -{\rm Gram}
\ \Phi \Bigr)^{-1}\,$ is also diagonal and equals 
$$\mbox{Diag}(f_p(\sigma_1), \dots, f_p(\sigma_N))\,,\quad\mbox{with}\quad f_p(x):=\frac{x^p}{1-x}\,,$$ the 
corresponding Euler-Lagrange  equations are then
$$D_{c,\Phi^{c,j}}\, \phi^{c,j}_k=f'_p(\sigma^{c,j}_k)\, \phi_k^{c,j}.$$
The numbers $\varepsilon^{c,j}_k:=f'_p(\sigma^{c,j}_k)$ are now 
explicit functions of $\phi^{c,j}_k$. Thus, $f_p$ being an increasing function,
 we automatically get $\varepsilon^{c,j}_k>0$.\smallskip

The third difficulty with DF, is that all critical points have an infinite Morse
index. This kind of
problem is often encountered in the theory of Hamiltonian systems and in certain elliptic PDEs.
One way of dealing with it is to use a concavity property of the functional,
to get rid of the ``negative directions", see e.g. \cite{Amann-79, Buffoni-Jeanjean-93, Buffoni-Jeanjean-Stuart-93, Castro-Lazer-76}.
This method was used in \cite{Esteban-Sere-99}. Doing so, we get a reduced functional $I_{c, \eta,p}$. A min-max argument gives us
Palais-Smale sequences $(\Phi_{n}^{c,\eta,p})_{n\geq1}$ for $I_{c,\eta,p}$ with a Morse index ``at most $j$" (up to an error which converges to 0 as $n\to\infty$),
thanks to \cite{Fang-Ghoussoub-92, Ghoussoub-93}. Moreover, adapting the arguments of \cite{Lions-87}, we prove that the
corresponding $\varepsilon^{c,j,\eta,p,n}_k$ are bounded away  from $c^2$. Finally, the assumptions made on $\,Z, N\,$ guarantee that
the $\varepsilon^{c,j,\eta,p,n}_k$ are also bounded away from $0$, uniformly on $\eta$, $p$ and $n$.
Then we pass to the limit $(\eta,p,n)\to (0,\infty,\infty)$, and get
the desired solutions of DF, with $0 < \varepsilon^{c,j}_k < c^2$.\smallskip

The fact that we recover at the limit $\,p\to +\infty\,$ the constraint $\,$Gram$\,\Phi=\un$ is a consequence of the  {\it a priori } estimates 
$$0<\varepsilon<\varepsilon^{c,j,\eta,p}_k= f'_p(\sigma^{c,j,\eta,p}_k)\,,$$
with $\varepsilon$ independent of $\,\eta, p$. The properties of the function $f_p$ and the above inequality imply that as $\,p\to +\infty\,$,
one necessarily has $\, \sigma^{c,j,\eta,p}\to 1$, which of course is equivalent to saying that in the limit there is no loss of charge: $\int_{\R^ 3}|\phi_k^{c,j}|^2=1$.

The method that we have just described can be generalized. In \cite{Buffoni-Esteban-Sere} an abstract version is provided, with applications to nonlinear periodic Schrödinger models arising in the physics of crystalline matter.

The concavity argument of \cite{Esteban-Sere-99} works only if  (\ref{unnatural}) holds. In his theorem, Paturel \cite{Paturel-00}  got rid of 
this assumption my making a finite dimensional reduction first. This allowed him to deal with finite Morse indices
again. \hfill $\square$

\subsection{Nonrelativistic limit and definition of the Dirac-Fock ``ground state"} 
As in the case of linear Dirac equations of Section 3, it is interesting here to see what is the nonrelativistic limit of the Dirac-Fock equations. We shall recover in the limit the Hartree-Fock equations \eqref{Hartree-Fock} presented above, for the two-dimensional upper component of the $\phi_i$'s (recall that $\phi_1,...,\phi_N$ are $\C^4$-valued functions), the lower component converging to zero.
 This  was proved rigorously in \cite{Esteban-Sere-01}. This result has been of importance to better understand the variational structure of the Dirac-Fock problem and in particular to obtain a good definition of an electronic ground-state energy, which is {\sl a priori} not clear because of the unboundedness of the Dirac-Fock energy. 

\begin{theorem}\label{non_rel_limit}  {\rm(Non-relativistic limit of the Dirac-Fock equations \cite{Esteban-Sere-01})} Let be $N < Z+1$.
Consider a sequence of numbers $c_n\to +\infty$ and a sequence
$(\Phi^n)_n$ of solutions of  (\ref{D-F}), i.e.
$\Phi^n=(\phi_1^n,...,\phi^n_N)$, each $\phi_k^n$ being
in $H^{1/2}(\BbR^3,\BbC^4)$, with ${\rm Gram}\;\Phi^n=\un$ and $D_{{c_n,\Phi^n}}\phi^n_k
=\varepsilon^n_k\phi^n_k$. Assume that the
multipliers $\varepsilon^n_k$, $\,k=1,\dots, N, \,$ satisfy
$$0 < \varepsilon < \varepsilon^{n}_1 \leq ... \leq
\varepsilon^{n}_N <
c^2-\varepsilon' , \;\hbox{ with $\varepsilon,\varepsilon' > 0$ independent of $n$}. $$
 Then for $n$ large enough, each $\phi_k^n$ is
in $H^{1}(\BbR^3,\BbC^4)$, and there exists a solution of the Hartree-Fock equations \eqref{Hartree-Fock},
$\bar \Phi = \left( \bar
\varphi_1, \cdots , \bar \varphi_N \right) \in H^1(\R^3,\C^2)^N$, with negative
multipliers,
$\bar{\lambda}_1, ...,
\bar{\lambda}_N\,$, such that, after extraction of a subsequence,
\bq \lim_{n\to\infty}(\varepsilon^n_k - (c_n)^2)=\bar{\lambda}_k,
     \label{(4)} \eq
\beq \phi_k^{n} = \left({\begin{array}{c}
\bar\varphi_k^{n}
\\ \chi_k^{n}\end{array}}\right) \longrightarrow_{_{_{\hspace{-6mm} n \rightarrow + \infty
}}} \left( {\begin{array}{c}\bar
\varphi_k
\\ 0\end{array}}\right)
   \ \; {\rm in \ } \ H^1 (\BbR^3,\BbC^{\,2})\times H^1 (\BbR^3,\BbC^{\,2}),
\label{(5)}
\eeq
\beq \left\Vert\chi_k^{n}  +\frac {i}{2c_n}(\sigma\cdot
\nabla)\bar\varphi_k^{n}\right\Vert_{_{L^2(\BbR^3,\BbC^{\,2})}}=\, O(1/(c_n)^3 ),
\label{(5bis)}
\eeq
for all $k = 1, ...,N$, and
\bq {\mathcal E}^{\nu,c_n}_{\rm DF}(\Phi^n)-Nc_n^2\quad
\longrightarrow_{_{_{\hspace{-7mm} n
\rightarrow + \infty }}} \quad {\mathcal E}_{\rm HF}(\bar
\Phi). \label{5ter}
\eq
\end{theorem}

The Hartree-Fock energy $\mathcal{E}_{\rm HF}$ appearing in \eqref{5ter} is the same as \eqref{DF_energy}, but with $D_c$ replaced by $-\Delta/2$ and $\C^4$ by $\C^2$. It can be proved \cite{Esteban-Sere-01} that the critical points constructed in Theorem \ref{thmDF} all satisfy the assumptions of Theorem \ref{non_rel_limit} for any fixed $j$. Therefore, all the $\Phi^{c,j}$ converge as $c\to\infty$ to a state whose upper component is a solution of the Hartree-Fock equations, and whose lower component vanishes. This result can even be made more precise in the case of one of the ``first" solutions (i.e. corresponding to $j=0$): the critical point $\Phi^{c,0}$ does not converge to \emph{any} solution of the HF equations, but actually to a Hartree-Fock \emph{ground state}, as stated in the following

\begin{theorem}\label{thmGS}{\rm(Non-relativistic limit of the Dirac-Fock ``ground state" \cite{Esteban-Sere-01})}
Assume that $N<Z+1$ and $\nu$ are fixed, and that $c_n\to\infty$. Then the critical point $\Phi^{c_n,0}$ constructed in Theorem \ref{thmDF} for $j=0$ satisfies
$$\lim_{c_n\to\infty}\left\{\mathcal{E}^{\nu,c}_{\rm DF}(\Phi^{c_n,0})- Nc_n^2\right\}=\min_{\substack{\bar\Phi\in H^1(\R^3,\C^2)^N\\{\rm Gram}\;\bar\Phi=\un}}\mathcal{E}_{\rm HF}(\bar\Phi).$$
Up to a subsequence, $(\Phi^{c_n,0})$ converges as $c_n\to\infty$ to $\left(^{\bar\Phi_0}_{0}\right)$ where $\bar\Phi_0$ is a minimizer of $\mathcal{E}_{\rm HF}$.

Furthermore, for $c_n$ large enough, the $\varepsilon_k^{c_n,0}$ are the first positive eigenvalues of $D_{c_n,\Phi^{c_n,0}}$:
\begin{equation}
\gamma_{\Phi^{c_n,0}}=\chi_{[0, \varepsilon_N^{c_n,0}]}\left(D_{c_n,\Phi^{c_n,0}}\right)
\label{no_unfilled_shell}
\end{equation}
\end{theorem}

\medskip

We notice that \eqref{no_unfilled_shell} means that the last level $\varepsilon_N^{c_n,0}$ is necessarily totally filled. In other words, similarly to the Hartree-Fock case \cite{Bach-Lieb-Loss-Solovej-94}, ``there are no unfilled shell in the Dirac-Fock theory for $c\gg1$".

\smallskip

\begin{figure}[h]
\input{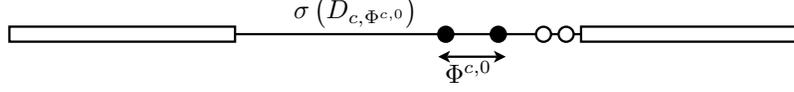}
\caption{For $c\gg1$, the Dirac-Fock `ground state' $\Phi^{c,0}$ contains the eigenfunctions associated with the $N$ first positive eigenvalues of the mean-field operator $D_{c,\Phi^{c,0}}$.}
\end{figure}

\smallskip

Although the Dirac-Fock functional $\mathcal{E}^{\nu,c}_{\rm DF}$ is not bounded-below, Theorem \ref{thmGS} allows to interpret the first min-max solution $\Phi^{c,0}$ (any of them, since there is no uniqueness) as an electronic ground state, since it converges to a Hartree-Fock ground state in the nonrelativistic limit. Actually, more has been proved in \cite{Esteban-Sere-01}:  $\Phi^{c,0}$ indeed minimizes the Dirac-Fock functional among all the $\Phi=(\phi_1,...,\phi_N)$ such that each $\phi_i$ belongs to the positive spectral subspace of the mean-field operator $D_{c,\Phi}$:

\begin{theorem}\label{M} {\rm(Variational interpretation of the Dirac-Fock ``ground state" \cite{Esteban-Sere-01})}  Assume that $N<Z+1$ and $\nu$ are fixed, and that  $c$ is sufficiently large.
Then $\Phi^{c, 0}$ is a solution of the following minimization problem:
\bq\label{minpro1}\inf\left\{{\mathcal E}^{\nu,c}_{\rm DF}(\Phi)\ |\ \Phi=(\phi_1,...,\phi_N),\ {\rm Gram}\,\Phi=
\un,\; \chi_{(-\infty, 0)}\left(D_{c,\Phi}\right)\,\Phi=0 \ \right\}\eq 
where $\,\chi_{(-\infty,0)}(D_{c,\Phi})\,$ denotes the negative spectral projector of the
operator $D_{c,\Phi}$, and 
$$\chi_{(-\infty, 0)}\left(D_{c,\Phi}\right)\,\Phi=0$$
means 
$$\chi_{(-\infty, 0)}\left(D_{c,\Phi}\right)\,\phi_k=0,$$
 for any $k=1,...,N$.
\end{theorem}
The interpretation of the theorem is the following: although the Dirac-Fock energy is unbounded from below, the critical points $\Phi^{c,0}$ are, for $c$ large enough, the minimizers of ${\mathcal E}^{\nu,c}_{\rm DF}$ on the set of functions satisfying the \emph{nonlinear} condition $\chi_{(-\infty, 0)}\left(D_{c,\Phi}\right)\,\Phi=0$. Calling these functions \emph{electronic}, we obtain that for $c$ large, $\Phi^{c, 0}$ is really an \emph{electronic ground state}. In particular, it is also a minimizer of $\mathcal{E}^{\nu,c}_{\rm DF}$ among all the solutions of Dirac-Fock equations with positive Lagrange multipliers.

\begin{remark}
Recently, explicit bounds on $c$ under which Theorem \ref{M} is valid have been provided by Huber and Siedentop \cite{Huber-Siedentop-06}.
\end{remark}

\section{The mean-field approximation in Quantum Electrodynamics}

In this last section, we want to present some progress that has been made recently concerning models from Quantum Electrodynamics (QED) in which, instead of trying to `avoid' the negative spectrum of $D_c$, the latter is reinterpreted as Dirac's vacuum and completely `incorporated' into the model. The somehow surprising consequence will be that, contrary to the previous sections, the energy functional will be \emph{bounded from below}. The price to pay is that one has to deal with \emph{infinitely many particles} (many of them being `virtual') instead of the finite number of electrons as previously. 

As mentioned in the introduction, Dirac interpreted the negative spectrum of his operator $D_c$ as  follows \cite{Dirac-30,Dirac-34a,Dirac-34b}. Since there exists no electron of negative kinetic energy, one has to find a way of avoiding the negative spectrum. This is done by assuming that the negative spectrum energies are all occupied by virtual electrons, one in each energy state, and that this (virtual) distribution of charge is not felt by the real particles on account on its uniformity. Then, the real electrons can in general only have positive energies due to the Pauli principle which prevents them to be in the same state as a virtual particle. 

Mathematically, Dirac's postulate is interpreted as follows: the states of the real free electrons necessarily belong to the positive spectral subspace $\gH^0_{+,c}=P_{+,c}^0\,\gH$ defined in Proposition \ref{R1}, and the vacuum (Dirac sea) is described by the infinite rank spectral projector $P^0_{-,c}$. Recall that in Hartree or Dirac-Fock theories, the density matrix $\gamma_\Phi$ of the $N$ electron state $\Psi=\phi_1\wedge\cdots\wedge\phi_N$ is precisely the orthogonal projector onto ${\rm span}(\phi_1,...,\phi_N)$, see, e.g. formulas \eqref{Slater} and \eqref{def_density_matrix}. Therefore, when one represents the Dirac sea by the projector $P^0_{-,c}$, one describes formally the vacuum as an infinite Slater determinant
\begin{equation}
\label{infinite_Slater}
\Omega^0=\phi_1\wedge\cdots\wedge\phi_i\wedge\cdots
\end{equation}
where $(\phi_i)_{i\geq1}$ is an orthonormal basis of $\gH^0_{-,c}=P^0_{-,c}\,\gH$.
Dirac's postulate is based on the important invariance by translation of the projector $P^0_{-,c}$, which is used to neglect the physical influence of this `constant background'. 

It was realized just after Dirac's discovery that, for consistency of the theory, the vacuum should not be considered as a totally virtual physical object which does not interact with the real particles.  Dirac himself \cite{Dirac-30,Dirac-34a,Dirac-34b} conjectured the existence of surprising physical effects as a consequence of his theory, which were then experimentally confirmed. First, the virtual electrons of the Dirac sea can feel an external field and they will react to this field accordingly, i.e. the vacuum will become \emph{polarized}. This polarization is then felt by the real particles and one therefore is led to consider a coupled system `Dirac sea + real particles'. From the experimental viewpoint, vacuum polarization plays a rather small role for the calculation of the Lamb shift of hydrogen but it is important for high-$Z$ atoms \cite{Mohr-Plunien-Soff-98} and it is even  a crucial physical effect for muonic atoms \cite{Foldy-Eriksen-54,Glauber-60}. Second, in the presence of strong external fields, the vacuum could react so importantly that an electron-positron pair can be spontaneously created \cite{Nenciu, RG,RGA,RMG}. 

The mathematical difficulties of a model aiming at describing both the Dirac sea and the real particles are important, for one has to deal at the same time with infinitely many particles (the real ones and the virtual ones of the Dirac sea). In the following, we present a Hartree-Fock (mean-field) type model for this problem, which has been mathematically studied by Hainzl, Lewin, Séré and Solovej \cite{HLS1,HLS2,HLS3,HLSo}. The model under consideration is inspired of an important physical article by Chaix and Iracane \cite{Chaix-Iracane-89} in which the possibility that a bounded-below energy could be obtained by adding vacuum polarization was first proposed. But the equations of this so-called Bogoliubov-Dirac-Fock model were already known  in Quantum Electrodynamics (QED) \cite{RGA}. For the sake of simplicity, we shall not explain how the model is derived from the QED Hamiltonian and we refer to \cite{HLS1,HLS2,HLS3,HLSo} and the review \cite{HLSS}. In the version studied in these works, the electromagnetic field is not quantized (photons are not considered). 

\medskip

Let us now describe \emph{formally} the mean-field approximation in QED, following mainly \cite{HLSo,HLSS}. The state of our system will be represented by an infinite rank projector $P$. This projector should be seen as the density matrix of an infinite Slater determinant made of an orthonormal basis of the subspace $P\gH$, like \eqref{infinite_Slater}. The projector $P$ describes not only Dirac's vacuum but the whole system consisting of the infinitely many virtual particles of the vacuum, together with the finitely many real particles (they could be electrons or positrons). Indeed, it is important to realize that in this model there is no \emph{a priori} possible distinction between the real and the virtual particles. For $N$ electrons, this would correspond to a decomposition of the form $P=P_{\rm vac} + \gamma$ where $\gamma$ is an orthogonal projector of rank $N$ satisfying $P_{\rm vac}\gamma=\gamma P_{\rm vac}=0$ (for $N$ positrons, this becomes $P=P_{\rm vac} - \gamma$). But there are infinitely many such decompositions for a given $P$ and a given $N$: it will only be for the final solution of our equation that this decomposition will be done in a natural way.

For the sake of simplicity, we take $c=1$ except at the very end of this section. We recall that in this case an additional parameter $\alpha=e^2$ appears, where $e$ is the charge of the electron. 
The energy in the state $P$ of our system in the presence of an external field $V=-\alpha\nu\ast|\cdot|^{-1}$ is then \emph{formally} given by \cite{HLSo}
\begin{equation}\label{def_energy_QED}
\mathcal{E}_{\rm QED}^\nu(P) := \mathcal{E}^\nu(P-1/2)
\end{equation}
where $\mathcal{E}^\nu$ is the Dirac-Fock functional expressed in term of the density matrix, see \eqref{DF_gamma},
\begin{multline}\label{formal_energy}
\mathcal{E}^\nu(\Gamma):= \tr(D_1\Gamma) -\alpha \int_{\R^3}\left(\nu\ast\frac{1}{|\cdot|}\right)(x)\rho_\Gamma(x)\,dx \\
+\frac{\alpha}{2}\iint_{\R^3\times\R^3}\frac{\rho_\Gamma(x)\rho_\Gamma(y)}{|x-y|}dx\,dy
- \frac{\alpha}{2}\iint_{\R^3\times\R^3}\frac{|\Gamma(x,y)|^2}{|x-y|}dx\,dy.
\end{multline}
The density of charge $\nu$ which creates the field $V$ can for instance represent a system of nuclei in a molecule. But in the following we shall not allow pointwise nuclei as we did for the Dirac-Fock model, and $\nu$ will essentially be an $L^1_{\rm loc}$ function. This is not important for pointwise particles do not exist in nature.

The subtraction of half the identity in \eqref{def_energy_QED} is a kind of \emph{renormalization} which was introduced by Heisenberg \cite{Hei} and has been widely used by Schwinger (see \cite[Eq. $(1.14)$]{Sch1}, \cite[Eq. $(1.69)$]{Sch2} and \cite[Eq. $(2.3)$]{Sch3}) as a necessity for a covariant formulation of QED. The importance of this renormalization will be clarified below.

Before we come to the problems of definition of the above energy (which are numerous), let us mention briefly how $\mathcal{E}_{\rm QED}^\nu$ is supposed to be used. There are two possibilities. If one wants to find the state of the vacuum alone in the field $V$ (no particle at all), then one has to minimize $P\mapsto \mathcal{E}^\nu(P-1/2)$ on the whole set of orthogonal projectors. When $\nu=0$, one should obtain the \emph{free vacuum}, a translation-invariant projector which is supposed to be physically unimportant. When $\nu\neq0$, one should obtain the \emph{polarized vacuum} in the presence of the field $V$. It can formally be seen that such a minimizer $P$ should be a solution of the following nonlinear equation:
\begin{equation}
\label{formal_eq_vac}
P=\chi_{(-\infty, 0)}\left(D_1+\alpha(\rho_{[P-1/2]}-\nu)\ast\frac{1}{|\cdot|}-\alpha\frac{(P-1/2)(x,y)}{|x-y|}\right).
\end{equation}
 If one wants to describe a system containing $N$ real particles (or more correctly of total charge $-eN$), then one has to minimize $P\mapsto \mathcal{E}^\nu(P-1/2)$ under the additional constraint
$\tr(P-1/2)=N$. Usually this will provide us with a projector $P$ which describes adequately the state of a system of charge $-eN$ interacting with Dirac's vacuum. It is formally solution of the following nonlinear equation
\begin{equation}
\label{formal_eq_mol}
P=\chi_{(-\infty, \mu]}\left(D_1+\alpha(\rho_{[P-1/2]}-\nu)\ast\frac{1}{|\cdot|}-\alpha\frac{(P-1/2)(x,y)}{|x-y|}\right),
\end{equation}
$\mu$ being a Lagrange multiplier due to the charge constraint. If $\nu$ is not too strong, the mean-field operator $$\cD:=D_1+\alpha(\rho_{[P-1/2]}-\nu)\ast\frac{1}{|\cdot|}-\alpha\frac{(P-1/2)(x,y)}{|x-y|}$$
will have exactly $N$ eigenvalues in $[0, \mu]$ (counted with their multiplicity). Therefore, there will be a natural decomposition
$$P=\chi_{(-\infty, 0)}\left(\cD\right)+\chi_{[0, \mu]}\left(\cD\right)=\chi_{(-\infty, 0)}\left(\cD\right) + \sum_{\ell=1}^N\phi_i\otimes\phi_i$$
where the $\phi_i$'s are solutions of
\begin{equation}
 \cD\phi_k=\varepsilon_k\,\phi_k,\quad \varepsilon_k>0,
\label{equation_DF_QED}
\end{equation}
which is the natural decomposition between the real and the virtual particles mentioned above. It will be shown below that Equations \eqref{equation_DF_QED} are very close to the Dirac-Fock equations \eqref{D-F}.

\medskip

The goal is now be to give a mathematical meaning to this program. 

First, one has to face an important difficulty: when $P$ is an orthogonal projector and $\gH$ is infinite dimensional, then $P-1/2$ is never compact. Therefore, none of the terms of \eqref{formal_energy} makes sense. In \cite{HLSo}, Hainzl, Lewin and Solovej tackled this problem in the following way: they considered a box $\cC_L:=[-L/2, L/2)^3$ of volume $L^3$ and replaced the ambient space $\gH$ by the space $L^2_{\rm per}(\cC_L)$ of $L^2$ functions on $\cC_L$ with periodic boundary conditions, with moreover a cut-off in Fourier space. Doing so, the problem becomes finite dimensional and everything makes perfect sense. Then, when no charge constraint is imposed, they took the thermodynamic limit $L\to\infty$ with the Fourier cut-off fixed and proved that the sequence of minimizers converges to a limiting projector $P$ satisfying equation \eqref{formal_eq_vac}. In this way, they could give a meaning to the minimization of the QED functional \eqref{def_energy_QED} and to the equation \eqref{formal_eq_vac}. Moreover, they also obtained the so-called Bogoliubov-Dirac-Fock (BDF) energy which attains its minimum at the limit state $P$. This energy has been studied by Hainzl, Lewin and Séré in \cite{HLS1,HLS2,HLS3}. In the following, we describe all these results in more detail.

Notice that the Fourier cut-off will \emph{not} be removed in this study. Indeed it is well-known that QED contains important divergences which are difficult to remove. Therefore $\gH$ will be replaced by the functional space
$$\gH_\Lambda:=\left\{f\in L^2(\R^3,\C^4),\ {\rm supp}(\widehat{f})\subset B(0,\Lambda)\right\}.$$

The ideas described in this section have recently been adapted \cite{Cances-et-al-07} to a nonlinear model of solid state physics describing nonrelativistic electrons in a crystal with a defect.

\subsection{Definition of the free vacuum}
We start by explaining how the free vacuum was constructed in \cite{HLSo}, in the case $\nu=0$. This is done by defining the energy \eqref{formal_energy} for projectors $P$ acting on the finite-dimensional space
\begin{eqnarray*}
\gH_\Lambda^L & := &\left\{f\in L^2_{\rm per}([-L/2, L/2)^3,\C^4),\ {\rm supp}(\widehat{f})\subset B(0,\Lambda)\right\}\\
 & = & \text{span}\left\{\exp(ik\cdot x),\ k\in(2\pi/L)\Z^3\cap B(0,\Lambda)\right\}.
\end{eqnarray*}
To define the energy properly, it is necessary to periodize the Coulomb potential as follows:
$$W_L(x)=\frac{1}{L^3}\left(\sum_{\substack{k\in(2\pi)\Z^3/L\\ k\neq0}}\frac{4\pi}{|k|^2}e^{ik\cdot x}+w L^2\right)\,,$$
where $w$ is some constant which is chosen such that $\min_{\,\cC_L} W_L=0$ for any $L$. The Dirac operator $D_1$ is also easily defined on $\gH_\Lambda^L$: it is just the multiplication of the Fourier coefficients by $(D_1(k))_{k\in 2\pi \Z^3/L}$. Then, one introduces
\begin{multline}
\label{energy_box}
\mathcal{E}_L^0(\Gamma):=\tr_{\gH_\Lambda^L}(D_1\Gamma) + 
+\frac{\alpha}{2}\iint_{\cC_L\times\cC_L}W_L(x-y)\rho_\Gamma(x)\rho_\Gamma(y)\,dx\,dy\\
- \frac{\alpha}{2}\iint_{\cC_L\times\cC_L}W_L(x-y)|\Gamma(x,y)|^2dx\,dy\,,
\end{multline}
for any self-adjoint operator $\Gamma$ acting on $\gH_\Lambda^L$. The kernel $\Gamma(x,y)$ of $\Gamma$ is easily defined since $\gH_\Lambda^L$ is finite-dimensional. Its density $\rho_{\Gamma}$ is then defined as $\rho_\Gamma(x):=\tr_{\C^4}(\Gamma(x,x))$. A translation-invariant operator $T$ acting on $\gH_\Lambda^L$ is by definition a multiplication operator in the Fourier domain. In this case, one has $T(x,y)=f(x-y)$ for some $f$ and therefore $\rho_T$ is constant. The identity of $\gH_\Lambda^L$, denoted by $I_\Lambda^L$  is an example of a translation-invariant operator.

The first result proved in \cite{HLSo} is the following:
\begin{theorem}\label{trans_inv} {\rm(QED mean-field minimizer in a box \cite{HLSo})} Assume that $0\leq \alpha< 4/\pi$,  $\Lambda>0$ and that $L$ is large enough. Then the functional $\mathcal{E}_L^0$ has a unique minimizer $\Gamma_L^0$ on the convex set
$$\mathcal{G}_\Lambda^L:=\left\{\Gamma\in\mathcal{L}(\gH_\Lambda^L),\ \Gamma^*=\Gamma,\ -I_\Lambda^L/2 \leq \Gamma \leq I_\Lambda^L/2\right\}.$$
It is invariant by translation and satisfies $\rho_{\Gamma_L^0}\equiv0$. Moreover, it takes the form $\Gamma_L^0=\cP_L^0-I_\Lambda^L/2$ where $\cP^0_L$ is an orthogonal projector on $\gH_\Lambda^L$.
\end{theorem}
Notice that in the definition of the variational set $\mathcal{G}_\Lambda^L$, we did not consider only operators taking the form $P-I_\Lambda^L/2$ , where $P$ is an orthogonal projector as suggested by \eqref{def_energy_QED}, but we indeed extended the energy functional to the convex hull of this set. But, as usual in Hartree-Fock type theories \cite{Lieb-81}, the global minimizer is always an extremal point, i.e. a state taking the form $P- I_\Lambda^L/2$.

Of course, it can easily be shown that $\cP^0_L$ satisfies an equation similar to \eqref{formal_eq_vac} with $\nu$ removed and $1/|x|$ replaced by $W_L$. But we do not give the details since we are more interested in the limit of $\cP^0_L$ as $L\to\infty$.

\medskip

To be able to state the thermodynamic limit correctly, one needs first to introduce the translation-invariant projector $\cP^0_-$ acting on $\gH_\Lambda$, which will be the limit of the sequence $(\cP^0_L)_L$. The identity of $\gH_\Lambda$ is denoted by $I_\Lambda$. We introduce
\begin{equation}
\mathcal{T}(A)=\frac{1}{(2\pi)^{3}}\int_{B(0,\Lambda)}\tr_{\C^4}[D^0(p)A(p)]dp-\frac{\alpha}{(2\pi)^5}\iint_{B(0,\Lambda)^2}\frac{\tr_{\C^4}[A(p)A(q)]}{|p-q|^2}dp\,dq
\label{def_fn_F}
\end{equation} 
for any $A$ belonging to the convex set
$$\mathcal{A}_\Lambda:=\left\{A\ \text{translation-invariant on } \gH_\Lambda,\ A^*=A,\ -I_\Lambda/2\leq A\leq I_\Lambda/2\right\}.$$
It will be shown in Theorem \ref{free_thermo} below that $\mathcal{T}$ represents the energy per unit volume of translation-invariant operators. For this reason, one now considers the minimization of $\mathcal{T}$ on $\mathcal{A}_\Lambda$. The following was proved in \cite{HLSo}:
\begin{theorem}{\rm (Definition of the free vacuum \cite{HLSo})}\label{free_vac}
Assume that $0\leq\alpha<4/\pi$ and $\Lambda>0$. Then $\mathcal{T}$ possesses a unique global minimizer $\Gamma^0$ on $\mathcal{A}_\Lambda$. It satisfies the self-consistent equation
\begin{equation}
\left\{\begin{array}{l}
\displaystyle\Gamma^0=-\frac{{\rm sgn}(\cD^0)}{2},\\
\displaystyle\cD^0=D_1-\alpha\frac{\Gamma^0(x,y)}{|x-y|}
\end{array} \right.
\label{scf_proj}
\end{equation} 
or, written in terms of the translation-invariant projector $\cP^0_-=\Gamma^0+I_\Lambda/2$,
\begin{equation}
\cP^0_-=\chi_{(-\infty, 0)}\left(\cD^0\right).
\label{scf_proj2}
\end{equation} 
Moreover, $\cD^0$ takes the special form, in the Fourier domain,
\begin{equation}
\cD^0(p)=g_1(|p|)\alp\cdot p +g_0(|p|)\beta
\label{form_new_D}
\end{equation} 
where $g_0$, $g_1\in L^\infty([0, \Lambda),\R)$ are such that $1\leq g_1(x)\leq g_0(x)$ 
for any $x\in[0, \Lambda)$, and therefore
\begin{equation}
|D_1(p)|^2\leq |\cD^0(p)|^2\leq g_0(|p|)|D_1(p)|^2.
\label{propD3}
\end{equation} 
\end{theorem} 
The self-consistent equation \eqref{scf_proj} has already been solved by Lieb and Siedentop in a different context \cite{Lieb-Siedentop-00}. They used a fixed point method only valid when $\alpha\log\Lambda\leq C$ for some constant $C$.

As shown by the next result, the negative spectral projector $\,\cP^0_-\,$ of the Dirac-type operator $\cD^0$ represents the \emph{free vacuum}, as it is the limit of the sequence $\cP_L^0$ when $L\to\infty$. An important property of $\Gamma^0$ showing the usefulness of the subtraction of half the identity in \eqref{def_energy_QED} is the following. Due to
$$\cP^0_-(p)-I_\Lambda(p)/2=\Gamma^0(p)=-\frac{g_1(|p|)\alp\cdot p +g_0(|p|)\beta}{2|\cD^0(p)|},$$
one infers
$$\tr_{\C^4}(\Gamma^0(p))=\tr_{\C^4}[(\cP^0_--I_\Lambda/2)(p)]=0\,,$$ 
for any $p\in B(0,\Lambda)$, the Pauli matrices being trace-less. This has the important consequence that the (constant) density of charge of the free vacuum vanishes:
$$\rho_{\Gamma^0}\equiv (2\pi)^{-3}\int_{B(0,\Lambda)}\tr_{\C^4}(\Gamma^0(p))\, dp=0,$$
which is physically meaningful. This formally means that
\begin{equation}
\text{``}\,\tr\left(\cP^0_- - I_\Lambda/2\right)=\int_{\R^3}\rho_{\Gamma^0}\, dx=0\,\text{''}.
\label{charge_nulle}
\end{equation}

 We notice that $\cP^0_-$ is \emph{not} Dirac's original choice $P^0_{-}$ (except when $\alpha=0$) because the interaction between the particles (the virtual and the real ones) is taken into account by the model. Notice also that Equation \eqref{scf_proj2} is exactly the same as \eqref{formal_eq_vac} with $\nu=0$, due to \eqref{scf_proj}.

As a consequence of \eqref{form_new_D}, the spectrum of $\cD^0$ is
$$\sigma(\cD^0)=\left\{\pm \sqrt{g_0(|p|)^2+g_1(|p|)^2|p|^2},\ p\in B(0,\Lambda)\right\}.$$
It has a gap which is greater than the one of $D^0$, by \eqref{propD3}:
\begin{equation}
1\leq m(\alpha):=\min\sigma(|\cD^0|).
\label{def_threshold}
\end{equation}
In \cite{HLS3}, it is proved that when $\alpha\ll1$, then $m(\alpha)=g_0(0)$ and conjectured this is true for any $0\leq \alpha<4/\pi$. Notice that the following expansion is known \cite{Lieb-Siedentop-00,HLSo}: $g_0(0)=1+\frac{\alpha}{\pi}\rm{arcsinh}(\Lambda)+O(\alpha^2)$.

\smallskip

\begin{figure}[h]
\input{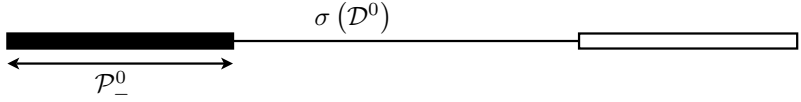}
\caption{The free vacuum $\cP^0_-$ is the negative spectral projector of the translation-invariant operator $\cD^0$.}
\end{figure}

\smallskip

We are now able to state the thermodynamic limit, as proved in \cite{HLSo}
\begin{theorem}{\rm (Thermodynamic limit in the free case \cite{HLSo})}\label{free_thermo} Assume that $0\leq\alpha<4/\pi$ and $\Lambda>0$. Then, one has 
$$\lim_{L\to\infty}\frac{\mathcal{E}_L^0(\Gamma_L^0)}{L^3} = \min_{\mathcal{A}_\Lambda}\mathcal{T}\,,$$
where we recall that $\Gamma_L^0$ is the unique minimizer of $\mathcal{E}_L^0$ defined in Theorem \ref{trans_inv}. Moreover, $\cP^0_L=\Gamma^0_L+I_\Lambda^L/2$ converges to $\cP^0_-$ in the following sense:
$$\lim_{L\to\infty}\norm{\cP^0_L-\cP^0_-}_{\gS_\infty(\gH_\Lambda^L)}=\lim_{L\to\infty}\sup_{p\in (2\pi\Z^3/L)\cap B(0,\Lambda)}|\cP^0_L(p)-\cP^0_-(p)|=0.$$
\end{theorem}

\subsection{The Bogoliubov-Dirac-Fock model}
Now that  the free vacuum $\cP^0_-$ has been correctly  defined, we will be able to introduce the Bogoliubov-Dirac-Fock (BDF) energy as studied in \cite{HLS1,HLS2,HLS3,HLSo}. Formally, it measures the energy \eqref{def_energy_QED} of a state $P$, relatively to the (infinite) energy of the free vacuum $\cP^0_-$. It will only depend on $P-\cP^0_-$ and reads formally:
\begin{eqnarray}
``\,\cE_{\rm BDF}^\nu(P-\cP^0_-) & = & \cE_{\rm QED}^\nu(P)-\cE_{\rm QED}^0(\cP^0_-)\nonumber\\
 & = & \cE^\nu(P-I_\Lambda/2)-\cE^0(\cP^0_--I_\Lambda/2)\label{formal_BDF}\\
 & = & \tr\left(\cD^0Q\right) - \alpha \iint_{\R^3\times\R^3}\frac{\rho_{Q}(x)\nu(y)}{|x-y|}dx\,dy\nonumber\\
  & & \!\!\!\!\!\!\!\!\!\!+\frac{\alpha}{2}\iint_{\R^3\times\R^3}\frac{\rho_{Q}(x)\rho_Q(y)}{|x-y|}dx\,dy-
  \frac{\alpha}{2}\iint_{\R^3\times\R^3}\frac{|Q(x,y)|^2}{|x-y|}dx\,dy\,",\label{formal_BDF2}
\end{eqnarray}
with $Q=P-\cP^0_-$. This new energy  looks again  like a Hartree-Fock type functional except that our main variable is $Q=P-\cP^0_-$, which measures the difference between our state and the (physically unobservable) translation-invariant free vacuum $\cP^0_-$. 

The energy \eqref{formal_BDF2} was introduced and studied by Chaix-Iracane in \cite{Chaix-Iracane-89} (see also Chaix-Iracane-Lions \cite{Chaix-Iracane-Lions-89}). An adequate mathematical formalism was then  provided by Bach, Barbaroux, Helffer and Siedentop \cite{Bach-Barbaroux-Helffer-Siedentop-99} in the free case $\nu=0$, and by Hainzl, Lewin and Séré \cite{HLS1,HLS2} in the external field case $\nu\neq0$. However, in all these works a simplified version was considered: $\cD^0$ and $\cP^0_-$ were replaced by Dirac's choice $D_1$ and $P^0_{-}$. As mentioned in \cite{HLSo}, although the choice of $\cP^0_-$ for the free vacuum is better physically, the two models are essentially the same from the mathematical point of view: the main results of \cite{Bach-Barbaroux-Helffer-Siedentop-99,HLS1,HLS2} can be easily generalized to treat the model in which $\cD^0$ and $\cP^0_-$ are used.

What is gained with \eqref{formal_BDF2} is that $Q$ can now be a compact operator (it will be Hilbert-Schmidt, indeed) and, thanks to the Fourier cut-off $\Lambda$, many of the terms in \eqref{formal_BDF2} will be mathematically well-defined. 
However, it will be necessary to generalize the trace functional to define correctly the kinetic energy $\tr\left(\cD^0Q\right)$. This is done by introducing the following space
$$\gS_1^{\cP^0_-}(\gH_\Lambda):= \left\{ Q\in \gS_2(\gH_\Lambda)\ |\ \cP^0_-Q\cP^0_-,\ \cP^0_+Q\cP^0_+\in \gS_1(\gH_\Lambda)\right\}\,,$$
with the  the usual notation 
$$\gS_p(\gH_\Lambda):=\{A\in\mathcal{L}(\gH_\Lambda),\ \tr(|A|^p)<\infty\}.$$
An operator $Q$ belonging to $\gS_1^{\cP^0_-}(\gH_\Lambda)$  is said to be \emph{$\cP^0_-$-trace class}. For any such $Q\in\gS_1^{\cP^0_-}(\gH_\Lambda)$, we then define its \emph{$\cP^0_-$-trace} as
$$\tr_{\cP^0_-}(Q):= \tr(\cP^0_-Q\cP^0_-) + \tr(\cP^0_+Q\cP^0_+).$$
Due to the fact that the free vacuum has a vanishing charge \eqref{charge_nulle}, $\tr_{\cP^0_-}(Q)$ can be interpreted as the charge of our state $P=Q+\cP^0_-$. We refer to \cite{HLS1} for interesting general properties of spaces of the form $\gS_1^{P'}(H)$ for any projector $P'$ and infinite-dimensional Hilbert space $H$.

Thanks to the cut-off in Fourier space, the charge density $\rho_Q$ of an operator $Q\in\gS_1^{\cP^0_-}$ is well-defined in $L^2(\R^3,\R)$, via
$$\widehat{\rho_Q}(k)=(2\pi)^{-3/2}\int_{\substack{|p+k/2|\leq L\\ |p-k/2|\leq L}}\tr_{\C^4}\left(\widehat{Q}(p+k/2,p-k/2)\right)\, dp.$$
Finally,  the following notation is  introduced
$$D(f,g)=4\pi\int\frac{\overline{\widehat{f}(k)}\widehat{g}(k)}{|k|^2}dk\,,$$
for any $(f,g)\in L^2(\R^3,\R)^2$, which coincides with $\iint_{\R^3\times\R^3}f(x)g(y)|x-y|^{-1}dx\,dy$ when $f$ and $g$ are smooth enough.
 
It is possible now to define the Bogoliubov-Dirac-Fock energy as \cite{HLS1,HLS2,HLSo}
\begin{equation}
\label{BDF1}
\cE^\nu_{\rm BDF}(Q):= \tr_{\cP^0_-}(\cD^0Q)-\alpha D(\rho_Q,\nu)+\frac{\alpha}{2}D(\rho_Q,\rho_Q)-\frac{\alpha}{2}\iint_{\R^6}\frac{|Q(x,y)|^2}{|x-y|}dx\,dy\,,
\end{equation} 
where 
\begin{equation}
Q\in  \mathcal{Q}_{\Lambda}:=\left\{Q\in\gS^{\cP^0_-}_1(\gH_\Lambda),\ -\cP^0_-\leq Q\leq \cP^0_+\right\}.
\label{def_Q_Lambda}
\end{equation} 
It is proved in \cite{HLS3} that any $Q\in\gS_1^{\cP^0_-}(\gH_\Lambda)$ automatically has its density $\rho_Q$ in the following so-called Coulomb space:
\begin{equation}
\cC=\left\{f\ |\ D(f,f)<\infty\right\}\,,
\label{def_C2}
\end{equation}
which is the natural space for defining the terms depending on $\rho_Q$ in \eqref{BDF1}.
Notice that the set $\mathcal{Q}_\Lambda$ is one more time the convex hull of our initial states $P-\cP^0_-$ where $P$ is an orthogonal projector on $\gH_\Lambda$.
The following was proved:
\begin{theorem}[The BDF energy is bounded-below \cite{Chaix-Iracane-Lions-89,Bach-Barbaroux-Helffer-Siedentop-99,HLS1,HLS2,HLSo}]\label{BDF_bound_below} Assume that $0\leq\alpha < 4/\pi$, $\Lambda>0$ and that $\nu\in\cC$. 

\smallskip

\noindent $(i)$ One has 
$$\forall Q\in\mathcal{Q}_\Lambda,\qquad \cE_{\rm BDF}^\nu(Q)+\frac\alpha2 D(\nu,\nu)\geq0\,,$$
and therefore $\cE_{\rm BDF}^\nu$ is bounded from below on $\mathcal{Q}_\Lambda$.

\smallskip

\noindent $(ii)$ If moreover $\nu=0$, then $\cE_{\rm BDF}^0$ is non-negative on $\mathcal{Q}_\Lambda$, $0$ being its unique minimizer.
\end{theorem} 

The boundedness from below of the BDF energy is an essential feature of the theory. It shows the usefulness of the 
inclusion of the vacuum effects in the model. The interpretation of $(ii)$ is the following: by \eqref{formal_BDF}, it proves that the free vacuum $\cP^0_-$ is the unique minimizer of the (formal) QED energy in the set of all the projectors $P$ which are such that $P-\cP^0_-\in\mathcal{Q}_\Lambda$. In the previous subsection (Theorem \ref{free_vac}), it was also proved that $\cP^0_-$ is the unique minimizer of the energy per unit volume $\mathcal{T}$. These are two different ways of giving a mathematical meaning to the fact that $\cP^0_-$ is the unique minimizer of the QED energy when no external field is present.

The case $\nu=0$, $(ii)$ in Theorem \ref{BDF_bound_below},  was proved by Bach, Barbaroux, Helffer and Siedentop \cite{Bach-Barbaroux-Helffer-Siedentop-99}. In this paper, the authors also study a relativistic model, but with vacuum polarization neglected (see Section \ref{ghgh} below). They were inspired of a paper by Chaix, Iracane and Lions \cite{Chaix-Iracane-Lions-89}. Then, it has been argued in \cite{HLS1, HLS2, HLSo} that the proof of the case $\nu\ne 0$,  $(i)$ in Theorem \ref{BDF_bound_below}, is a trivial adaptation of  \cite{Bach-Barbaroux-Helffer-Siedentop-99}.

\medskip

Now that $\cE_{\rm BDF}^\nu$ has been shown to be bounded-below, it is natural to try to minimize it. Actually, we shall be interested in two minimization problems. The first is the global minimization of $\cE_{\rm BDF}^\nu$ in the whole set $\mathcal{Q}_\Lambda$. As mentioned above, a global minimizer of the mean-field QED energy $\cE_{\rm QED}^\nu$ and therefore of the BDF energy $\cE_{\rm BDF}^\nu$ (they formally differ by an infinite constant !) is interpreted as the \emph{polarized vacuum} in the external electrostatic field $-\alpha\nu\ast|\cdot|^{-1}$. If one wants to describe a system of charge $-eN$, one has to minimize $\cE_{\rm BDF}^\nu$ in the $N$th charge sector:
$$\mathcal{Q}_\Lambda(N):=\left\{Q\in\mathcal{Q}_\Lambda\ |\ \tr_{\cP^0_-}(Q)=N\right\}.$$
These two minimization problems will be tackled in the following two subsections.

Let us also notice that the boundedness from below of the BDF energy \eqref{BDF1} has been used by Hainzl, Lewin and Sparber in \cite{HLSp} to prove the existence of \emph{global-in-time} solutions to the time-dependent nonlinear equation associated with the BDF functional:
\begin{equation}
i\frac{d}{dt}{P}(t)=[\cD_{(P(t)-\cP^0_-)}\, , \, P(t)]
\label{time_BDF}
\end{equation}
where $[\cdot,\cdot]$ is the usual commutator and
$$\cD_{Q}:=\cD^0+\alpha(\rho_{Q}-\nu)\ast\frac{1}{|\cdot|}-\alpha\frac{Q(x,y)}{|x-y|}$$
(indeed, like in \cite{HLS1,HLS2}, the model for which $\cP^0_-$ and $\cD^0$ are replaced by $P^0_-$ and $D_1$ was considered, but the proof holds similarly in the case of \eqref{time_BDF}). Global solutions are shown to exist for \emph{any} initial orthogonal projector
$$P(0)\in\cP^0_- + \gS_2(\gH_\Lambda).$$
For nonlinear Dirac theories, it is usually quite difficult to prove the existence of global solutions (see the comments in Section \ref{time_dep}). Here, the proof is highly simplified by the important property that the BDF energy $\cE_{\rm BDF}^\nu$, which is conserved along solutions of \eqref{time_BDF}, is bounded-below and coercive.

\medskip

Let us now give the proof of Theorem \ref{BDF_bound_below}, which is a simple adaptation of arguments of \cite{Bach-Barbaroux-Helffer-Siedentop-99}.

\medskip

\noindent{\bf Proof of Theorem \ref{BDF_bound_below}.} Let be $Q\in\mathcal{Q}_\Lambda$, and therefore satisfying the operator inequality 
\begin{equation}
-\cP^0_-\leq Q\leq \cP^0_+.
\label{inegalite_Q}
\end{equation}
It is easily proved that \eqref{inegalite_Q} is equivalent to
\begin{equation}
Q^2\leq Q^{++}-Q^{--}
\label{inegalite_Q2}
\end{equation}
where $Q^{++}:=\cP^0_+Q\cP^0_+$ and $Q^{--}:=\cP^0_-Q\cP^0_-$. This now implies that
\begin{equation}
0\leq \tr(|\cD^0|Q^2)\leq \tr_{\cP^0_-}(\cD^0Q)
\label{inegalite_Q3}
\end{equation}
which shows that the kinetic energy is non-negative. We now use Kato's inequality \eqref{kato-ineg} and Equation  \eqref{propD3} to obtain
$$\iint_{\R^3\times\R^3}\frac{|Q(x,y)|^2}{|x-y|}dx\,dy\leq \frac\pi2\, \tr(|D_1|Q^2)\leq \frac\pi2\, \tr(|\cD^0|Q^2).$$
Together with \eqref{inegalite_Q3}, this implies
$$\cE^\nu_{\rm BDF}(Q)\geq \left(1-\alpha\frac\pi4 \right)\tr(|\cD^0|Q^2) -\frac\alpha2 D(\nu,\nu),$$
which easily ends the proof of Theorem \ref{BDF_bound_below}.\hfill$\square$\\

\subsection{Global minimization of $\cE_{\rm BDF}^\nu$: the polarized vacuum.} The existence of a global minimizer of $\cE_{\rm BDF}^\nu$ has been proved by Hainzl, Lewin and Séré, first in \cite{HLS1} by a fixed-point argument valid only when $\alpha\sqrt{\log\Lambda}\leq C_1$ and $\alpha D(\nu,\nu)^{1/2}\leq C_2$, and then by a global minimization procedure in \cite{HLS2}, valid for any cut-off $\Lambda$ and $0\leq\alpha<4/\pi$. The precise statement of the latter is the following:
\begin{theorem}[Definition of the polarized vacuum \cite{HLS1,HLS2,HLSo}]\label{BDF_min} Assume that $0\leq\alpha<4/\pi$, $\Lambda>0$ and that $\nu\in\cC$. Then $\cE_{\rm BDF}^\nu$ possesses a minimizer $\bar Q$ on $\mathcal{Q}_\Lambda$ such that $\cP=\bar Q+\cP_-^0$ is an orthogonal projector satisfying the self-consistent equation
\begin{equation}
\cP  =  \chi_{(-\infty, 0)}\left(\cD_{\bar Q}\right),\label{scf_Q_vide1}
\end{equation}
\begin{eqnarray}
\cD_{\bar Q} & = &  \cD^0+\alpha\left(\rho_{\bar Q}-\nu\right)\ast\frac{1}{|\cdot|}-\alpha\frac{\bar Q(x,y)}{|x-y|}\label{scf_Q_vide2}\\
 & = & D_1+\alpha\left(\rho_{[\bar \cP_- - 1/2]}-\nu\right)\ast\frac{1}{|\cdot|}-\alpha\frac{(\bar \cP_- - 1/2)(x,y)}{|x-y|}.\label{scf_Q_vide3}
\end{eqnarray} 
Additionally, if $\alpha$ and $\nu$ satisfy
\begin{equation}
0\leq\alpha\frac\pi4\left\{1-\alpha\left(\frac\pi2 \sqrt{\frac{\alpha/2}{1-\alpha\pi/4}}+\pi^{1/6}2^{11/6}\right)D(\nu,\nu)^{1/2}\right\}^{-1}\leq1,
\label{condition_uniqueness}
\end{equation} 
then this global minimizer $\bar Q$ is unique and the associated polarized vacuum is neutral, i.e. $\bar Q\in\mathcal{Q}_\Lambda(0)$:
\begin{equation}
\tr_{\cP^0_-}(\bar Q)=\tr_{\cP^0_-}(\bar \cP_--\cP^0_-)=0.
\label{neutral}
\end{equation}
\end{theorem}

The proof consists in showing that $\cE_{\rm BDF}^\nu$ is lower semi-continuous for the weak-$\ast$ topology of $\mathcal{Q}_\Lambda$. For this purpose, one shows that, in the electron-positron field, any mass escaping to infinity takes away a positive energy. This is the so-called dichotomy case of the concentration-compactness principle \cite{Lions-84}. To prove \eqref{neutral}, one first shows that $\tr_{\cP^0_-}(\bar \cP_--\cP^0_-)$ is always an integer, then one applies a continuation argument.

Notice that  the definition \eqref{scf_proj} of $\cD^0$ has been used  to obtain \eqref{scf_Q_vide3} from \eqref{scf_Q_vide2}. Of course, equations \eqref{scf_Q_vide1} and \eqref{scf_Q_vide3} are exactly the one we wanted to solve in the beginning \eqref{formal_eq_vac}. For not too strong external densities $\nu$, a neutral vacuum is necessarily obtained, as shown by \eqref{neutral}. But in general, a charged polarized vacuum could be found.

\smallskip

\begin{figure}[h]
\input{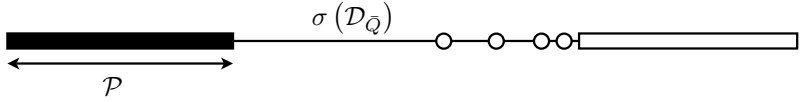}
\caption{The polarized vacuum $\cP$ in the presence of the external density $\nu$ is the negative spectral projector of the mean-field operator $\cD_{\bar Q}$.}
\end{figure}

\smallskip

In \cite{HLSo}, a thermodynamic limit was considered as for the free case $\nu=0$, to justify the formal computation \eqref{formal_BDF2} when $\nu\neq0$. As before, the QED energy is well-defined in a box $\cC_L=[-L/2, L/2)^3$ with periodic boundary conditions and a cut-off in Fourier space by
\begin{equation}
\label{energy_box2}
\mathcal{E}_L^\nu(\Gamma):=\mathcal{E}_L^0(\Gamma)-\alpha\iint_{\cC_L\times\cC_L}W_L(x-y)\rho_\Gamma(x)\nu_L(y)\, dx\,dy,
\end{equation}
where
\begin{equation}
\nu_L(x)=\frac{(2\pi)^{3/2}}{L^3}\sum_{k\in(2\pi\Z^3)/L}\widehat{\nu}(k)e^{ik\cdot x},
\label{def_n_L}
\end{equation} 
(for simplicity, it is assumed that $\widehat \nu$ is a smooth function). Then the following was proved in \cite{HLSo}.
\begin{theorem}[Thermodynamic limit with external field \cite{HLSo}]\label{thermo2} Assume that $0\leq\alpha<4/\pi$, $\Lambda>0$, $\nu\in\mathcal{C}$ and that $\widehat{\nu}$ is continuous on $B(0,\Lambda)$. Then for any $L$, $\cE_L^\nu$ possesses a minimizer $\Gamma_L=\cP_L-I_\Lambda^L/2$ on $\mathcal{G}_\Lambda^L$ where $\cP_L$ is an orthogonal projector, and one has 
\begin{equation}
\lim_{L\to\infty}\left\{\cE_L^\nu(\Gamma_L)-\cE_L^0(\Gamma^0_L)\right\} =\min\left\{\cE_{\rm BDF}^\nu(Q),\ Q\in\mathcal{Q}_\Lambda\right\}.
\label{limit_thermo2}
\end{equation} 
Moreover, up to a subsequence, $Q_L(x,y):=(\Gamma_L-\Gamma_L^0)(x,y)=(\cP_L-\cP^0_L)(x,y)$ converges uniformly on compact subsets of $\R^6$ to $\bar Q(x,y)$, a minimizer of $\cE_{\rm BDF}^\nu$ on $\mathcal{Q}_\Lambda$. 
\end{theorem} 

\medskip

Since all the previous results hold for any fixed $\Lambda$, it would be natural to consider the limit $\Lambda\to\infty$. This was done in \cite{HLS2}  where it is argued that this limit is highly unphysical: the vacuum polarization density \emph{totally cancels the external density $\nu$}. In QED, this ``nullification" of the theory as the cut-off $\Lambda$ diverges was first suggested by Landau \cite{Lan84} and later studied by Pomeranchuk {\it et al.} \cite{Pom}.

Notice that in QED, the procedure of renormalization is often used to formally remove the cut-off $\Lambda$ and the divergence of the theory. It consists in assuming that the parameter $\alpha$ is not the physical one but the \emph{bare} one. The physical $\alpha_{\rm phys}\simeq 1/137$ is related to $\alpha$ by a formula of the form
$$\alpha_{\rm phys}\simeq\frac{\alpha}{1+2\alpha\log\Lambda/(3\pi)}.$$
When $\alpha_{\rm phys}$ is fixed at its physical value, then necessarily $\Lambda\leq 10^{280}$, meaning that the large $\Lambda$ limit should not be considered in principle.
In the case where the exchange term is neglected in the energy (the so-called \emph{reduced} BDF model), this renormalization procedure has been studied in detail in \cite{HLS2}.

\subsection{Minimization of $\cE_{\rm BDF}^\nu$ in charge sectors} The previous subsection was devoted to the global minimization of the BDF energy. We now mention some results that have been obtained in \cite{HLS3} for the minimization with a charge constraint. 
It is believed that the charge constrained BDF model can be obtained as the thermodynamical limit of the full QED model in a fixed charge sector and posed in a box with periodic boundary conditions, but this has not been shown yet.

Due to the charge constraint and like for the Hartree-Fock model for instance, minimizers will not always exist for the BDF functional: it depends whether the external electrostatic potential created by the charge distribution $\nu$ is strong enough to be able to bind the $N$ particles in the presence of the Dirac sea. On the other hand, it must not be too strong otherwise electron-positron pairs could appear. 
 
 We start with a general result proved in \cite{HLS3} providing the form of a minimizer, if it exists. To this end, we introduce the minimum energy in the $N$th charge sector:
\begin{equation}
E^\nu(N):= \inf\left\{\cE^\nu_{\rm BDF}(Q)\ |\ Q\in\mathcal{Q}_\Lambda,\ \tr_{\cP^0_-}(Q)=N\right\}.
\label{def_min_BDF}
\end{equation}
In principle $N$ could be any real number, but here, for simplicity, we shall restrict ourselves to integers. 
\begin{theorem}[Self-Consistent Equation of a BDF Minimizer \cite{HLS3}] Let be $0\leq\alpha<4/\pi$, $\Lambda>0$, $\nu\in\cC$ and $N\in\Z$. Then any minimizer $Q$ solution of the minimization problem \eqref{def_min_BDF}, if it exists, takes the form $Q=P-\cP^0_-$ where
\begin{equation}
P=\chi_{(-\infty, \mu]}(\cD_Q)=\chi_{(-\infty, \mu]}\left(\cD^0+\alpha(\rho_Q-\nu)\ast1/|\cdot|-\alpha\frac{Q(x,y)}{|x-y|}\right),
\label{scf_eq_min_BDF}
\end{equation}
for some $\mu\in[-m(\alpha),m(\alpha)]$.
\end{theorem}

Recall that $m(\alpha)$ is the threshold of the free operator $\cD^0$ defined in \eqref{def_threshold}. We remark that \eqref{scf_eq_min_BDF} implicitly means that the last eigenvalue below $\mu$ of the mean-field operator $\cD_Q$ is necessarily totally filled. As already mentioned in the Dirac-Fock case, this is a general fact for Hartree-Fock type theories \cite{Bach-Lieb-Loss-Solovej-94}.
For a minimizer of the form \eqref{scf_eq_min_BDF} and when $N,\mu>0$, it is natural to consider the decomposition
$$P=\Pi+\chi_{[0\,, \,\mu]}(\cD_Q),$$
where $\Pi$ is the polarized Dirac sea:
$$\Pi:= \chi_{(-\infty\,, \,0)}(\cD_Q).$$

For not too strong external potentials, the vacuum will be neutral, {\it i.e.}
$$\tr_{\cP^0_-}(\Pi-\cP^0_-)=0\,,$$
and therefore $\chi_{[0\,, \,\mu]}(\cD_Q)$ will be a projector of rank $N$:
$$\chi_{[0, \mu]}(\cD_Q)=\sum_{n=1}^N\phi_n\otimes\phi_n^*,\quad \cD_Q\phi_n=\varepsilon_n\phi_n\,,$$
where $\varepsilon_1\leq\cdots\leq\varepsilon_N$ are the first $N$ positive eigenvalues of $\cD_Q$ counted with their multiplicity. Notice that
\begin{multline}
\label{dec_mean_field}
\cD_Q=D_1+\alpha(\rho_\Phi-\nu)\ast\frac{1}{|\cdot|}-\alpha\frac{\gamma_\Phi(x,y)}{|x-y|}\\
+\alpha\rho_{[\Pi-I_\Lambda/2]}\ast\frac{1}{|\cdot|}-\alpha\frac{(\Pi-I_\Lambda/2)(x,y)}{|x-y|},
\end{multline}
where
$$\gamma_\Phi:=\chi_{[0, \mu]}(\cD_Q)=\sum_{n=1}^N\phi_n\otimes\phi_n^*,\qquad \rho_\Phi(x):=\tr_{\C^4}(\gamma_\Phi(x,x))=\sum_{n=1}^N|\phi_n(x)|^2.$$
In the first line of \eqref{dec_mean_field}, the Dirac-Fock operator associated with $(\phi_1,...,\phi_N)$ appears, see \eqref{def_DF_mean_field_op}. This shows that the electronic orbitals $\phi_i$ are solutions of a Dirac-Fock type equation in which the mean-field operator $\cD_{\Phi}$ is perturbed by the (self-consistent) potentials of the Dirac sea $\Pi-I_\Lambda/2$. In practice, these potentials are small, and the DF equations are a good approximation of the BDF equations for the electronic orbitals. But the energy functionals behave in a completely different way: as we have seen, the DF energy is strongly indefinite while the BDF energy is bounded below. The Dirac-Fock model is thus interpreted as a \emph{non-variational} approximation of the mean-field model of no-photon QED \cite{Chaix-Iracane-89}.\smallskip

\begin{figure}[h]
\input{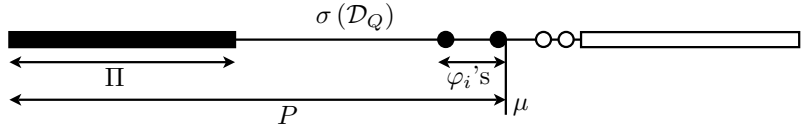}
\caption{Decomposition of the system `vacuum + $N$ electrons' for the solution $P=\Pi+\gamma_\Phi$ in the $N$th charge sector.}
\end{figure}

\smallskip

Concerning the existence of a minimizer, solution of \eqref{scf_eq_min_BDF}, the following result was proved in \cite{HLS3}: 
\begin{theorem}[Binding Conditions and Existence of a BDF Minimizer \cite{HLS3}]\label{HVZ} Let be $0\leq\alpha<4/\pi$, $\Lambda>0$, $\nu\in\cC$ and $N\in\Z$. Then the following two assertions are equivalent:

\medskip

\noindent $(H_1)$ \qquad $\displaystyle E^\nu(N) < \min\left\{ E^\nu(N-K)+E^0(K),\ K\in\Z\setminus\{0\}\right\}.$

\medskip

\noindent $(H_2)$ {Each minimizing sequence $(Q_n)_{n\geq1}$ for $E^\nu(N)$ is precompact in $\mathcal{Q}_\Lambda$ and converges, up to a subsequence, to a minimizer $Q$ of $E^\nu(N)$. }
\end{theorem}

Conditions like $(H_1)$ appear classically when analyzing the compactness properties of minimizing sequences, for instance by using the concentration-compactness principle of P.-L. Lions \cite{Lions-84}. They are also very classical for linear models in which the bottom of the essential spectrum has the form of the minimum in the right hand side of $(H_1)$, as expressed by the HVZ Theorem \cite{H,V,Zhislin-60}. 
Assume $N>0$ for simplicity. When $0<K\leq N$, $(H_1)$ means that it is not favorable to let $K$ electrons escape to infinity, while keeping $N-K$ electrons near the nuclei. When $K<0$, it means that it is not favorable to let $\vert K\vert$ positrons escape to infinity, while keeping $N+\vert K\vert$ electrons near the nuclei. When $K>N$, it means that it is not favorable to let $K$ electrons escape to infinity, while keeping $K-N$ positrons near the nuclei. 
When $\alpha$ is small enough and $N>0$, it was shown in \cite{HLS3} that the separation of electron-positron pairs is not energetically favorable, so that one just needs to check $(H_1)$ for $K=1,2,...,N$.

To prove the existence of a minimizer, one can therefore prove that $(H_1)$ holds. Two situations in which $(H_1)$ is true have been provided by Hainzl, Lewin and Séré in \cite{HLS3}. The first one is the case of weak coupling $\alpha\ll1$ and $\alpha\nu=\bar\nu$ fixed (the charge $N$ is also fixed). The following was proved:

\begin{theorem}[Existence of a minimizer in the weak coupling limit \cite{HLS3}]
\label{exists_weak_limit} Assume that $\Lambda>0$, that $N$ is a non-negative integer, and that $\bar\nu\in\cC$ is such that 
\begin{enumerate}
\item the spectrum $\sigma(D^0-\bar\nu\ast|\cdot|^{-1})$ contains at least $N$ positive eigenvalues below 1,
\item ${\rm ker}(D^0-t\bar\nu\ast|\cdot|^{-1})=\{0\}$ for any $t\in[0, 1]$.
\end{enumerate}
Then $(H_1)$ holds in Theorem \ref{HVZ} for $\alpha$ small enough and $\alpha\nu=\bar\nu$, and therefore there exists a minimizer $Q_\alpha$ of $E^{\bar\nu/\alpha}(N)$. 
It takes the form
\begin{equation}
Q_\alpha=\chi_{(-\ii,0]}\left(\cD_{Q_\alpha}\right)-\cP^0_-+\chi_{(0,\mu_\alpha]}\left(\cD_{Q_\alpha}\right):=Q_\alpha^{\rm vac}+\sum_{i=1}^N|\phi_i^\alpha\rangle\langle\phi_i^\alpha|
\label{form_Q_weak_coupling}
\end{equation}
\begin{equation}
\cD_{Q_\alpha}\phi_i^\alpha=\varepsilon_i^\alpha\phi_i^\alpha
\label{form_Q_weak_coupling2}
\end{equation}
where $\varepsilon_1^\alpha\leq\cdots\leq \varepsilon_N^\alpha$ are the $N$ first positive eigenvalues of $\cD_{Q_\alpha}$.
Finally, for any sequence $\alpha_n\to0$, $(\phi_1^{\alpha_n},...,\phi_N^{\alpha_n})$ converges (up to a subsequence) in $\gH_\Lambda$ to $(\phi_1,...,\phi_N)$ which are $N$ first eigenfunctions of $D^0-\bar\nu\ast|\cdot|^{-1}$ and $Q_{\alpha_n}^{\rm vac}$ converges to $\chi_{(-\ii;0)}\left(D^0-\bar\nu\ast|\cdot|^{-1}\right)-P^0_-$ in $\gS_2(\gH_\Lambda)$.
\end{theorem}

The second situation provided in \cite{HLS3} is the case of the non-relativistic regime $c\gg1$. To state the result correctly, we reintroduce the speed of light $c$ in the model (of course, we shall then take $\alpha=1$). The expression of the energy and the definition of the free vacuum $\cP^0_-$ (which of course then depends on $c$ and the ultraviolet cut-off $\Lambda$) are straightforward. We denote by $E^\nu_{\alpha,c,\Lambda}(N)$ the minimum energy of the BDF functional depending on the parameters $(\alpha,c,\Lambda)$. The following was proved:

\begin{theorem}[Existence of a minimizer in the non-relativistic limit \cite{HLS3}] Assume that $\alpha=1$ and that the ultraviolet cut-off is $\Lambda=\Lambda_0 c$ for some fixed $\Lambda_0$. Let be $\nu\in\cC\cap L^1(\R^3,\R^+)$ with $\int_{\R^3}\nu=Z$, and $N$ a positive integer which is such that $Z>N-1$. Then, for $c$ large enough, $(H_1)$ holds in Theorem \ref{HVZ} and therefore there exists a minimizer $Q_c$ for $E^\nu_{1,c,\Lambda_0 c}(N)$. It takes the following form:
$$Q_c= \chi_{(-\infty, 0)}(\cD_Q)-\cP^0_-+\chi_{[0, \mu_c)}(\cD_Q)=Q_c^{\rm vac} +\sum_{i=1}^N|\phi_i^c\rangle\langle\phi_i^c| \,,$$
and one has
$$\lim_{c\to\infty}\left\{E^\nu_{1,c,\Lambda_0 c}(N)-Ng_0(0)\right\}= \min_{\substack{\bar\Phi\in H^1(\R^3,\C^2)^N\\{\rm Gram}\;\bar\Phi=\un}}\mathcal{E}_{\rm HF}(\bar\Phi).$$

Moreover, for any sequence $c_n\to\infty$, $(\phi_1^{c_n},..., \phi_N^{c_n})$ converges  in $H^1(\R^3,\C^4)^N$ (up to a subsequence) towards $(\phi_1,...,\phi_N)$ with $\phi_i=\left(^{\bar\phi_i}_{0}\right)$, and where $\bar\Phi_0=(\bar\phi_1,...,\bar\phi_N)$ is a global minimizer of the Hartree-Fock energy.
\end{theorem}

We notice that this result is very similar to Theorem \ref{thmGS} providing the convergence of the Dirac-Fock `ground state' in the non-relativistic limit. 

\subsection{Neglecting Vacuum Polarization: Mittleman's conjecture}\label{ghgh}
In view of the complications introduced by the Dirac sea, some authors \cite{Bach-Barbaroux-Helffer-Siedentop-98,Bach-Barbaroux-Helffer-Siedentop-99,Barbaroux-Farkas-Helffer-Siedentop-05} considered an approximate model in which the vacuum polarization is neglected, in the spirit of a paper by Mittleman \cite{Mittleman-81}. In this subsection, we shall keep the speed of light $c$ and therefore take $\alpha=1$.

In the vacuum case, a possible way to describe Mittleman's approach is first to write that the global minimizer $\cP-1/2$ constructed in Theorem \ref{BDF_min} is formally a solution of the following tautological max-min principle:
\begin{equation}
e=\sup_{\substack{P\\ P^2=P}}\ \inf_{\substack{\gamma\\ -P\leq \gamma\leq 1-P}}\left\{\cE_{\rm QED}^\nu(P+\gamma)-\cE_{\rm QED}^\nu(P)\right\}.
\label{Tauto_vac}
\end{equation}
Indeed taking $\gamma=0$, one finds that $e\leq0$. Saying that $\cP$ is a global QED minimizer exactly means that
$$\min_{\substack{\gamma\\ -\cP\leq \gamma\leq 1-\cP}}\left\{\cE_{\rm QED}^\nu(\cP+\gamma)-\cE_{\rm QED}^\nu(\cP)\right\}=0.$$
Then, the idea is to approximate \eqref{Tauto_vac} by neglecting the vacuum polarization terms coming from $P$. One has formally
\begin{multline}
\cE_{\rm QED}^\nu(P+\gamma)-\cE_{\rm QED}^\nu(P)= \tr(D^P\gamma)-  \iint_{\R^3\times\R^3}\frac{\rho_{\gamma}(x)\nu(y)}{|x-y|}dx\,dy\\
  +\frac{1}{2}\iint_{\R^3\times\R^3}\frac{\rho_{\gamma}(x)\rho_\gamma(y)}{|x-y|}dx\,dy-
  \frac{1}{2}\iint_{\R^3\times\R^3}\frac{|\gamma(x,y)|^2}{|x-y|}dx\,dy
\end{multline}
with
$$D^P:=D_c+\rho_{[P-1/2]}\ast\frac{1}{|\cdot|}-\frac{(P-1/2)(x,y)}{|x-y|}.$$
Neglecting the vacuum polarization potentials then simply amounts to replacing $D^P$ by the free Dirac operator $D_c$. The following max-min principle was studied by Bach, Barbaroux, Helffer and Siedentop in \cite{Bach-Barbaroux-Helffer-Siedentop-99}, inspired by Mittleman \cite{Mittleman-81}:
\begin{equation}
e_{\rm Mitt}^{\nu,c}=\sup_{\substack{P,\\ P^2=P=P^*}}\ \inf_{\substack{\gamma\in\gS_1(\gH_\Lambda),\\ -P\leq \gamma\leq 1-P}}\cE^{\nu,c}_P(\gamma)
\label{Mitt_vac}
\end{equation}
where $\cE_P^{\nu,c}$ is defined as 
\begin{equation}
\cE^{\nu,c}_P(\gamma):=\tr(D_c\gamma)- D(\nu,\rho_\gamma)+\frac{1}{2}D(\rho_\gamma,\rho_\gamma)-\frac1 2\iint_{\R^6}\frac{|\gamma(x,y)|^2}{|x-y|}dx\,dy
\end{equation}
on the set depending on $P$:
$$\Gamma(P):= \{\gamma\in\gS_1(\gH),\ |\nabla|\gamma\in\gS_1(\gH),\ -P\leq \gamma\leq 1-P\}.$$
The advantage of this formulation is that, since the vacuum polarization has been neglected, no divergence problem is encountered and \eqref{Mitt_vac} can be studied without any Fourier cut-off (i.e. $\gH=H^{1/2}(\R^3,\C^4)$), and with pointwise external charges (i.e. $\nu=Z\delta_0$ where $\delta_0$ is the Dirac distribution at the point $0\in\R^3$). The following was proved in \cite{Bach-Barbaroux-Helffer-Siedentop-99}:
\begin{theorem}[Mittleman Principle for the Vacuum \cite{Bach-Barbaroux-Helffer-Siedentop-99}]
Assume that $\nu=Z\delta_0$ for some $Z\geq0$ and that $c>0$ satisfies $4c(1-2Z/c)/\pi\geq 1$. Then 
$$\bar P_c:=\chi_{(-\infty, 0)}(D_c-Z/|x|)$$
is the unique solution of Mittleman's max-min principle
\begin{equation}
e_{\rm Mitt}^{\nu,c}=\sup_{P\in\mathcal{S}}\ \inf_{\gamma\in\Gamma(P)}\cE^{\nu,c}_P(\gamma),
\label{Mitt_vac2}
\end{equation}
where $\mathcal{S}$ denotes the set of all the orthogonal projectors $P$ which are such that $P$ and $1-P$  leave the domain of $D_c-Z/|x|$ invariant. Moreover, $e^{\nu,c}_{\rm Mitt}=0$.
\end{theorem}
As a consequence, when vacuum polarization is neglected, the Dirac sea is represented by the negative spectral projector of the operator $D_c-Z/|x|$. 

The $N$ electron case was studied by Mittleman in \cite{Mittleman-81}. His main idea was to justify the validity of the Dirac-Fock approximation by a type of max-min principle from QED in which vacuum polarization is neglected. Let us introduce the following max-min principle:
\begin{equation}
e_{\rm Mitt}^{\nu,c}(N)=\sup_{P\in\mathcal{S}}\ \inf_{\substack{\gamma\in\Gamma(P)\\ \gamma P=P\gamma,\\ \tr(\gamma)=N}}\cE^{\nu,c}_P(\gamma).
\label{Mitt_mol}
\end{equation}
The interpretation is that $\gamma$ represents the $N$ electrons, whereas $P$ is the Dirac sea. Notice that in this interpretation, the real particles are artificially separated from the virtual electrons of the Dirac sea. However, it was  believed that \eqref{Mitt_mol} could be a simpler problem with interesting practical implications. Remark that compared to \eqref{Mitt_vac}, we have added in \eqref{Mitt_mol} the condition that $\gamma$ commutes with $P$, i.e. it cannot contain off-diagonal terms. Without this requirement, it is known \cite{Aschbacher} that (\ref{Mitt_mol})
cannot give any solution of the Dirac-Fock equations.

Mittleman's conjecture consists in saying that the max-min principle \eqref{Mitt_mol} is attained by a solution to the Dirac-Fock system. More precisely, we follow \cite{Barbaroux-Esteban-Sere-05} and state it as:

\medskip

\noindent \textbf{Mittleman's conjecture.} \textit{A solution of \eqref{Mitt_mol} is given by a pair
$$(P,\gamma)=(\chi_{(-\infty,0)}(D_{c,\Phi}),\gamma_\Phi)$$
where $\Phi$ is a solution of the Dirac-Fock equation $D_{c,\Phi}\phi_i=\varepsilon_i\phi_i$ with $\varepsilon_i>0$, and where $D_{c,\Phi}$ is the Dirac-Fock mean-field operator defined in \eqref{def_DF_mean_field_op}.}

\medskip

Mittleman's conjecture for molecules has been investigated in \cite{Barbaroux-Farkas-Helffer-Siedentop-05,Barbaroux-Esteban-Sere-05}.  In \cite{Barbaroux-Farkas-Helffer-Siedentop-05} Barbaroux, Farkas, Helffer and Siedentop have studied  the minimization problem on $\Gamma(P)$ in \eqref{Mitt_mol} for a fixed $P$. Under suitable conditions, they proved that $\gamma$ is an electronic density matrix $\gamma=\sum_{i=1}^N|\phi_i\rangle\langle\phi_i|$, where $(\phi_1,...,\phi_N)$ is a solution of the projected Dirac-Fock equations $(1-P)D_{c,\Phi}(1-P)\phi_i=\varepsilon_i\phi_i$. Then they further proved that if  the energy is stationary with respect to variations of the projector $P$, then the unprojected Dirac-Fock are obtained. But they were unable to prove the existence of such a state.

In \cite{Barbaroux-Esteban-Sere-05}, Barbaroux, Esteban and Séré investigated the validity of Mittleman's conjecture by a perturbation argument. Namely they added a parameter $\tau$ in front of the interaction terms as follows:
\begin{equation}
\cE^{\nu,c,\tau}_P(\gamma):=\tr(D_c\gamma)- D(\nu,\rho_\gamma)+\frac{\tau}{2}D(\rho_\gamma,\rho_\gamma)-\frac\tau 2\iint_{\R^6}\frac{|\gamma(x,y)|^2}{|x-y|}dx\,dy.
\end{equation}
Here $\nu$ is a fixed positive and smooth radial function with compact support and $\int_{\R^3}\nu=1$.
They studied Mittleman's conjecture in the regime $c\gg1$ and $\tau\ll1$. We emphasize that by a scaling argument this physically corresponds to assuming $\alpha\ll1$ and $Z\gg1$ with $\alpha Z\ll1$. We denote by $\cE_{\rm DF}^{\nu,c,\tau}$ the Dirac-Fock functional which is easily defined with these parameters, and by $\Phi^{c,\tau,0}$ the DF solution obtained by the corresponding versions of Theorems \ref{thmDF} and \ref{M} for $j=0$.

When $c>1$, it is known that $D_c-\nu\ast|\cdot|^{-1}$ is essentially self-adjoint on $L^2(\R^3,\C^4)$ and that its spectrum is as follows:
$$\sigma(D_c-\nu\ast|\cdot|^{-1})=(-\infty, -c^2]\cup\{\mu_1(c)<\mu_2(c)\cdots\}\cup [c^2, \infty),$$
where $\lim_{i\to\infty}\mu_i(c)=c^2$.
We denote by $(N_i(c))_{i\geq1}$ the multiplicities of the $\mu_i(c)$'s. The following was proved in \cite{Barbaroux-Esteban-Sere-05}:
\begin{theorem}[Validity and non validity of Mittleman's conjecture \cite{Barbaroux-Esteban-Sere-05}]\label{Mitt_bof} Assume that $\nu$ is a fixed positive and smooth radial function with compact support and such that $\int_{\R^3}\nu=1$.
 
If $N=\sum_{i=1}^IN_i(c)$ for some $I\geq1$ and some fixed $c>1$, then Mittleman's conjecture is \textbf{true} for $\tau$ small enough: one has
$$e_{\rm Mitt}^{\nu,c,\tau}(N)=\cE_{\rm DF}^{\nu,c,\tau}(\Phi^{c,\tau,0})\,,$$
where $\Phi^{c,\tau,0}$ is  any of the solutions obtained in Theorem \ref{thmDF} in the case  $j=0$. Moreover, the optimal projector for the $\sup$ part of \eqref{Mitt_mol} is $P=\chi_{(-\infty, 0)}(D_{c,\Phi^{c,\tau,0}})$. 
 
 If $N=\sum_{i=1}^IN_i(c)+1$ for some $I\geq1$ and $c$ large enough, then Mittleman's conjecture is \textbf{wrong} when $\tau>0$ is small enough: there is no solution $\Phi$ of the Dirac-Fock equations with positive multipliers such that the pair
$\left(\chi_{(-\infty,0)}(D_{c,\Phi}),\gamma_\Phi\right)$
realizes Mittleman's max-min principle \eqref{Mitt_mol}.
\end{theorem}

One can consider a weaker version of Mittleman's conjecture which consists in only comparing energy levels and not the solutions themselves.

\medskip

\noindent \textbf{Weaker Mittleman's conjecture.} \textit{One has}
$$e_{\rm Mitt}^{\nu,c,\tau}(N)=\cE_{\rm DF}^{\nu,c,\tau}(\Phi^{c,\tau,0}).$$

\medskip

When $N=1$ and for $c\gg1$, $\tau\ll1$, it is known by Theorem \ref{Mitt_bof} that Mittleman's conjecture is wrong. But Barbaroux, Helffer and Siedentop proved in \cite{Barbaroux-Helffer-Siedentop-06} that the weaker conjecture is indeed true. 

\begin{theorem}[Validity of the weaker Mittleman conjecture for $N=1$ \cite{Barbaroux-Helffer-Siedentop-06}]\label{Mitt_bofbof} Assume that $\nu=\delta_0$. Then for $c$ large enough and $\tau$ small enough, the weaker Mittleman conjecture is true:
$$e_{\rm Mitt}^{\nu,c,\tau}(N)=\cE_{\rm DF}^{\nu,c,\tau}(\Phi^{c,\tau,0}).$$
The sup-inf \eqref{Mitt_mol} is realized by the pair 
$$\left( \bar P_c, \gamma\right)\quad \text{with}\quad  \bar P_c=\chi_{(-\infty,0)}(D_c-1/|x|),\quad \gamma=|\phi^c\rangle\langle\phi^c|$$
where $\phi^c$ is any eigenvector of
\begin{equation}
 D_c-1/|x|
\label{egalite_MF_1_e}
\end{equation}
with eigenvalue $e_{\rm Mitt}^{\nu,c,\tau}(N)$.
\end{theorem}

In \cite{Barbaroux-Helffer-Siedentop-06}, an explicit condition on $\tau$ and $c$ is provided. Notice the equality 
$$D_{c,\phi}\phi^c=D_{c,0}\phi^c=(D_c-1/|x|)\phi^c=e_{\rm Mitt}^{\nu,c,\tau}(N)\phi^c$$
which is very specific to the one-electron case, and means that the electron does not ``see" its own Coulomb field. In particular $\bar P_c\phi^c=0$.

For $N\geq2$, the question remains completely open.



\end{document}